\documentclass[18pt, twocolumn, times, trackchanges]{aastex631}

\usepackage{color}
\usepackage{longtable}
\usepackage{threeparttable}
\usepackage{ulem}
\usepackage[nolist]{acronym}
\usepackage{array}
\usepackage{amsmath}
\usepackage{float}
\usepackage{graphicx}
\usepackage{natbib}
\usepackage{hyperref}
\usepackage{xspace}
\usepackage{lipsum} 
\usepackage{CJKutf8} 

\usepackage{tabularx} 
\usepackage{booktabs}
\usepackage{multirow}
\usepackage{booktabs} 

\hypersetup{linkcolor=red,citecolor=blue,filecolor=cyan,urlcolor=magenta}

\newcommand{\angstrom}{\ \textup{\AA}}

\newcolumntype{C}[1]{>{\centering\arraybackslash}p{#1}}

\newcommand{\pagefootlink}[2][*]{%
    \footnote[0]{\url{#2}}
    \textsuperscript{%
        \hyperlink{footnote.\the\value{footnote}}{#1}%
    }%
}

\newcounter{footlinkcounter}
\newcommand{\autofootlink}[1]{%
    \stepcounter{footlinkcounter}%
    \footnote[0]{\url{#1}}%
    \textsuperscript{%
        \hyperlink{footnote.\the\value{footnote}}{\thefootlinkcounter}%
    }%
}

\begin{acronym}
	\acro{LAMOST}{Large Sky Area Multi-Object Fiber Spectroscopic Telescope}
     \acro{VMP}{very metal-poor}
     \acro{EMP}{extremely metal-poor}
     \acro{GSE}{Gaia-Sausage-Enceladus}
     \acro{HDS}{High Dispersion Spectrograph}
     \acro{S/N}{signal-to-noise ratio}
     \acro{LTE}{local thermodynamic equilibrium}
     \acro{NLTE}{non-local thermodynamic equilibrium}
     \acro{EWs}{equivalent widths}
     \acro{EW}{equivalent width}
     \acro{TAME}{Tool for Automatic Measurement of Equivalent-width}
     \acro{$s$-process}{slow neutron-capture process}
     \acro{$r$-process}{rapid neutron-capture process}
     \acro{$i$-process}{intermediate neutron-capture process}
     \acro{CEMP}{carbon-enhanced metal-poor}
     \acro{LEPP}{light-element primary process}
     \acro{HFS}{hyperfine structure}
     \acro{IS}{isotopic shifts}
     \acro{HEW}{high-entropy wind}
     \acro{WP}{waiting point}
     \acro{AGB}{asymptotic giant branch}
     \acro{NSMs}{neutron star mergers}
     \acro{NSM}{neutron star merger}
     \acro{BC}{bolometric correction}
     \acro{MRDSNe}{magnetorotationally driven jet supernovae}
     \acro{MRDSN}{magnetorotationally driven jet supernova}
     \acro{CCSNe}{core-collapse supernovae}
     \acro{CCSN}{core-collapse supernova}
     \acro{MAD}{median absolute deviation}
     \acro{MADs}{median absolute deviations}
     \acro{AIC}{Akaike information criterion}
     \acro{GMM}{Gaussian Mixture Model}
     \acro{HR}{Hertzsprung-Russell}
     \acro{FDU}{first dredge-up}
     \acro{TDU}{third dredge-up}
     \acro{PIE}{proton ingestion event}
     \acro{TP}{thermal pulse}
     \acro{TPs}{thermal pulses}
\end{acronym}

\begin{document}

\title{Four-hundred Very Metal-poor Stars Studied with LAMOST and Subaru. IV. Heavy Elements Enhanced Stars}

\date{\today}

\correspondingauthor{Haining Li}
\email{lhn@nao.cas.cn}

\author[0009-0000-8769-3142]{Yangming Lin}
\affiliation{CAS Key Laboratory of Optical Astronomy, National Astronomical Observatories, Chinese Academy of Sciences\\ 
Beijing 100101, People's Republic of China}
\affiliation{School of Astronomy and Space Science, University of Chinese Academy of Sciences\\
No.19(A) Yuquan Road, Shijingshan District, Beijing 100049, People's Republic of China}

\author[0000-0002-0389-9264]{Haining Li}
\affiliation{CAS Key Laboratory of Optical Astronomy, National Astronomical Observatories, Chinese Academy of Sciences\\ 
Beijing 100101, People's Republic of China}

\author[0000-0002-8975-6829]{Wako Aoki}
\affiliation{National Astronomical Observatory of Japan, 2-21-1 Osawa, Mitaka, Tokyo 181-8588, Japan}
\affiliation{Astronomical Science Program, The Graduate University for Advanced Studies, SOKENDAI, 2-21-1 Osawa, Mitaka, Tokyo 181-8588, Japan}

\author[0009-0001-0604-072X]{Ruizheng Jiang}
\affiliation{CAS Key Laboratory of Optical Astronomy, National Astronomical Observatories, Chinese Academy of Sciences\\ 
Beijing 100101, People's Republic of China}
\affiliation{School of Astronomy and Space Science, University of Chinese Academy of Sciences\\
No.19(A) Yuquan Road, Shijingshan District, Beijing 100049, People's Republic of China}

\author[0000-0002-8077-4617]{Tadafumi Matsuno}
\affiliation{Astronomisches Rechen-Institut, Zentrum für Astronomie der Universität Heidelberg, Mönchhofstraße 12-14, 69120 Heidelberg, Germany}

\author[0009-0002-0124-9492]{Zhenyu He}
\affiliation{School of Physics, Peng Huanwu Collaborative Center for Research and Education, and International Research Center for Big-Bang \\
Cosmology and Element Genesis, Beihang University, Beijing 100191, China}
\affiliation{ShenYuan Honors College, Beihang University, Beijing 100191, China}

\author[0009-0008-1319-1084]{Ruizhi Zhang}
\affiliation{CAS Key Laboratory of Optical Astronomy, National Astronomical Observatories, Chinese Academy of Sciences\\ 
Beijing 100101, People's Republic of China}

\author[0000-0002-1126-9289]{Zhuohan Li}
\affiliation{School of Space Science and Technology, Shandong University, Weihai 264209, People’s Republic of China}

\author[0000-0001-6653-8741]{Satoshi Honda}
\affiliation{Nishi-Harima Astronomical Observatory, Center for Astronomy, University of Hyogo, 407-2,\\
 Nishigaichi, Sayo-cho, Sayo, Hyogo 679-5313, Japan}

\author[0000-0002-8980-945X]{Gang Zhao}
\affiliation{CAS Key Laboratory of Optical Astronomy, National Astronomical Observatories, Chinese Academy of Sciences\\ 
Beijing 100101, People's Republic of China}
\affiliation{School of Astronomy and Space Science, University of Chinese Academy of Sciences\\
No.19(A) Yuquan Road, Shijingshan District, Beijing 100049, People's Republic of China}

\begin{abstract}

We perform a homogeneous analysis of 82 heavy-element-enhanced metal-poor stars from the\\ LAMOST/Subaru sample, including 64 $r$-process-enhanced stars, 2 limited-$r$ stars, and 16 stars exhibiting significant or moderate carbon enhancement (comprising 9 $s$-process-enhanced, 2 $r$-process-enhanced, 2 $r$/$s$, and 3 $r$+$s$ stars). Systematic differences in $\mathrm{[X/Eu]}$ between $r$-I and $r$-II stars, together with their distinct dynamical properties, suggest that $r$-II stars likely formed in low-mass dwarf galaxies where contamination from additional nucleosynthetic channels was minimal. We have also discovered a number of peculiar objects. We identify an extremely metal-poor $r$-II star, J1158+0734 ($\mathrm{[Fe/H]}$ = -2.93), whose enhanced Zn abundance ($\mathrm{[Zn/Fe]}$ = +0.67) is best explained by yields from a high-energy, massive core-collapse supernova. The surface abundances of $s$-process-enhanced stars exhibit pronounced dispersion, consistent with enrichment from AGB progenitors with diverse properties. Notably, one star, J2256+0215, reveals compelling evidence for additional evolutionary mixing, as indicated by its unusually low carbon abundance ($\mathrm{[C/Fe]}$ = +0.08) and extremely low $^{12}$C/$^{13}$C ratio of 2.33. Additionally, we identify 3 $r$+$s$ stars whose abundance patterns reveal combined contributions from both the $r$- and $s$-processes. We further propose a modified diagnostic scheme  for the identification of CEMP-$r$+$s$ stars, defined as: -0.2 $\leqslant$ $\mathrm{[La/Nd]}$ $\leqslant$ 0.2 and -0.5 $\leqslant$ $\mathrm{[Eu/Nd]}$ $\leqslant$ 0.0.

\end{abstract}

\keywords{\href{http://astrothesaurus.org/uat/1577}{Stellar abundances (1577)}; \href{https://astrothesaurus.org/uat/1131}{Nucleosynthesis (1131)}; \href{http://astrothesaurus.org/uat/1324}{R-process (1324)}; \href{https://astrothesaurus.org/uat/1419}{S-process (1419)}; \href{https://astrothesaurus.org/uat/2105}{CEMP stars (2105)}; \href{https://astrothesaurus.org/uat/226}{Chemically peculiar stars (226)}; \href{http://astrothesaurus.org/uat/2178}{Galactic archaeology (2178)}; \href{http://astrothesaurus.org/uat/1060}{Milky Way stellar halo (1060)}}

\section{Introduction} \label{sec:intro}

Elements heavier than Fe are predominantly produced via neutron-capture processes. Although proton-capture ($p$-process) and neutrino-induced ($\nu$-process, $\nu p$-process) reactions can contribute to the formation of a small fraction of heavy isotopes in explosive environments, neutron capture remains the dominant nucleosynthetic mechanism for heavy elements. Based on the relative rates of neutron capture and $\beta$-decay, neutron-capture processes are classified into the \ac{$s$-process}, \ac{$i$-process}, and \ac{$r$-process}.

The \ac{$s$-process}, characterized by a neutron capture rate significantly slower than $\beta$-decay, operates at relatively low neutron densities ($N_n \approx 10^5 \sim 10^{10} \, \text{cm}^{-3}$) over extended timescales ($ 10^3 \sim 10^4 \, \text{years} $). This process predominantly occurs during the \ac{TP} phase of \ac{AGB} stars, where a weakly mixed convective boundary region forms between the He- and H-burning shells. Protons from the H-burning shell are transported into the He-intershell, where they are captured by $^{12}$C to form a thin $^{13}$C pocket. When temperatures reach $\sim 10^8 \, \text{K}$, the $^{13}\text{C}(\alpha, n)^{16}\text{O}$ reaction is activated, releasing free neutrons and driving \ac{$s$-process} nucleosynthesis \citep{1995ApJ...440L..85S, 1998ApJ...497..388G, 2010MNRAS.404.1529B}. 

In contrast, the \ac{$i$-process} exhibits a distinct neutron production mechanism. In low-metallicity \ac{AGB} stars during \ac{TPs}, convective overshooting can extend the mixing region to the base of the H-burning shell, triggering a \ac{PIE} \citep{1977ApJ...212..149C, 2016ApJ...831..171H, 2019ApJ...887...11H, 2022A&A...667A.155C, 2024A&A...684A.206C}. This leads to the reaction chain $^{12}\text{C}(p, \gamma)^{13}\text{N}(\beta^+)^{13}\text{C}$, resulting in neutron densities six orders of magnitude higher ($N_n \approx 10^{13} \sim 10^{16} \, \text{cm}^{-3}$) than those in the \ac{$s$-process}, with nucleosynthetic timescales as short as days to years. This fundamental difference enhances the production of heavy elements near and beyond the second peak (Z $\geq$ 56) \citep{2018ARNPS..68..237F}. Beyond low- to  intermediate mass \ac{AGB} stars, potential astrophysical sites for the \ac{$i$-process} include super-AGB stars \citep{2015MNRAS.446.2599D, 2016MNRAS.455.3848J} and rapidly accreting white dwarfs \citep{2017ApJ...834L..10D, 2019MNRAS.488.4258D}. 


The \ac{$r$-process} requires extreme neutron densities ($ N_n > 10^{20} \, \text{cm}^{-3} $) and proceeds on second-scale timescales, driving nucleosynthesis pathways far from the $\beta$-stability valley. Due to its extreme astrophysical environment, the astrophysical site of the \ac{$r$-process} has long been debated. The detection of the \ac{NSM} event GW170817 in 2017 and its associated kilonova AT2017gfo \citep{2017PhRvL.119p1101A, 2017ApJ...848L..13A, 2017LRR....20....3M, 2017PASJ...69..102T, 2017ApJ...851L..21V} provided the first direct evidence that NSMs are a viable \ac{$r$-process} site \citep{2017Natur.551...64A, 2017ApJ...848L..19C, 2023arXiv230209188F}. However, the long merger timescales ($ 10^7 \sim 10^8 \, \text{years}$; \citealp{2012ApJ...759...52D, 2018A&A...615A..91B, 2018MNRAS.474.2937C, 2019MNRAS.490.3740N}) of binary systems struggle to account for the observed \ac{$r$-process} enrichment in extremely metal-poor stars ($\rm \left [ \mathrm{Fe}/\mathrm{H} \right ] < $ -3; \citealp{2005ARA_A..43..531B}) in the Milky Way, suggesting the existence of prompt \ac{$r$-process} sources in the early Universe \citep{2023ApJ...943L..12K}, such as \ac{MRDSNe} \citep{2012ApJ...750L..22W, 2015ApJ...810..109N, 2017ApJ...836L..21N} or collapsar jets \citep{1999ApJ...524..262M,2019Natur.569..241S, 2024ApJ...966L..37H}. However, these astrophysical sites have not been conclusively confirmed by direct observational evidence. Therefore, the frequency of occurrence, yield variations of r-process nucleosynthesis in different astrophysical environments, and their relative importance across different cosmic epochs remain critically important open questions.

Low-mass \ac{VMP} stars ($\rm \left [ \mathrm{Fe}/\mathrm{H} \right ] < $ -2.0) serve as "fossil records" of early cosmic chemical evolution, having formed in interstellar gas clouds enriched by only one or a few nucleosynthetic events \citep{2005ARA_A..43..531B, 2015ARA&A..53..631F, 2018ARNPS..68..237F}. These Population II stars provide unique observational tracers for studying galactic chemical evolution and neutron-capture processes. Spectral screening techniques utilizing Ca II H and K line strengths at both low-resolution (R $\sim$ 400) and medium-resolution (R $\sim$ 2000) have demonstrated efficiency in identifying metal-poor stars ($\rm \left [ \mathrm{Fe}/\mathrm{H} \right ] < $ -1.0). The HK objective-prism survey \citep{1992AJ....103.1987B}, together with subsequent large-scale spectroscopic surveys --- including the Hamburg/ESO Survey \citep{2008A&A...484..721C}, RAVE \citep{2010ApJ...724L.104F}, SDSS \citep{2017arXiv171103234K}, SEGUE \citep{2009AJ....137.4377Y}, and the \ac{LAMOST} survey \citep{2006ChJAA...6..265Z, 2012RAA....12..723Z, 2012RAA....12..735D} --- have dramatically expanded the number of known metal-poor stars, resulting in hundreds of thousands of candidates. In parallel, wide-field photometric surveys, such as SkyMapper \citep{2007PASA...24....1K}, the Pristine survey based on narrow-band Ca H$\&$K photometry \citep{2017MNRAS.471.2587S, 2018MNRAS.481.3838S, 2024A&A...692A.115M}, and the Stellar Abundances and Galactic Evolution Survey (SAGES; \citealt{2023ApJS..268....9F, 2024ApJS..273...12H, 2025ApJS..277...19G}), have further improved the efficiency of identifying very and extremely metal-poor stars.

Utilizing such an extensive sample, distinct subclasses of heavy-element-enhanced stars have been identified. The seminal discovery of CS 22892–052 \citep{1994ApJ...431L..27S, 2003ApJ...591..936S} marked a pivotal advancement --- an \ac{EMP} star ($\left [ \mathrm{Fe}/\mathrm{H} \right ]  $ = -3.12) from the HK survey exhibiting extraordinary \ac{$r$-process} enhancement with $\left [ \mathrm{Eu}/\mathrm{Fe} \right ]  $ = 1.65 (approximately dozens of times higher than that in solar). Subsequent systematic investigations, notably the "First Stars" program \citep{2002A&A...387..560H} and HERES survey \citep{2004A&A...428.1027C, 2005A&A...439..129B}, have identified dozens of \ac{$r$-process}-enhanced stars through VLT/UVES spectroscopic observations. Based on their europium enrichment levels, these stars are classified into $r$-I and $r$-II stars \citep{2005ARA_A..43..531B, 2020ApJS..249...30H}, as summarized in Table \ref{tab:heavy_classification}. Recent studies have further identified \ac{$r$-process}-enhanced stars with $\left [ \mathrm{Eu}/\mathrm{Fe} \right ] > $ 2.0, named as $r$-III stars \citep{2020ApJ...898...40C, 2024ApJ...971..158R}. The origin of these europium abundance variations remains contentious --- whether they reflect birth environment differences \citep{2025ApJ...990..125H} or distinct \ac{$r$-process} nucleosynthesis sources.

Conversely, \citet{2006ApJ...643.1180H, 2007ApJ...666.1189H} discovered two chemically peculiar stars (HD 122563 and HD 88609), designated as "limited-$r$" stars, which exhibit pronounced enhancement in light neutron-capture elements ($\left [ \mathrm{Sr}/\mathrm{Ba} \right ] > $ 0.5, $\left [ \mathrm{Sr}/\mathrm{Eu} \right ] > $ 0.0) but depleted heavy \ac{$r$-process} signatures ($\left [ \mathrm{Eu}/\mathrm{Fe} \right ] < $ 0.3). Their abundance patterns suggest nucleosynthetic origins either from the weak \ac{$r$-process} \citep{2001ApJ...554..578W} or the \ac{LEPP} \citep{2004ApJ...601..864T}. The R-Process Alliance has significantly advanced this field through multi-telescope observations (including GTC/HORuS and Magellan/MIKE high-resolution follow-ups), increasing the confirmed sample to approximately 580 \ac{$r$-process}-enhanced stars and 50 limited-$r$ stars \citep{2018ApJ...858...92H,2018ApJ...868..110S,2020ApJ...898..150E,2020ApJS..249...30H,2024ApJS..274...39B}.

A substantial population of metal-poor stars in the Galactic halo exhibit carbon enrichment ($\rm \left [ \mathrm{C}/\mathrm{Fe} \right ] \geq $ 0.7, \citealt{2007ApJ...655..492A}), characterized by prominent molecular absorption features of CH, CN, and C$_2$ in their spectra. Recent studies of large samples indicate that the fraction of \ac{CEMP} stars increases significantly with decreasing metallicity. Specifically, the fraction rises to $\sim 40\%$ for $[\mathrm{Fe/H}] \lesssim -3$ and reaches $\sim 75\%$ for $[\mathrm{Fe/H}] \lesssim -4$ \citep{2013ApJ...762...27Y, 2013AJ....146..132L, 2013AJ....145...13A, 2014ApJ...797...21P, 2021MNRAS.505.1239A, 2022MNRAS.515.4082A, 2024BSRSL..93..406G}, underscoring the importance of understanding their origins. Based on their surface heavy-element abundance, \ac{CEMP} stars are further divided into distinct subclasses: \ac{CEMP}-no stars show no significant neutron-capture element enhancement, \ac{CEMP}-$s$ stars exhibit clear enhancement of \ac{$s$-process} elements, \ac{CEMP}-$r$ stars demonstrate pronounced \ac{$r$-process} elements enrichment, while CEMP-$r$/$s$ and CEMP-$r$+$s$ stars display concurrent enhancement of both $s$- and $r$-process elements. These different neutron-capture signatures imply different nucleosynthetic origins. Observational evidence suggests that \ac{CEMP}-$r$/$s$ stars predominantly follow \ac{$i$-process} nucleosynthesis patterns \citep{2016ApJ...831..171H, 2019ApJ...887...11H, 2022A&A...667A.155C, 2024A&A...684A.206C}, whereas \ac{CEMP}-$r$+$s$ stars are better explained by combined $r$- and $s$-process enrichment scenarios \citep{2018ApJ...862..174G, 2020A&A...641A.135S}. However, the classification criteria and astrophysical origins of these subclasses remain highly debated, particularly given the limited sample sizes --- only about 41 confirmed \ac{CEMP}-$r$/$s$ stars \citep{2021A_A...649A..49G} and two reported \ac{CEMP}-$r$+$s$ stars in the literature.



The investigation of heavy-element-enhanced stars necessitates large, homogeneously analyzed samples of metal-poor stars to resolve the key outstanding questions in this field. By analyzing the large collection of low-resolution spectroscopic data (R $\sim$ 1800) from the \ac{LAMOST} survey (DR1 through DR5), we identified over 15,000 \ac{VMP} star candidates through the methods of template matching of observed flux and line index matching \citep{2018ApJS..238...16L}. Between 2014 and 2017, we obtained high-resolution follow-up spectra (R = 36,000, $\lambda$ = 4000-6800 $\angstrom$) for 385 bright targets (g $<$ 13.0 mag) using the Subaru Telescope/\ac{HDS} \citep{2002PASJ...54..855N}, as reported in \citet{2022ApJ...931..146A}. Subsequent systematic analysis by \citet{2022ApJ...931..147L} determined atmospheric parameters and abundances of 22 species for this \ac{VMP} sample, while \citet{2024ApJ...966..174Z} investigated the chemo-dynamical properties of this sample through friends-of-friends clustering algorithms.
Building upon these foundations, this study focuses on metal-poor stars exhibiting heavy-element enhancements within this sample, aiming to provide new observational constraints on nucleosynthetic processes through detailed analysis of abundance patterns across distinct heavy-element-enhanced categories. This paper is structured as follows: Section \ref{sec:abundance} details the methodology for deriving the abundances of neutron-capture elements, Section \ref{sec:results} presents the discussion of the abundance results, including (1) the number of different heavy-element-enhanced stars based on established classification criteria, (2) comparison with literature values, and (3) the abundance characteristics of different types of heavy-element-enhanced stars. The conclusions and summary are presented in Section \ref{subsec: conclusion}.


\section{Data} \label{sec:abundance}

\begin{figure}[htbp] 
\centering
\includegraphics[width=0.5\textwidth,keepaspectratio]{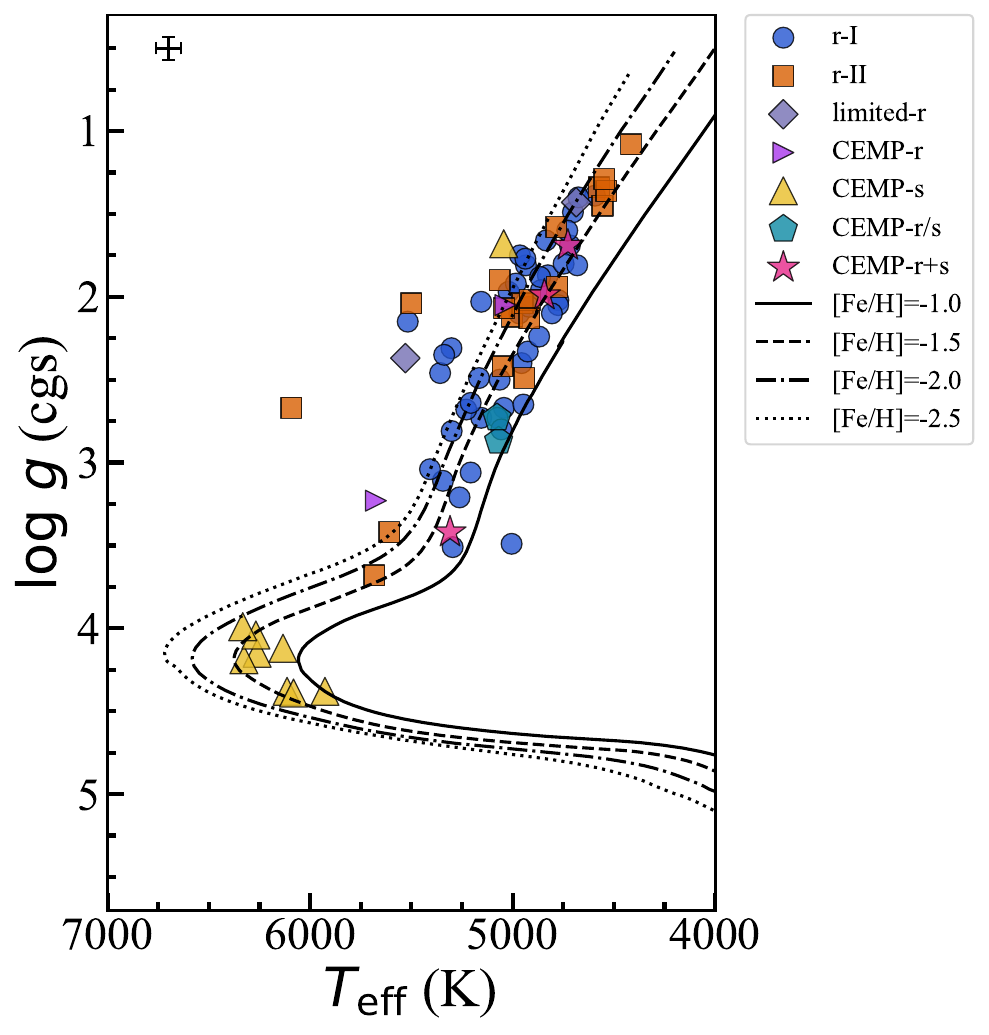}
\caption{The distribution of $T_{\mathrm{eff}}$ versus $\log g$ for the sample stars, with Dartmouth isochrones overplotted \citep{2008ApJS..178...89D}. The solid, dashed, dot-dashed, and dotted lines represent isochrones with an age of 12.5 Gyr and $\rm \left [ \mathrm{\alpha}/\mathrm{Fe} \right ]$ = 0.4, corresponding to metallicities $\rm \left [ \mathrm{Fe}/\mathrm{H} \right ]$ = -1.0, -1.5, -2.0, and -2.5, respectively. The different subclasses of heavy-element-enhanced stars are represented by distinct colors and symbols. The average 1$\sigma$ uncertainties in $T_{\mathrm{eff}}$ and $\log g$ for the sample stars are indicated by error bars in the upper-left corner of the figure.}
\label{fig:HR_diagram}
\end{figure}

The sample analyzed in this study is drawn from the \ac{LAMOST}/Subaru \ac{VMP} star sample \citep{2022ApJ...931..147L}, with the aim of investigating metal-poor stars exhibiting heavy-element enhancements. The sample selection process consists of two steps. First, an initial sample is identified based on the abundances reported by \citet{2022ApJ...931..147L}, following the classification criteria outlined in Table \ref{tab:heavy_classification}. Subsequently, the stellar classifications are confirmed based on the updated abundance data obtained in this work, thereby ensuring that all selected stars satisfy the criteria presented in Table \ref{tab:heavy_classification}. Table \ref{tab:atmospheric parameters} provides the final stellar sample along with the corresponding atmospheric parameters. The distribution of $T_{\rm eff}$ versus $\log g$ for the stellar sample is shown in Figure \ref{fig:HR_diagram}, and is overlaid with four 12.5 Gyr Dartmouth isochrones \citep{2008ApJS..178...89D} at enhanced $\alpha$-element abundance ($[\alpha/\mathrm{Fe}] = 0.4$), covering metallicities from $[\mathrm{Fe}/\mathrm{H}] = -2.50$ to $-1.00$ in steps of 0.5 dex. A more detailed discussion of the stellar classification is provided in Section \ref{subsec:classification of the sample star}.

\begin{table*}[t]
\scriptsize
\centering
\caption{Classification criteria of heavy-element-enhanced stars}
\renewcommand{\arraystretch}{1.4} 
\begin{tabular}{cccc}
\hline\hline
Type & Definition & $N$ & Reference \\
\hline
$\mathrm{r-I}$ & $0.3 \leq [\mathrm{Eu}/\mathrm{Fe}] \leq 0.7$ and $[\mathrm{Ba}/\mathrm{Eu}] < 0.0$ 
& 45 & \cite{2005ARA_A..43..531B, 2020ApJS..249...30H} \\
\hline
$\mathrm{r-II}$ & $[\mathrm{Eu}/\mathrm{Fe}] > 0.7$ and $[\mathrm{Ba}/\mathrm{Eu}] < 0.0$ 
& 19 & \cite{2005ARA_A..43..531B, 2020ApJS..249...30H} \\
\hline
$\mathrm{limited-r}$ & $[\mathrm{Eu}/\mathrm{Fe}] < 0.3$, $[\mathrm{Sr}/\mathrm{Ba}] > 0.5$, and $[\mathrm{Sr}/\mathrm{Eu}] > 0.0$ 
& 2 & \cite{2018ARNPS..68..237F} \\
\hline
$\mathrm{CEMP-r}$ & $[\mathrm{C}/\mathrm{Fe}] \geq 0.7$, $[\mathrm{Eu}/\mathrm{Fe}] > 0.7$, and $[\mathrm{Ba}/\mathrm{Eu}] < 0.0$ 
& 1 & \cite{2005ARA_A..43..531B,2007ApJ...655..492A, 2020ApJS..249...30H} \\
\hline
\multirow{3}{*}{$\mathrm{CEMP-s}$} 
& $[\mathrm{C}/\mathrm{Fe}] \geq 0.7$ and $[\mathrm{Ba}/\mathrm{Fe}] \geq 1.0$ 
& \multirow{3}{*}{8} & \multirow{3}{*}{\cite{2021A_A...649A..49G}} \\
& (1) $[\mathrm{Eu}/\mathrm{Fe}] < 1.0$, $[\mathrm{Ba}/\mathrm{Eu}] > 0.0$ and/or $[\mathrm{La}/\mathrm{Eu}] > 0.5$ & & \\
& (2) $[\mathrm{Eu}/\mathrm{Fe}] \geq 1.0$, $[\mathrm{Ba}/\mathrm{Eu}] > 1.0$ and/or $[\mathrm{La}/\mathrm{Eu}] > 0.7$ & & \\
\hline
\multirow{2}{*}{$\mathrm{CEMP-r/s}$} 
& $[\mathrm{C}/\mathrm{Fe}] \geq 0.7$, $[\mathrm{Ba}/\mathrm{Fe}] \geq 1.0$, and $[\mathrm{Eu}/\mathrm{Fe}] \geq 1.0$ 
& \multirow{2}{*}{1} & \multirow{2}{*}{\cite{2021A_A...649A..49G}} \\
& (1) $0.0 \leq [\mathrm{Ba}/\mathrm{Eu}] \leq 1.0$ and/or $0.0 \leq [\mathrm{La}/\mathrm{Eu}] \leq 0.7$ & & \\
\hline
$\mathrm{CEMP-r+s}$ & $[\mathrm{C}/\mathrm{Fe}] \geq 0.7$, $-0.2 \leq [\mathrm{La}/\mathrm{Nd}] \leq 0.2$, and $-0.5 \leq [\mathrm{Eu}/\mathrm{Nd}] \leq 0.0$ 
& 2 & This study \\
\hline
$\mathrm{CEMP^{*}}$\textsuperscript{a} & $0.4 < [\mathrm{C}/\mathrm{Fe}] < 0.7$ & 4 & ... \\
\hline
\end{tabular}
\begin{tablenotes}
\item Note. 
\textsuperscript{a}Stars within our sample exhibit moderate carbon enhancement but display pronounced enrichment in heavy elements.
\end{tablenotes}
\label{tab:heavy_classification}
\end{table*}

\begin{table*}[t]
\scriptsize
\centering
\caption{Atmospheric parameters of the sample stars}
\label{tab:atmospheric parameters}
\begin{tabular}{cccccccccc}
\hline
\hline
Name & 
$[\mathrm{Fe}/\mathrm{H}]$ & 
$T_{\rm eff}$ & 
$\log g$ & 
$\xi_{\rm t}$ &
$\sigma_{[\mathrm{Fe}/\mathrm{H}]}$ & 
$\sigma_{T_{\rm eff}}$ & 
$\sigma_{\log g}$ & 
$\sigma_{\xi_{\rm t}}$ &
Class \\
 & (dex) & (K) & (dex) & (km\,s$^{-1}$) & (dex) & (K) & (dex) & (km\,s$^{-1}$) \\
\hline
J0003+1556 & $ -2.32 $ & $ 5302 $ & $ 2.31 $ & $ 2.11 $ & $ 0.11 $ & $ 45 $ & $ 0.08 $ & $ 0.12 $ &  $r$-I \\
J0006+0123 & $ -2.27 $ & $ 5675 $ & $ 3.23 $ & $ 1.52 $ & $ 0.11 $ & $ 53 $ & $ 0.04 $ & $ 0.08 $ & $ r$-I  \\
J0040+2729 & $ -2.81 $ & $ 4574 $ & $ 1.34 $ & $ 2.16 $ & $ 0.11 $ & $ 78 $ & $ 0.07 $ & $ 0.07 $ & $ r$-II  \\
J0041+3633 & $ -2.37 $ & $ 5155 $ & $ 2.03 $ & $ 2.05 $ & $ 0.11 $ & $ 74 $ & $ 0.06 $ & $ 0.06 $ & $ r$-I  \\
J0055+1857 & $ -2.34 $ & $ 5043 $ & $ 2.67 $ & $ 1.50 $ & $ 0.09 $ & $ 27 $ & $ 0.04 $ & $ 0.06 $ & $ r$-I  \\
J0119-0121 & $ -3.19 $ & $ 6262 $ & $ 4.15 $ & $ 1.69 $ & $ 0.16 $ & $ 164 $ & $ 0.06 $ & $ 0.30 $ & CEMP-$s$ \\
\hline
\end{tabular} 
\begin{tablenotes}
\item Note. (This table is available in its entirety in machine-readable form.)
\end{tablenotes} 
\end{table*}

The abundances of neutron capture elements in our sample stars are derived from \ac{EW} measurements or spectral synthesis, with the results presented in Table \ref{tab:element abundances}. These abundances are expressed in three forms: absolute abundances, denoted as $\log\epsilon(\mathrm{X}) = \log_{10}(N_{\mathrm{X}}/N_{\mathrm{H}}) $ + 12, and relative abundances, represented as $\rm \left [ \mathrm{X}/\mathrm{H} \right ] = \log_{10}(N_{\mathrm{X}}/N_{\mathrm{H}}) - \log_{10}(N_{\mathrm{X}}/N_{\mathrm{H}})_{\odot} $ and $\rm \left [ \mathrm{X}/\mathrm{Fe} \right ] = \log_{10}(N_{\mathrm{X}}/N_{\mathrm{Fe}}) - \log_{10}(N_{\mathrm{X}}/N_{\mathrm{Fe}})_{\odot} $.  The line list utilized in this study is compiled from numerous literature sources \citep{2013A&A...558A..36C, 2014A&A...569A..43M, 2017A&A...607A..91H, 2022ApJ...931..147L}, with atomic parameters updated using the VALD database \citep{2015PhyS...90e4005R, 2019ARep...63.1010P}. \ac{EW} measurements are performed using the \ac{TAME} \citep{2012MNRAS.425.3162K}, which applies theoretical Gaussian/Voigt profile fitting to normalized observed spectra. The abundances are derived using an updated version of the abundance analysis code MOOG \citep{1973PhDT.......180S}, incorporating Rayleigh scattering as coherent and isotropic scattering \citep{2011AJ....141..175S}. Specifically, the abundances are derived from the \ac{EW} measurements using the MOOG "abfind" driver. In cases where an element has fewer than three measurable spectral lines or when the standard deviation of abundances from multiple lines exceeds 0.15, spectral synthesis is performed using the MOOG "synth" driver. The analysis employs 1D plane-parallel, hydrostatic model atmospheres from the $\alpha$-enhanced ATLAS NEWODF grid \citep{2003IAUS..210P.A20C}, assuming \ac{LTE} and excluding convective overshooting effects. The molecular and blended atomic line lists used in spectral synthesis are generated with the LINEMAKE code \citep{2021RNAAS...5...92P}, an open-source line list generator. We verified that the atomic parameters in these lists are consistent with the VALD database. Subsequent test calculations confirmed that any discrepancies yield abundance differences that fall well within our quoted uncertainties.

\begin{table*}[t]
\scriptsize
\centering
\caption{Elemental abundances of the sample stars}
\begin{tabular}{cccccccccc}
\hline
\hline
Species & 
$\log\epsilon_\odot $ & 
$\log \epsilon $ & 
$ \rm \left [\mathrm{X}/\mathrm{H} \right ]$ & 
$\rm \left [\mathrm{X}/\mathrm{Fe} \right ]$ &
$ \sigma_{stat} $ &
$ \sigma_{sys} $ &
$ \sigma_{tot} $ &
$ \sigma_{\rm \left [\mathrm{X}/\mathrm{Fe} \right ]} $ &
$N$ \\
\hline
\multicolumn{10}{c}{J0003+1556} \\  
\hline
$\rm \mathrm{Sr}\ \uppercase\expandafter{\romannumeral 2} $ & $ 2.87 $ & $ 1.29 $ & $ -1.58 $ & $ 0.65 $ & $ 0.11 $ & $ 0.02 $ & $ 0.11 $ & $ 0.16 $ & $ 1 $  \\
$\rm \mathrm{Y}\ \uppercase\expandafter{\romannumeral 2}  $ & $ 2.21 $ & $ -0.15$ & $ -2.36 $ & $-0.13 $ & $ 0.04 $ & $ 0.04 $ & $ 0.06 $ & $ 0.12 $ & $ 7$   \\
$\rm \mathrm{Zr}\ \uppercase\expandafter{\romannumeral 2} $ & $ 2.58 $ & $ 0.80 $ & $ -1.78 $ & $ 0.45 $ & $ 0.08 $ & $ 0.04 $ & $ 0.09 $ & $ 0.14 $ & $ 2$   \\
$\rm \mathrm{Ba}\ \uppercase\expandafter{\romannumeral 2} $ & $ 2.18 $ & $ 0.16 $ & $ -2.02 $ & $ 0.21 $ & $ 0.07 $ & $ 0.07 $ & $ 0.10 $ & $ 0.15 $ & $ 5 $  \\
$\rm \mathrm{La}\ \uppercase\expandafter{\romannumeral 2} $ & $ 1.10 $ & $ -0.76$ & $ -1.86 $ & $ 0.37 $ & $ 0.08 $ & $ 0.04 $ & $ 0.09 $ & $ 0.14 $ & $ 2$   \\
$\rm \mathrm{Ce}\ \uppercase\expandafter{\romannumeral 2} $ & $ 1.58 $ & $ -0.34$ & $ -1.92 $ & $ 0.31 $ & $ 0.11 $ & $ 0.04 $ & $ 0.12 $ & $ 0.16 $ & $ 1$   \\
$\rm \mathrm{Pr}\ \uppercase\expandafter{\romannumeral 2} $ & $ 0.72 $ & $ ...$ & $ ... $ & $ ... $ & $ ... $ & $ ... $ & $ ... $ & $ ... $ & $ ... $  \\
$\rm \mathrm{Nd}\ \uppercase\expandafter{\romannumeral 2} $ & $ 1.42 $ & $ -0.26$ & $ -1.68 $ & $ 0.55 $ & $ 0.05 $ & $ 0.04 $ & $ 0.06 $ & $ 0.13 $ & $ 6$   \\
$\rm \mathrm{Sm}\ \uppercase\expandafter{\romannumeral 2} $ & $ 0.96 $ & $ -0.67$ & $ -1.63 $ & $ 0.60 $ & $ 0.08 $ & $ 0.04 $ & $ 0.09 $ & $ 0.14 $ & $ 2$   \\
$\rm \mathrm{Eu}\ \uppercase\expandafter{\romannumeral 2} $ & $ 0.52 $ & $ -1.06$ & $ -1.58 $ & $ 0.65 $ & $ 0.08 $ & $ 0.06 $ & $ 0.10 $ & $ 0.15 $ & $ 2 $  \\
$\rm \mathrm{Gd}\ \uppercase\expandafter{\romannumeral 2} $ & $ 1.07 $ & $ ...$ & $ ... $ & $ ... $ & $ ... $ & $ ... $ & $ ... $ & $ ... $ & $ ...$   \\
$\rm \mathrm{Dy}\ \uppercase\expandafter{\romannumeral 2} $ & $ 1.10 $ & $ -0.45$ & $ -1.55 $ & $ 0.68 $ & $ 0.11 $ & $ 0.05 $ & $ 0.12 $ & $ 0.16 $ & $ 1 $   \\
$\rm \mathrm{Ho}\ \uppercase\expandafter{\romannumeral 2} $ & $ 0.48 $ & $ ...$ & $ ... $ & $ ... $ & $ ... $ & $ ... $ & $ ... $ & $ ... $ & $ ... $  \\
$\rm \mathrm{Hf}\ \uppercase\expandafter{\romannumeral 2} $ & $ 0.85 $ & $ ...$ & $ ... $ & $ ... $ & $ ... $ & $ ... $ & $ ... $ & $ ... $ & $ ... $  \\
$\rm \mathrm{Os}\ \uppercase\expandafter{\romannumeral 1} $ & $ 1.40 $ & $ ...$ & $ ... $ & $ ... $ & $ ... $ & $ ... $ & $ ... $ & $ ... $ & $ ... $  \\
$\rm \mathrm{Pb}\ \uppercase\expandafter{\romannumeral 1} $ & $ 1.75 $ & $ ... $ & $ ... $ & $ ... $ & $ ... $ & $ ... $ & $ ... $ & $ ... $ & $ ... $  \\
\hline
\end{tabular} 
\begin{tablenotes}
\item Note. 
(This table is available in its entirety in machine-readable form.)
\end{tablenotes} 
\label{tab:element abundances}
\end{table*}




\subsection{Elemental abundances} \label{subsec:neutron-capture elements}

\begin{figure*}[t]
\centering
\includegraphics[width=0.9\textwidth,keepaspectratio]{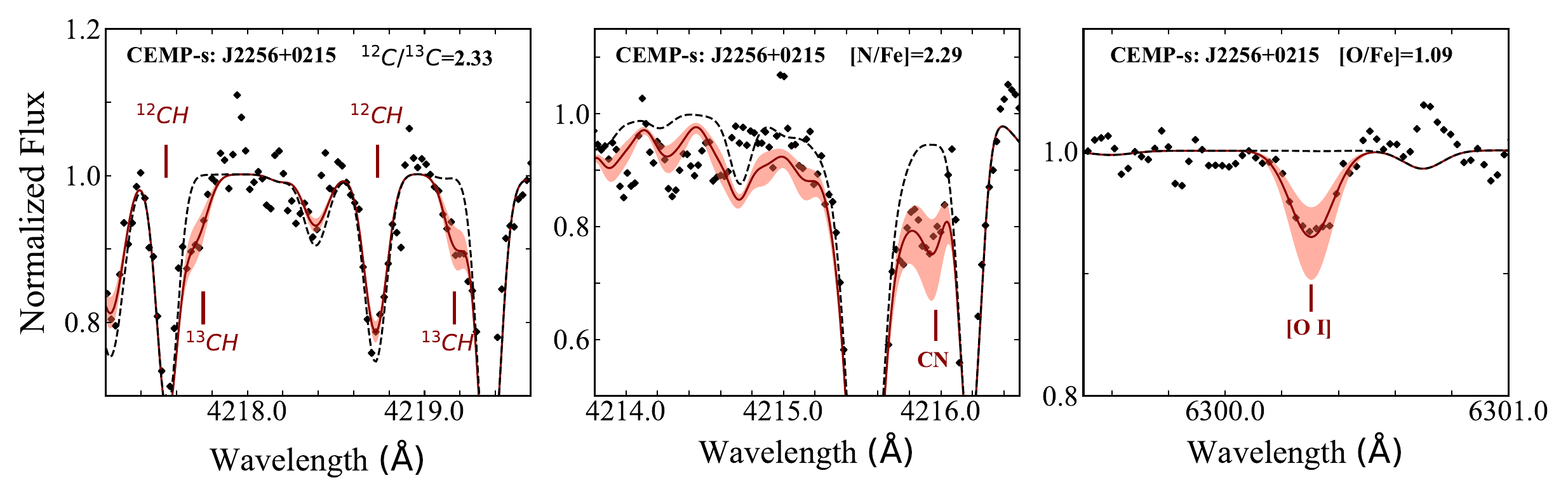}
\caption{Spectral synthesis of the $^{12}$CH and $^{13}$CH around 4215 $\angstrom$, CN $\lambda$4215, and [O I] $\lambda$6300 features for the \ac{CEMP}-$s$ star J2256+0215. The observed spectra are shown as black data points. The best-fits are represented by solid red lines. For the $^{12}$CH and $^{13}$CH spectral synthesis, the red shaded regions indicate $^{12}$C/$^{13}$C isotope ratios of 80/20 and 60/40, respectively. In other panels, the red shaded areas correspond to the $\pm$0.2 dex uncertainties associated with the best-fit. Spectral synthesis calculations excluding the elements of interest are shown as black dashed lines.
\label{fig:J2256+0215_spectral_synthesis}}
\end{figure*}

\citet{2022ApJ...931..147L} reported the stellar atmospheric parameters and light-element abundances for the stars in our sample, and we therefore directly adopt their published values. However, the surface chemical composition of stars undergoes significant variation during stellar evolution. Owing to the CNO cycle, \ac{FDU} mixing, and additional mixing along the red giant branch, sub-giant and red giant branch stars typically exhibit carbon depletion and nitrogen enhancement \citep{1995ApJ...453L..41C, 2000A&A...354..169G, 2014ApJ...797...21P}. This effect is particularly critical for the identification and analysis of \ac{CEMP} stars, requiring evolutionary corrections to their carbon abundances in order to ensure accurate classification \citep{2014ApJ...797...21P}. To properly account for these evolutionary effects, we additionally measure the abundances of key diagnostic elements in \ac{CEMP} stars: nitrogen abundances derived from the CN molecular band at 4215 $\angstrom$, oxygen abundances from the forbidden [O I] lines at 6300 $\angstrom$ and 6363 $\angstrom$, and carbon isotopic ratios determined from $^{12}$CH and $^{13}$CH molecular features in the 4217–4230 $\angstrom$ spectral region. All results are presented in Table \ref{tab:CEMP_stars}. In addition, Figure \ref{fig:J2256+0215_spectral_synthesis} illustrates the spectral synthesis of these elements for the $s$-process enhanced star J2256+0215.

Furthermore, the effects of \ac{HFS} and \ac{IS} on the abundance determination are taken into account for Sr, Ba, La, Pr, Eu, and Pb. In the calculations of synthetic spectra, the isotope ratios for different elements are taken from \citet{2020MNRAS.491.1832P}. Specifically, the solar \ac{$r$-process} isotope ratios are adopted for \ac{$r$-process}-enhanced metal-poor stars and limited-r stars, while solar \ac{$s$-process} isotope ratios are applied to \ac{CEMP}-$s$ stars. Since the solar heavy-element abundance pattern represents the combined outcome of both $r$- and \ac{$s$-process} nucleosynthesis, we adopt solar isotopic ratios for \ac{CEMP}-$r$+$s$ stars and \ac{CEMP}-$r$/$s$ stars that display simultaneous enhancements from both neutron-capture processes. The solar reference abundances are taken from \citet{2009ARA&A..47..481A}. In the following, we describe in detail the derivation of neutron-capture element abundances for our sample and the spectral lines used in the analysis.

$Strontium$, $yttrium$, and $zirconium$. The abundance of Sr is derived from the $\mathrm{Sr} \;\rm \uppercase\expandafter{\romannumeral 2}$ lines at 4077, 4161 and 4215 $\angstrom$, with \ac{HFS} and \ac{IS} adopted from \citet{1983HyInt..15..177B} and \citet{1990PhRvC..41.2883B, 1990PhRvC..42.2754B}.The resonance lines $\mathrm{Sr} \;\rm \uppercase\expandafter{\romannumeral 2}$ 4077 and 4215 are detectable in all sample stars, typically yielding Sr abundances with standard deviations below 0.1 dex when derived from these two lines. In contrast, the $\mathrm{Sr} \;\rm \uppercase\expandafter{\romannumeral 2}$ 4161 transition is significantly weaker and is only detectable in three stars within our sample. For Y, most stars exhibit detectable absorption in at least five  $\mathrm{Y} \;\rm \uppercase\expandafter{\romannumeral 2}$ lines, and we directly adopt the abundance results derived from \ac{EW} measurements. The Zr abundance is determined from 1 to 6 $\mathrm{Zr} \;\rm \uppercase\expandafter{\romannumeral 2}$ lines per star, with most stars exhibiting consistent abundances ($\sigma \sim$ 0.15 dex) across different $\mathrm{Zr} \;\rm \uppercase\expandafter{\romannumeral 2}$ transitions.

$Barium$, $lanthanum$, and $cerium$. The abundance of Ba is derived from five $\mathrm{Ba} \;\rm \uppercase\expandafter{\romannumeral 2}$ lines, with the \ac{HFS} in spectrum synthesis referenced from \citet{1988ZPhyA.329..407W}, \citet{1985ZPhyA.321..215V}, and \citet{1993JPhB...26.4289V}. 
The Ba II lines at $\lambda$4554 and $\lambda$4934 exhibit greater saturation effects and show stronger sensitivity to isotopic ratios assumed in the calculation compared to the lines at $\lambda$5853, $\lambda$6141, $\lambda$6496. When using solar \ac{$r$-process} isotope ratios, the Ba abundances derived from these two lines are systematically lower by more than 0.2 dex compared to those obtained with solar \ac{$s$-process} isotope ratio. The La abundance is derived through spectral synthesis of 12 La II lines, incorporating \ac{HFS} constants from \citet{2001ApJ...556..452L}, \citet{2008JPhB...41u5004F}, and \citet{2008JPhB...41w5002F}. In contrast to $\mathrm{Ba} \;\rm \uppercase\expandafter{\romannumeral 2}$ transitions, $\mathrm{La} \;\rm \uppercase\expandafter{\romannumeral 2}$ lines show relatively weaker absorption features and are dominated by the $^{139}\mathrm{La}$ isotope in solar material. For Ce, we detect an average of six Ce II lines per star in our sample and derive the abundances through \ac{EW} measurements.

$Praseodymium$, $neodymium$, and $samarium$. The Pr abundance is derived from spectral synthesis of nine $\mathrm{Pr} \;\rm \uppercase\expandafter{\romannumeral 2}$ lines, with particular attention to \ac{HFS} effects \citep{1989PhyS...39..694G, 2001PhyS...64..455I, 2002CaJPh..80..557R} and potential line blending. Within our sample, the $\mathrm{Pr} \;\rm \uppercase\expandafter{\romannumeral 2}$ line at 4179 $\angstrom$ demonstrates the highest detection rate, followed by $\mathrm{Pr} \;\rm \uppercase\expandafter{\romannumeral 2}$ transitions at 4143 $\angstrom$ and 4222 $\angstrom$. However, the $\mathrm{Pr} \;\rm \uppercase\expandafter{\romannumeral 2}$ 4179 feature shows significant blending with Cr II and V I, as well as with Nd II on its red wing. The line list includes 31 $\mathrm{Nd\ II}$ and 15 $\mathrm{Sm\ II}$ transitions. On average, about eight $\mathrm{Nd\ II}$ lines and four $\mathrm{Sm\ II}$ lines are detected per star. In most cases, the standard deviation among the abundances derived from different lines remains small ($\sigma < $ 0.15 dex).

\begin{figure*}[t]
\centering
\includegraphics[width=0.9\textwidth,keepaspectratio]{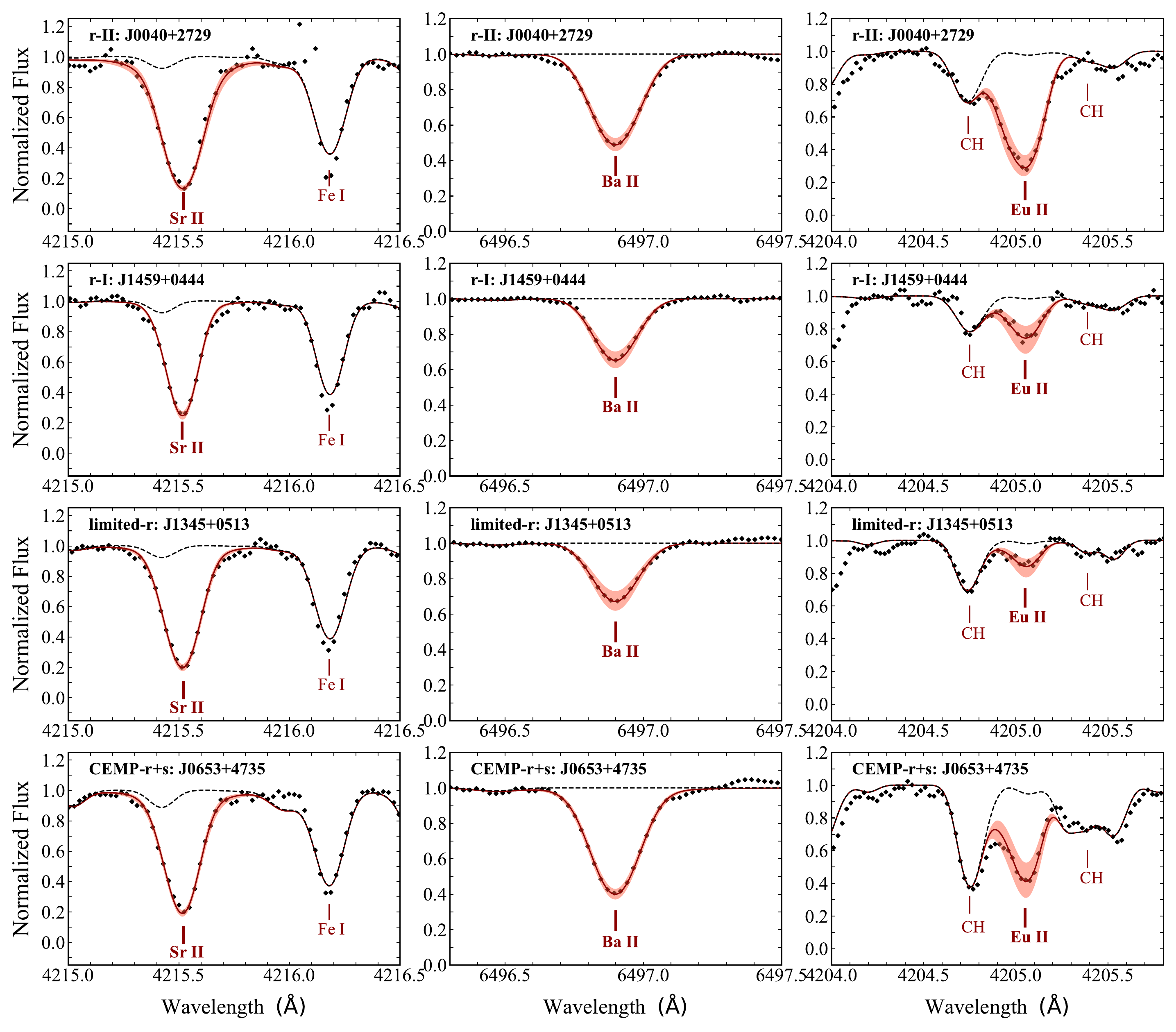}
\caption{Spectral synthesis of the Sr II $\lambda$4215, Ba II $\lambda$6496, and Eu II $\lambda$4205 features for the $r$-II star J0040+2729, the $r$-I star J1459+0444, the limited-$r$ star J1345+0513, and (d) the CEMP-$r$+$s$ star J0653+4735. The observed spectra are represented by black data points. The best fits are shown as solid red lines, with associated uncertainties of 0.2 dex indicated by the shaded red regions. The spectral synthesis excluding the elements of interest are displayed as black dashed lines.
\label{fig:spectral synthesis}}
\end{figure*}

$Europium$, $gadolinium$, $dysprosium$, $holmium$ and $hafnium$. As Eu is a representative \ac{$r$-process} element, 95 $\%$ of its abundance in solar material originates from the \ac{$r$-process}. We derived the Eu abundance through spectral synthesis from $\mathrm{Eu} \;\rm \uppercase\expandafter{\romannumeral 2}$ 4129 $\angstrom$ and 4205 $\angstrom$ lines, with \ac{HFS} and \ac{IS} adopted from \citet{2001ApJ...563.1075L}. The $\mathrm{Eu\ II}$ 4129 $\angstrom$ line is blended with $\mathrm{Sc\ I}$ 4129.752 $\angstrom$, $\mathrm{Ti\ II}$ 4129.724 $\angstrom$, $\mathrm{Fe\ II}$ 4129.692 $\angstrom$, as well as CN and CH molecular features. However, under metal-poor conditions, the impact of these blends on the abundance analysis is negligible. Notably, the left wing of $\mathrm{Eu} \;\rm \uppercase\expandafter{\romannumeral 2}$ 4205 line is significantly blended with a CH molecular band, as shown in Figure \ref{fig:spectral synthesis}, requiring careful evaluation when incorporating this feature into the final abundance determination. The Gd abundance is derived from six $\mathrm{Gd} \;\rm \uppercase\expandafter{\romannumeral 2}$ lines, with spectral synthesis employed when necessary to ensure the reliability of the abundance determination. The Dy abundance is derived from three $\mathrm{Dy} \;\rm \uppercase\expandafter{\romannumeral 2}$ lines at 4073 $\angstrom$, 4077 $\angstrom$, and 4103 $\angstrom$. The $\lambda$4077 line appears on the right wing of $\mathrm{Sr} \;\rm \uppercase\expandafter{\romannumeral 2}$ 4077 $\angstrom$, while the $\lambda$4103 line is located redward of the H$\delta$ line at 4101.76 $\angstrom$. Although the strong H$\delta$ absorption affects the local continuum level near this line, spectral synthesis yields abundances consistent with those derived from the Dy II 4077 $\angstrom$ line within the measurement uncertainties (0.15 dex). Within our sample, weak absorption features are detected for $\mathrm{Ho} \;\rm \uppercase\expandafter{\romannumeral 2}$ $\lambda$4152 in one star and $\mathrm{Hf} \;\rm \uppercase\expandafter{\romannumeral 2}$ $\lambda$4093 in six stars.


$Osmium$ and $lead$. As the third peak of the \ac{$r$-process}, the abundance of Os is derived through three lines: $\mathrm{Os} \;\rm \uppercase\expandafter{\romannumeral 1}$ $\lambda$4135, $\lambda$4260, and $\lambda$4420. The left wing of the Os I $\lambda$4260 is affected by $\mathrm{Fe} \;\rm \uppercase\expandafter{\romannumeral 1}$ $\lambda$4260, while the Os I $\lambda$4420 line shows significant blending with Sm, Nd and CH molecular features. The Pb abundance is derived from the $\mathrm{Pb} \;\rm \uppercase\expandafter{\romannumeral 1}$ $\lambda$4057 line, incorporating \ac{HFS} and \ac{IS} effects \citep{2000PhRvA..63a2516B}. For \ac{CEMP}-s stars, the \ac{$s$-process} isotopic ratio of Pb at low-metallicity is adopted \citep{2003A&A...404..291V}, as the enhanced neutron-to-seed ratio in metal-poor environments preferentially produces heavier elements through the \ac{$s$-process} nucleosynthesis pathway.

For the actinides (such as Th and U), no corresponding absorption features are detected in our sample. The strongest absorption lines of Th and U are located at $\mathrm{Th\ II}$ 4019.129 $\angstrom$ and $\mathrm{U\ II}$ 3859.571 $\angstrom$, respectively, which are not covered in our spectra. Detecting these features requires high-resolution, high signal-to-noise, blue-end spectra \citep{2023ApJ...948..122S, 2025ApJ...984L..43L}.


\subsection{Abundance Uncertainties} \label{subsec:abundance Uncertainties}
The total uncertainties ($\sigma_{\mathrm{tot}}$) in elemental abundances determinations arise from two primary sources: statistical errors ($\sigma_{\mathrm{stat}}$) associated with spectral measurements and systematic errors ($\sigma_{\mathrm{sys}}$) related to the determination of  atmospheric parameters. These distinct uncertainty components are presented separately in columns (6)-(8) of Table \ref{tab:element abundances}. The $\sigma_{\mathrm{tot}}$ is determined through quadrature summation of $\sigma_{\mathrm{stat}}$ and $\sigma_{\mathrm{sys}}$, expressed as: 

\begin{equation}
\label{eqn:sigma_tot}
\sigma_{\text{tot}} = \sqrt{\sigma_{\text{stat}}^2 + \sigma_{\text{sys}}^2}. 
\end{equation}

The $\sigma_{\mathrm{stat}}$ component primarily incorporates errors originating from \ac{EW} measurements, continuum normalization, and line blending effects. For a given element $X$, $\sigma_{\mathrm{stat}}$ is quantitatively estimated as:

\begin{equation}
\footnotesize
\label{eqn:sigma_stat}
\sigma_{\mathrm{stat}} = 
\begin{cases} 
\dfrac{\sigma_X}{\sqrt{N}}, & N \geq 4 \\[10pt]
\max\left( \dfrac{\sigma_X}{\sqrt{N}}, \dfrac{\sigma_{\mathrm{Fe\,\textrm{I}}}}{\sqrt{N}} \right), & N < 4
\end{cases}
\end{equation}
where $\sigma_{X}$ represents the standard deviation of abundances derived from individual spectral lines and $N$ denotes the number of measured lines. This statistical error can be significantly minimized through the measurement of a substantial number of spectral lines. For neutron-capture elements such as Y, Ba, La, Ce, Nd, and Sm, most stellar spectra in our sample typically exhibit more than four detectable lines.

The $\sigma_{\mathrm{sys}}$ component primarily arises from uncertainties in atmospheric parameters,  including effective temperature ($T_{\rm eff}$), surface gravity ($\log \textit{g}$), metallicity ($\rm \left [ \mathrm{Fe}/\mathrm{H} \right ]$), and microturbulent velocity ($\xi$). To estimate $\sigma_{\mathrm{sys}}$, each atmospheric parameter is independently varied by its respective 1$\sigma$ uncertainty, and the corresponding abundance variations are combined in quadrature, as shown in the following equation:

\begin{equation}
\footnotesize
\begin{aligned}
\sigma_{\text{sys}} = \bigg[ & 
    \left(\frac{\partial \log \epsilon}{\partial T_{\text{eff}}}\Delta T_{\text{eff}}\right)^2 + 
    \left(\frac{\partial \log \epsilon}{\partial \log g}\Delta \log g\right)^2 \\
    & + \left(\frac{\partial \log \epsilon}{\partial \text{[Fe/H]}}\Delta \text{[Fe/H]}\right)^2 + 
    \left(\frac{\partial \log \epsilon}{\partial \xi}\Delta \xi\right)^2 \bigg]^{1/2}
\end{aligned}
\label{eq:sigma_sys}
\end{equation}
Notably, sensitivity analyses reveal that elemental abundances demonstrate significantly greater dependence on $T_{\rm eff}$ variations compared to other atmospheric parameters.

\section{Results and Discussion} \label{sec:results}

\subsection{Classification of the sample stars} \label{subsec:classification of the sample star}


\begin{table*}[t]
\centering
\footnotesize
\caption{Carbon, nitrogen, and oxygen abundances, $^{12}$C/$^{13}$C ratios, and radial-velocity statistics for the CEMP stars in our sample. The last column lists their CEMP subclasses; asterisks (*) denote stars that exhibit moderately enhanced carbon abundances ($0.4 < [\mathrm{C}/\mathrm{Fe}]_{\mathrm{corr}} < 0.7$) after evolutionary corrections are applied.}
\label{tab:CEMP_stars}
\begin{tabular}{lcccccccccccccc}
\toprule
\hline
Name & [Fe/H] & [C/Fe] & [C/Fe]$_{\text{corr}}$ & $^{12}$C/$^{13}$C & [N/Fe] & [O/Fe] & [Ba/Eu] & RV mean & $\sigma$ & $n$ & $p(\chi^{2})$ & $z$ & Binary & Class \\
     & (dex)  & (dex)  & (dex)                  &                   & (dex)  & (dex)  &         & (km\,s$^{-1}$) & (km\,s$^{-1}$) &     &             &        &       \\
\midrule
J0006+0123 & -2.27 & 0.74 & 0.75 & $>$6.69 & $<$2.74 & ... & -1.05 & -65.89 & 0.19 & 3 & 0.622 & ... & No & CEMP-$r$ \\


J2333+0505 & -2.43 & 0.50 & 0.55 & 11.5 & $<$1.96 & ... & -0.92 & -136.92 & 0.08 & 2 & 0.685 & 0.406 & No & CEMP*-$r$ \\

J0119-0121 & -3.19 & 2.42 & 2.42 & $>$10.11 & $<$3.58 & ... & ... & -71.54 & 7.87 & 9 & 0.000 & ... & Yes & CEMP-$s$ \\

J0824+3025 & -2.72 & 2.21 & 2.21 & ... & ... & ... & ... & 18.33 & 2.64 & 2 & 0.000 & 5.296 &Yes & CEMP-$s$ \\

J0949+2707 & -2.90 & 2.00 & 2.00 & ... & ... & ... & ... & -104.61 & 30.17 & 3 & 0.000 & ... &Yes & CEMP-$s$ \\

J1003-0358 & -2.18 & 1.31 & 1.31 & ... & ... & ... & ... & 98.99 & 0.29 & 2 & 0.148 & 1.448 & No & CEMP-$s$ \\

J1011+3524 & -2.25 & 1.42 & 1.42 & $>$32.33  & $<$2.10 & 2.88 & ... & 54.19 & 0.38 & 3 & 0.158 & ... & No & CEMP-$s$ \\

J1020+4046 & -2.94 & 2.37 & 2.37 & 10.11 & $<$2.64 & ... & ... & -91.16 & 2.92 & 2 & 0.000 & 5.862 & Yes & CEMP-$s$ \\

J1044-0358 & -2.13 & 1.31 & 1.31 & ... & ... & ... & ... & 90.51 & 0.11 & 2 & 0.568 & 0.572 & No & CEMP-$s$ \\

J1434+3540 & -1.91 & 0.72 & 0.72 & 8.09 & 1.85 & ... & ... & -201.16 & 1.63 & 3 & 0.000 & ... & Yes & CEMP-$s$ \\

J2256+0215 & -2.27 & 0.08 & 0.44 & 2.33 & 2.29 & 1.09 & 1.02 & -161.05 & 0.29 & 2 & 0.137 & 1.486 & No & CEMP*-$s$ \\

J1551+2040 & -2.48 & 0.75 & 0.76 & 3.55 & ... & ... & 0.51 & -74.10 & 0.00 & 2 & 0.994 & 0.008 & No & CEMP-$r$/$s$ \\

J0446+2124 & -2.82 & 0.56 & 0.57 & 4.56 & ... & ... & 0.23 & -62.89 & 0.19 & 2 & 0.705 & 0.379 & No & CEMP*-$r$/$s$ \\

J2114-0616 & -2.55 & 0.64 & 0.90 & 5.25 & 1.69 & 0.90 & 0.62 & -161.11 & 0.24 & 4 & 0.002 & ... & Yes & CEMP-$r$+$s$ \\

J2257+3859 & -2.65 & 1.26 & 1.26 & 15.67 & ... & ... & 0.21 & -374.80 & ... & 1 & ... & ... & ... & CEMP-$r$+$s$ \\

J0653+4735 & -2.41 & 0.55 & 0.60 & 8.09 & 0.79 & 0.78 & 0.09 & 76.08 & 0.36 & 5 & 0.508 & ... & No & CEMP*-$r$+$s$ \\

\hline
\end{tabular}
\end{table*}

Based on the elemental abundances derived from our analysis, we systematically reclassify the sample stars. Following conventional methodologies, metallicity ($\rm \left [ \mathrm{Fe}/\mathrm{H} \right ]$) and carbon abundances ($\rm \left [ \mathrm{C}/\mathrm{Fe} \right ]$) serve as primary diagnostics for identifying metal-poor stars and their \ac{CEMP} subclasses \citep{2005ARA_A..43..531B}. To investigate nucleosynthetic origins, we utilize Ba and Eu as tracers of the \ac{$s$-process} and \ac{$r$-process}, respectively, enabling the differentiation of stars with distinct enrichment histories. For \ac{$r$-process}-enhanced stars, while previous studies adopted an empirical threshold of $\rm \left [ \mathrm{Eu}/\mathrm{Fe} \right ] = 1.0$ to distinguish between $r$-I and $r$-II stars \citep{2005ARA_A..43..531B}, recent work by \citet{2020ApJS..249...30H} has statistically refined this criterion. Through $k$-medoids clustering and Gaussian mixture modeling of abundance data from the R-Process Alliance, they established $\rm \left [ \mathrm{Eu}/\mathrm{Fe} \right ] = 0.7$ as the optimal threshold for $r$-II classification. Consequently, this study applies the updated criterion ($\rm \left [ \mathrm{Eu}/\mathrm{Fe} \right ] > 0.7$) to identify $r$-II and \ac{CEMP}-$r$ stars. Stars exhibiting anomalous light neutron-capture element ratios such as elevated $\rm \left [ \mathrm{Sr}/\mathrm{Ba} \right ]$ and $\rm \left [ \mathrm{Sr}/\mathrm{Eu} \right ]$, are flagged as candidates for the 'limited-r' process \citep{2018ARNPS..68..237F}. \ac{CEMP} stars are further classified (\ac{CEMP}-$s$, \ac{CEMP}-$r$/$s$) based on their surface heavy-element abundances, according to the classification criteria established by \citet{2021A_A...649A..49G}. Table \ref{tab:heavy_classification} additionally includes a subset of stars classified as CEMP-$r$+$s$. \citet{2018ARNPS..68..237F} proposed empirical selection criteria of 0.0 $ < \rm \left [ \mathrm{Ba}/\mathrm{Eu} \right ] < $  +0.5 and -1.0 $ < \rm \left [ \mathrm{Ba}/\mathrm{Pb} \right ] < $ -0.5 to identify objects exhibiting mixed $r$- and $s$-process signatures. Our analysis reveals that these criteria do not effectively identify \ac{CEMP}-$r$+$s$ stars in the current sample. Therefore, this study introduces more robust selection standards to improve the reliability of identification, as discussed in Section \ref{subsec:CEMP-r/s and CEMP-r+s stars}.  


Our final catalog comprises 82 heavy-element-enhanced metal-poor stars, including 45 $r$-I stars, 19 $r$-II stars, 2 limited-$r$ stars, 1 \ac{CEMP}-$r$ star, 8 \ac{CEMP}-$s$ stars, 1 \ac{CEMP}-$r$/$s$ star, and 2 candidate \ac{CEMP}-$r$+$s$ stars. In addition, our sample includes four stars with moderate carbon abundances ($0.4 < \mathrm{[C/Fe]} < 0.7$) that do not strictly meet the formal classification criteria for \ac{CEMP} stars. However, given that their neutron-capture abundance patterns closely resemble those of specific \ac{CEMP} subclasses, we classify these objects as moderately CEMP stars. For clarity, these stars are explicitly marked with asterisks (*) in Table~\ref{tab:heavy_classification}, Table ~\ref{tab:atmospheric parameters}, and Table~\ref{tab:CEMP_stars}. This notation highlights their close connection to the classic \ac{CEMP} subclasses regarding heavy-element enrichment, while retaining the information about their only moderate carbon enhancement. Unless otherwise specified, subsequent discussions and figures include these moderately carbon-enhanced stars within the \ac{CEMP} heavy-element-enhanced subcategories.


The Hertzsprung-Russell diagram shown in Figure \ref{fig:HR_diagram} reveals a clear dichotomy, with r-process-enhanced stars, limited-r stars, \ac{CEMP}-$r$ stars, as well as both \ac{CEMP}-$r$/$s$ and \ac{CEMP}-$r$+$s$ stars predominantly occupying the sub-giant and red giant branches, whereas \ac{CEMP}-$s$ stars are predominantly found on the main sequence and turn-off region. This can be interpreted as evidence that \ac{CEMP}-$s$ stars are formed via mass transfer from companion stars, and the accreted material has not yet been significantly diluted by convection during their subsequent evolution to the red giant branch.


\subsection{Comparison with Literature} \label{subsec:comparison with Literature}

\begin{figure*}[t]
\centering
\includegraphics[width=0.9\textwidth,keepaspectratio]{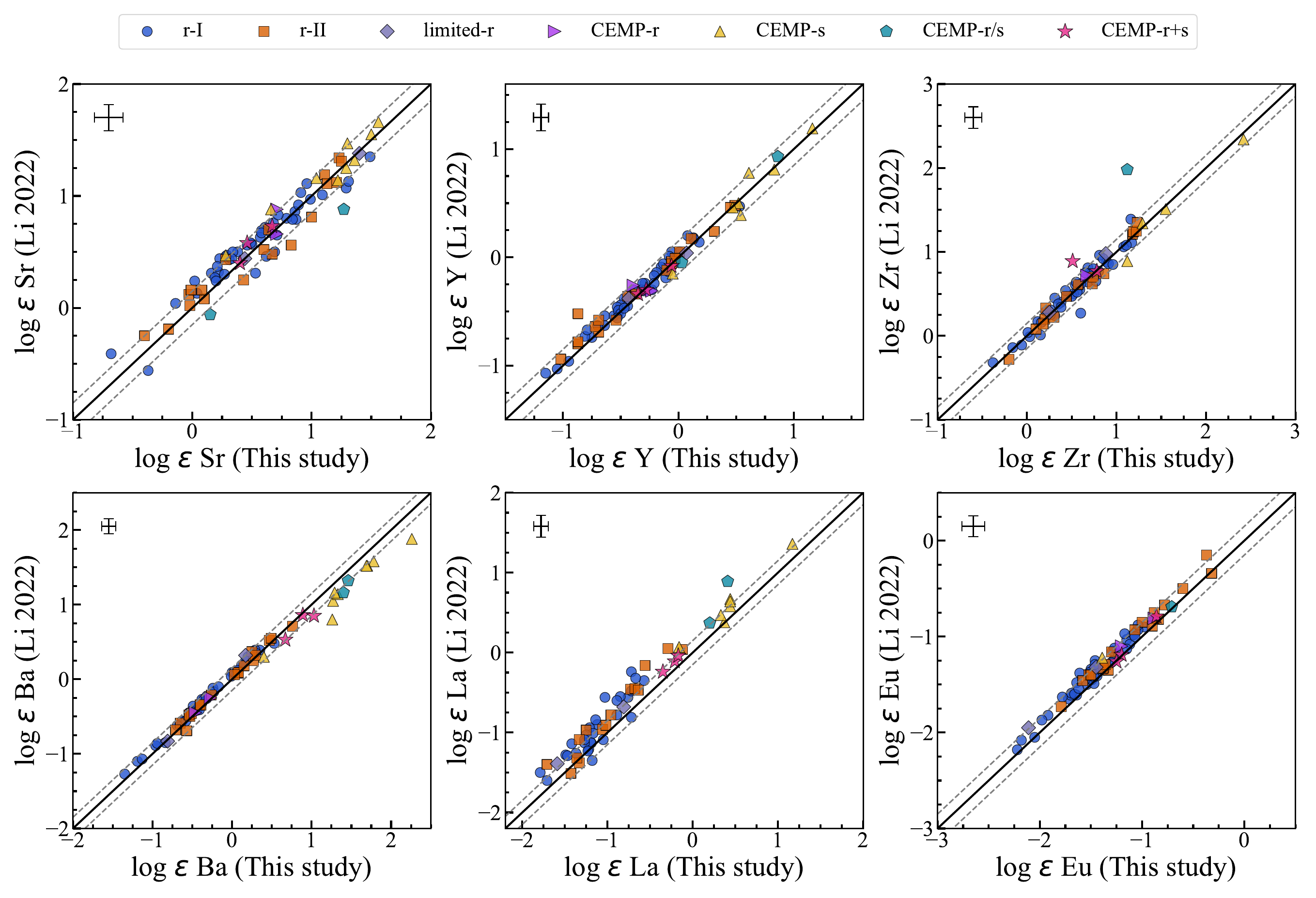}
\caption{Comparison of Sr, Y, Zr, Ba, La, and Eu abundances derived in this work with those reported by \citet{2022ApJ...931..147L}. The solid line indicates perfect agreement between our measurements and \citet{2022ApJ...931..147L} values, while the dashed lines represent the $\pm$ 0.15 dex deviation thresholds. The mean 1$\sigma$ uncertainties for each elemental abundance are displayed in the upper-left corner of the corresponding panel.
\label{fig:abundance compare}}
\end{figure*}

\citet{2022ApJ...931..147L} reported neutron-capture element abundances (Sr, Y, Zr, Ba, La, and Eu) for this sample. As shown in Figure \ref{fig:abundance compare}, our new abundance measurements exhibit excellent agreement with their values for Y, Zr, and Ba in most stars, falling within the typical measurement uncertainties of $\pm$0.15 dex. However, systematic discrepancies emerge for the Ba abundances of CEMP-$s$, CEMP-$r$/$s$ and CEMP-$r$+$s$ stars, with our results systematically higher by approximately 0.15 dex. This discrepancy stems primarily from differing assumptions regarding isotopic ratios: we employ solar \ac{$s$-process} isotopic ratios for \ac{CEMP}-$s$ stars and solar isotopic ratios for \ac{CEMP}-$r$/$s$ and \ac{CEMP}-$r$+$s$ stars, whereas \citet{2022ApJ...931..147L} apply \ac{$r$-process} ratios uniformly. As demonstrated in Section \ref{subsec:neutron-capture elements}, through spectrum synthesis tests, this fundamental difference in assumption can introduce variations exceeding 0.2 dex in derived Ba abundances, particularly for the Ba II $\lambda$ 4554 and $\lambda$ 4934 features. 

For Sr, La, and Eu, our derived abundances exhibit systematic offsets of 0.1-0.2 dex toward lower values compared to \citet{2022ApJ...931..147L}. These discrepancies primarily arise from methodological differences in spectral analysis. Our study employed the 'synth' driver in MOOG for spectrum synthesis to derive the abundances, whereas \citet{2022ApJ...931..147L} utilized the 'blends' driver, which determines abundances through \ac{EW} measurements. This distinction becomes particularly significant for severely blended or asymmetric spectral features. A prominent example is the Eu II $\lambda$ 4205 line, whose left wing exhibits strong contamination from the CH molecular band. This blending effect depresses the entire wing profile and introduces profile asymmetry, leading to an overestimation of the \ac{EW}. Consequently, the \ac{EW}-based 'blends' driver yields systematically higher Eu abundances by approximately 0.15 dex compared to the synthesis approach. For La, the differences arise from both line selection and blending effects: while \citet{2022ApJ...931..147L} primarily used four La II lines, our analysis incorporated more additional transitions, enabling more robust abundance determinations with reduced uncertainties ($\sigma < 0.1 $ dex).

\begin{figure*}[t]
\centering
\includegraphics[width=0.9\textwidth,keepaspectratio]{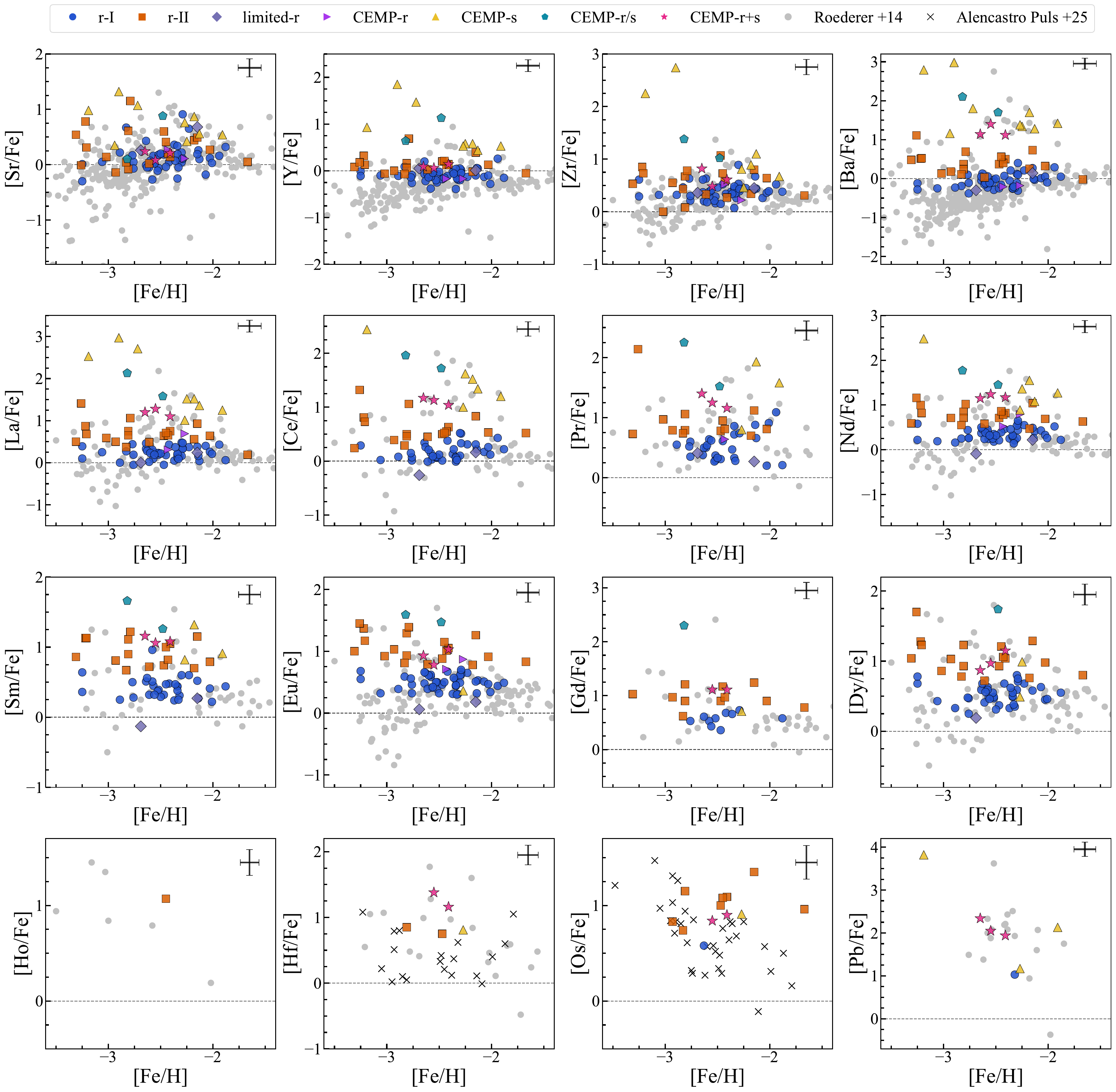}
\caption{$\rm \left [ \mathrm{X}/\mathrm{Fe} \right ]$ versus $\rm \left [ \mathrm{Fe}/\mathrm{H} \right ]$ abundance ratios for the sample stars, with comparative data from \citet{2014AJ....147..136R} (gray filled circles) and \citet{2025A&A...693A.294A} (black crosses) overplotted. The solar reference values ($\rm \left [ \mathrm{X}/\mathrm{Fe} \right ]$ = 0) are indicated by the gray dashed line.
\label{fig:X_Fe_Fe_H}}
\end{figure*}

Figure \ref{fig:X_Fe_Fe_H} presents the $\rm \left [ \mathrm{X}/\mathrm{Fe} \right ]$ versus $\rm \left [ \mathrm{Fe}/\mathrm{H} \right ]$ distributions for our stellar sample, with comparative data from \citet{2014AJ....147..136R} and \citet{2025A&A...693A.294A}. As expected from our sample selection, the abundances of most neutron-capture elements in our sample are systematically higher than those reported in the literature. Notably, we observe that the abundances of the first-peak elements (Sr, Y, Zr) show overlapping distributions among $r$-I, $r$-II, \ac{CEMP}-$r$, and limited-$r$ stars, while \ac{CEMP}-$s$, \ac{CEMP}-$r$/$s$ and \ac{CEMP}-$r$+$s$ stars exhibit systematically elevated abundances. The abundance stratification becomes even more pronounced in the lanthanide region (57 $\leq$ Z $\leq$ 71): \ac{CEMP}-$s$ and \ac{CEMP}-$r$/$s$ stars occupy the highest $\rm \left [ \mathrm{X}/\mathrm{Fe} \right ]$ region, followed by \ac{CEMP}-$r$+$s$ stars and $r$-II stars, while \ac{CEMP}-$r$, $r$-I and limited-$r$ stars show the lowest enhancement. The significant scatter in $\rm \left [ \mathrm{X}/\mathrm{Fe} \right ]$ observed across the entire metallicity range reflects the diverse nucleosynthetic origins of heavy elements in these metal-poor stars. For \ac{$r$-process}-enhanced stars, the surface abundances of heavy elements primarily originate from NSMs or specific types of core-collapse supernovae. The considerable dispersion in surface $\left [r/\mathrm{Fe} \right ]$ ratios at low metallicities reflects both the infrequency of \ac{$r$-process} events —-- where only one main \ac{$r$-process} event may occur per 100-1000 ordinary supernovae —-- and the inhomogeneous mixing of the interstellar medium in the early Milky Way \citep{2021RvMP...93a5002C}. Conversely, the surface elemental abundances of \ac{CEMP}-$s$ stars primarily result from \ac{$s$-process} nucleosynthesis in low- to medium-mass \ac{AGB} companion stars. Particularly noteworthy are two \ac{CEMP}-$s$ stars in our sample that exhibit extreme enhancement in barium abundance ($\rm \left [ \mathrm{Ba}/\mathrm{Fe} \right ] > $  +2.0), which simultaneously display the highest observed abundances of most \ac{$s$-process} elements (Ba, La, Ce and Pb). This suggests variations in both the initial masses of their progenitor \ac{AGB} stars and the efficiency of \ac{$s$-process} nucleosynthesis. Furthermore, the observed abundances in \ac{CEMP}-$s$ stars may be influenced by two main factors: binary orbital parameters, which affect mass transfer efficiency, and differences in the mixing and dilution processes within stellar envelopes.


\subsection{$r$-process-enhanced stars} \label{sec:r-process enhanced star}

\subsubsection{Abundance pattern fitting} \label{subsec:abundance pattern}

\begin{figure*}[t]
\centering
\includegraphics[width=0.6\textwidth,keepaspectratio]{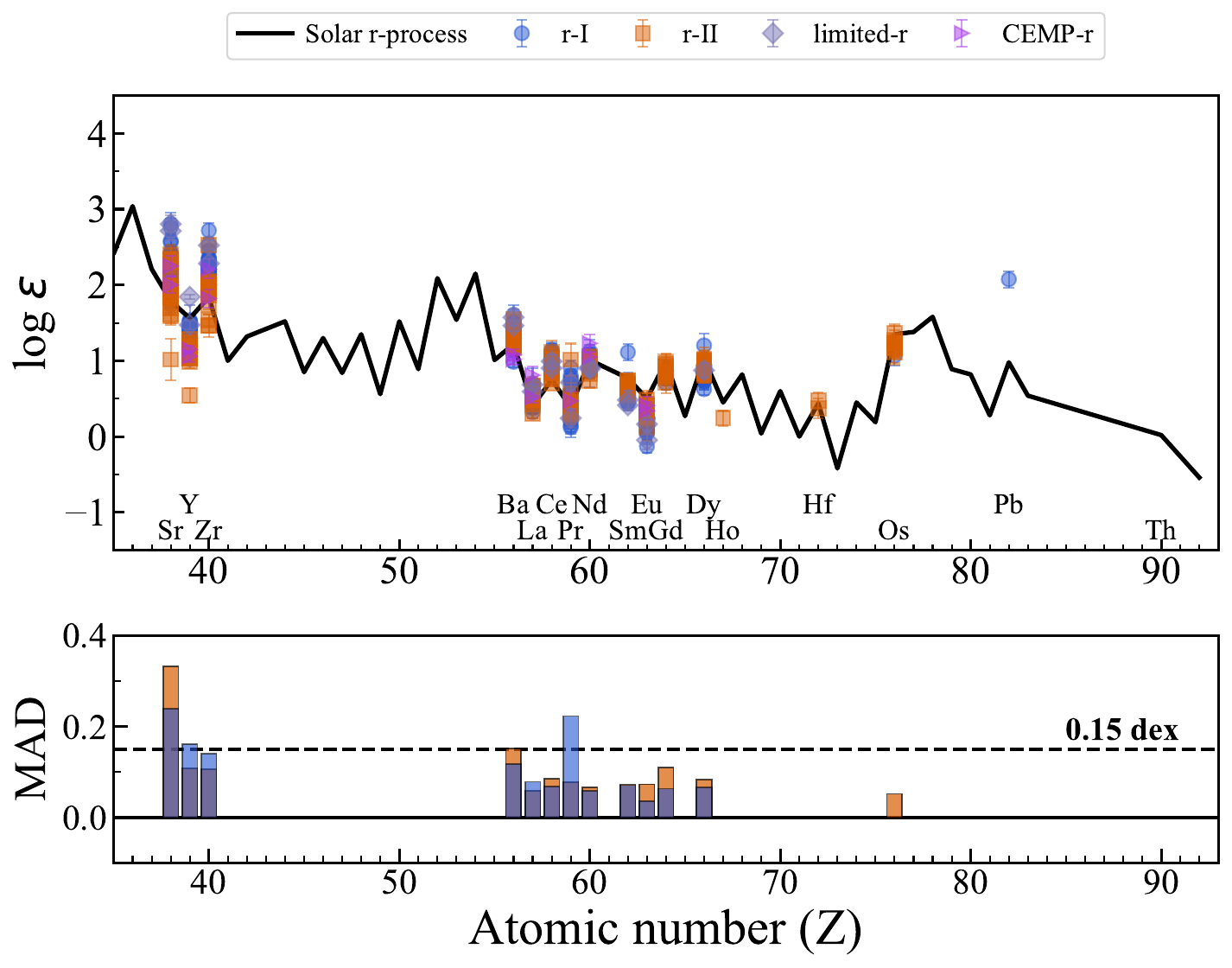}
\caption{TOP panel: Abundance patterns of \ac{$r$-process}-enhanced stars. The solid black line represents the solar \ac{$r$-process} abundance pattern. The blue circles, yellow squares, grayish blue diamonds and purple triangles with error bars correspond to $r$-I, $r$-II, limited-$r$ and \ac{CEMP}-$r$ stars, respectively. Bottom panel: \ac{MAD} of scaled abundances for different elements in the $r$-I and $r$-II stars. The black solid and dashed lines indicate \ac{MAD} values of 0 and 0.15 dex, respectively.
\label{fig:solar r-process pattern}}
\end{figure*}

This section focuses on the $r$-process-enhanced stars in our sample --- including the $r$-I, $r$-II, limited-$r$, and CEMP-$r$ categories --- with their neutron-capture element abundance patterns presented in Figure \ref{fig:solar r-process pattern}. The abundances of these stars are scaled to the solar \ac{$r$-process} using the following formula:


\begin{equation}
\log\epsilon(X) = \log\epsilon(X)_{*} + O_{r}
\label{eq:abundance_fitting1}
\end{equation}

\begin{equation}
O_r = \frac{\displaystyle\sum_{i}^{N}\frac{(\log\epsilon(X)_{r} - \log\epsilon(X)_{*})}{\sigma_{X}^2} }{\displaystyle\sum_{i}^{N} \frac{1}{\sigma_{X}^2}}
\end{equation}
where $\log\epsilon(X)_{*}$ and $\sigma_{X}$ represent the observed abundance and its associated uncertainty for element $X$ in each star, respectively, and $\log\epsilon(X)_{r}$ denotes the corresponding solar \ac{$r$-process} residuals abundance from \citet{2020MNRAS.491.1832P}. The parameter $O_r$ represents a scaling coefficient for the observed stellar abundances. It is derived from the weighted abundance difference between the solar \ac{$r$-process} pattern and the observed abundances for $N$ elements with atomic number Z $\geqslant$ 56. 

The abundances of light neutron-capture element (Z = 38, 39, 40) in \ac{$r$-process}-enhanced stars exhibit significant deviations from the solar \ac{$r$-process} abundance pattern. As shown in the upper panel of Figure \ref{fig:solar r-process pattern}, most stars display elevated abundances of Sr and Zr relative to solar \ac{$r$-process} values, while Y abundances remain systematically lower. This indicates a more complex origin for these elements and suggests that the solar \ac{$r$-process} may not be a suitable reference for these light neutron-capture elements \citep{2023arXiv230209188F}. Furthermore, we note that NLTE corrections for Y become particularly significant at low metallicities. In metal-poor red giants, when $\mathrm{[Fe/H]} \leqslant$ -3, the NLTE corrections for low-excitation-potential (EP) Y II lines can exceed 0.5 dex \citep{2023MNRAS.525.3718S}. However, among the nine Y II lines employed in this study, seven possess medium EP values, for which the NLTE corrections in metal-poor red giants generally remain below 0.05 dex. Consequently, the mean NLTE correction for $\log\epsilon(Y)$ in our sample is approximately +0.15 dex.

In our sample, the abundance variations of light neutron-capture elements among different stars exceed 1 dex, with limited-$r$ stars exhibiting the most pronounced first-peak element enhancements, a characteristic that correlates with their relative deficiency in heavy neutron-capture elements. In $r$-I and $r$-II stars, the \ac{MADs} of the Sr scaled abundances --- obtained by applying Equation \ref{eq:abundance_fitting1} --- exceed 0.2 dex. This value is significantly larger than the dispersion observed for the heavy neutron-capture elements. In contrast, the scaled Y and Zr abundances in these stars show significantly lower dispersion (MAD $\leq$ 0.13 dex). This trend has been demonstrated by \citet{2023MNRAS.524.5607S}, who showed that $\rm \left [ \mathrm{Sr}/\mathrm{Fe} \right ]$ exhibits markedly greater scatter than $\rm \left [ \mathrm{Y}/\mathrm{Fe} \right ]$ and $\rm \left [ \mathrm{Zr}/\mathrm{Fe} \right ]$ with increasing $\rm \left [ \mathrm{Eu}/\mathrm{Fe} \right ]$. These observations imply a stronger nucleosynthetic coupling between Y and Zr and heavy neutron-capture elements in \ac{$r$-process}-enhanced stars. \citet{2022ApJ...936...84R} conducted a detailed analysis of eight metal-poor stars with relatively complete light neutron-capture element measurements, showing that when normalized to the solar \ac{$r$-process} Zr abundance, the star-to-star dispersion among the abundances of Se, Sr, Y, Zr, Nb, Mo, and Te decreases substantially (MAD $\leq$ 0.13 dex). This consistency suggests the existence of an astrophysical site producing a universal abundance pattern for light neutron-capture elements. Potential sites for this process include: (1) neutrino-driven wind in core-collapse supernovae \citep{2010ApJ...722..954R, 2011ApJ...731....5A, 2020EPJWC.22701009A}; (2) electron-capture supernovae with progenitor masses of 8–10 M$_\odot$ \citep{2006A&A...450..345K, 2008A&A...485..199J, 2011ApJ...726L..15W}; or (3) quark deconfinement supernovae \citep{2020ApJ...894....9F}. The characteristically high electron fraction ($Y_e$) in these environments would suppress the efficient production of heavier \ac{$r$-process} nuclides.

\begin{figure*}[t]
\centering
\includegraphics[width=0.9\textwidth,keepaspectratio]{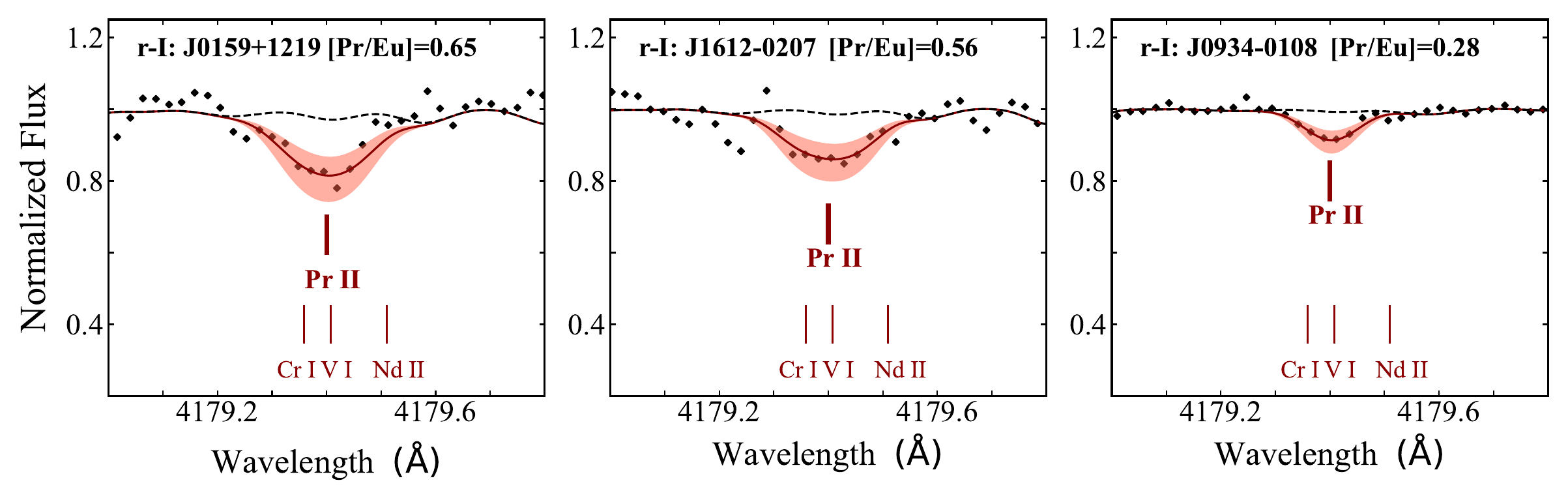}
\caption{Spectral synthesis of the Pr II $\lambda$4179 for three $r$-I stars. The observed spectra are represented by black data points. The best fits are shown as solid red lines, with associated uncertainties of 0.2 dex indicated by the shaded red regions. The spectral synthesis excluding the elements of interest are displayed as black dashed lines.
\label{fig:Hihg_Pr_spectral_synthesis}}
\end{figure*}

In contrast, the heavy neutron-capture elements (Z $\geq$ 56) exhibit a relatively uniform abundance pattern that closely matches the solar \ac{$r$-process} abundances, consistent with their predominant production through the main \ac{$r$-process}. However, Ba and Pr abundances demonstrate notable dispersion compared to other heavy neutron-capture elements. This scatter in Pr abundances primarily originates from eight \ac{$r$-process}-enhanced stars exhibiting anomalously high Pr abundances, with their average $\rm \left [ \mathrm{Pr}/\mathrm{Eu} \right ]$ value reaching 0.47. This starkly contrasts with the solar \ac{$r$-process} range of $\rm \left [ \mathrm{Pr}/\mathrm{Eu} \right ]_r$ = -0.31 to -0.18. This phenomenon has also been observed by \citet{2025ApJ...985..250X} in their study of \ac{$r$-process}-enhanced stars within the \ac{GSE}. It is noteworthy that the Pr abundances in these stars are mainly derived from the Pr II line at 4179 $\angstrom$. Spectral synthesis of the Pr II $\lambda$4179 feature in three representative stars (Figure \ref{fig:Hihg_Pr_spectral_synthesis}) reveals potential explanations for the elevated abundances: (1) unidentified line-blending effects influencing the Pr abundance measurements; or (2) additional nucleosynthetic contributions that enhance Pr production through currently unidentified mechanisms.

In the Galactic halo, \ac{$r$-process}-enhanced stars with $\rm \left [ \mathrm{Fe}/\mathrm{H} \right ]$ $>$ -2.6 could exhibit \ac{$s$-process} contamination \citep{2004ApJ...617.1091S}, demonstrating systematically enhanced barium abundances relative to stars with pure \ac{$r$-process} patterns. To quantify the \ac{$s$-process} contribution to elemental abundances in these stars, we employ the following formulation:

\begin{equation}
\label{eqn:fitting model}
\log\epsilon(X) = \log_{10} (10^{\log\epsilon(X)_r+O_r}  + 10^{\log\epsilon(X)_s+O_s})
\end{equation}

\begin{equation}
\label{eqn:contribution ratios}
F_{s} = \frac{1}{N} \displaystyle\sum_{i}^{N}\frac{10^{\log\epsilon(X)_s+O_s}}{10^{\log\epsilon(X)}}
\end{equation}
where $\log\epsilon(X)_{r}$ and $\log\epsilon(X)_{s}$ denote the abundance of element $X$ in the solar \ac{$r$-process} and solar \ac{$s$-process} \citep{2020MNRAS.491.1832P}, respectively, and $O_r$ and $O_s$ represent their corresponding dilution factors, which are derived through $\chi^{2}$ minimization between the model and observations. The parameter $F_s$ represents the average fractional contribution of the \ac{$s$-process} to the abundance of elements with Z $\geq$ 56.

\begin{figure}[h!]
\centering
\includegraphics[width=0.5\textwidth,keepaspectratio]{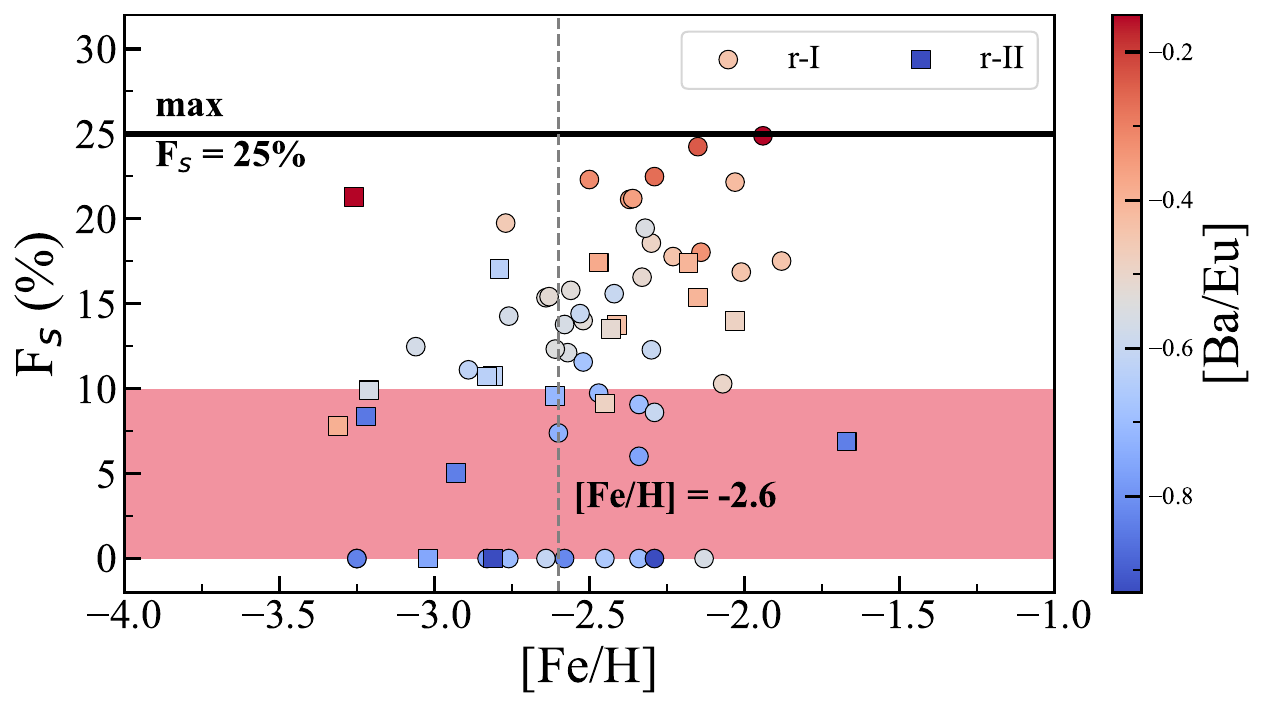}
\caption{The variation of \ac{$s$-process} contribution fraction ($F_s$) as a function of $\rm \left [ \mathrm{Fe}/\mathrm{H} \right ]$ in $r$-I and $r$-II stars. The color scale represents the corresponding $\rm \left [ \mathrm{Ba}/\mathrm{Eu} \right ]$ values for individual stars. The red shaded area indicates the region where 0 $\leqslant$ $F_s$ $\leqslant$ 0.10, encompassing stars that we consider to be essentially free from \ac{$s$-process} contamination, as discussed in the main text.
\label{fig:Fs}}
\end{figure}

Figure \ref{fig:Fs} shows the distribution of $F_s$ as a function of $\mathrm{[Fe/H]}$ for $r$-I and $r$-II stars in our sample. Over the metallicity range -3.5 $ < \mathrm{[Fe/H]} < $ -1.5, we observe an increase in $F_s$ with increasing metallicity, a trend consistent with the delayed contribution of \ac{$s$-process} nucleosynthesis. In the early universe, compared to prompt enrichment \ac{$r$-process} sources such as \ac{MRDSNe} or collapsars, low- and intermediate-mass \ac{AGB} stars only contribute significantly to chemical enrichment at higher metallicities owing to their longer evolutionary timescales.

Statistical analysis further reveals a significant positive correlation between $F_s$ and $\mathrm{[Ba/Eu]}$ ($r$ = 0.83, $p$ $<$ 0.001), indicating that stars with greater \ac{$s$-process} contributions systematically exhibit higher $\mathrm{[Ba/Eu]}$ ratios, confirming $\mathrm{[Ba/Eu]}$ is a good indicator. Meanwhile, only a small fraction of stars in the sample ($\sim21\%$) retain pure solar \ac{$r$-process} patterns ($F_s$ = 0), while the remaining $\sim79\%$ exhibit values in the range 0 $< F_s \leqslant $ 0.25. It should be emphasized that $F_s$ $>$ 0 does not necessarily imply a definite \ac{$s$-process} contribution, as the $F_s$ calculation is based solely on the solar \ac{$r$-process} and \ac{$s$-process} patterns given by \citet{2020MNRAS.491.1832P}. However, the $\mathrm{[Ba/Eu]}$ ratio of the solar \ac{$r$-process} varies across literature studies, ranging from approximately -0.81 to -0.90 \citep{2008ARA&A..46..241S, 2014ApJ...787...10B, 2020MNRAS.491.1832P}. In addition, insufficient deposition of fission fragments may lead to a Eu deficiency, which can also cause a mild elevation in $\mathrm{[Ba/Eu]}$ ( $\mathrm{[Ba/Eu]}_{*}$ - $\mathrm{[Ba/Eu]}_{pure-r}$ $<$ 0.2 dex; \citealp{2023Sci...382.1177R}). Therefore, considering the uncertainties in the $r$-process, we suggest that a star can be reasonably interpreted as having \ac{$s$-process} contamination only when $\mathrm{[Ba/Eu]}$ $\geqslant$ -0.65, which corresponds to $F_s$ $\geqslant$ 10$\%$.


Notably, our sample contains several extremely metal-poor stars ($\rm \left [ \mathrm{Fe}/\mathrm{H} \right ] < $ -3.0) that exhibit detectable \ac{$s$-process} contributions, most prominently J1641+0334 ($\rm \left [ \mathrm{Fe}/\mathrm{H} \right ] $ = -3.2) with $\rm \left [ \mathrm{Ba}/\mathrm{Eu} \right ]$ = -0.34 and $F_s$ = 20.4$\%$. This unexpected finding implies the operation of \ac{$s$-process} production mechanisms in the early Universe, potentially facilitated by rotating massive stars \citep{2012A&A...538L...2F, 2016MNRAS.456.1803F, 2017A&A...607L...3C}. At low metallicity ($[\mathrm{Fe/H}] \lesssim -2.5$) and high rotational velocities ($v_{\mathrm{rot}} \gtrsim 300$--500 km s$^{-1}$), rotation strongly enhances the mixing between the hydrogen- and helium-burning shells, promoting the CNO cycle to produce primary $^{14}$N. Through the reaction chain $^{14}\mathrm{N}(\alpha,\gamma)^{18}\mathrm{F}(e^{+}\nu_{e})^{18}\mathrm{O}(\alpha,\gamma)^{22}\mathrm{Ne}$, additional $^{22}$Ne is synthesized, and activation of the neutron source reaction $^{22}\mathrm{Ne}(\alpha,n)^{25}\mathrm{Mg}$ efficiently triggers the \ac{$s$-process} chain.

\begin{figure}[h!]
\centering
\includegraphics[width=0.4\textwidth,keepaspectratio]{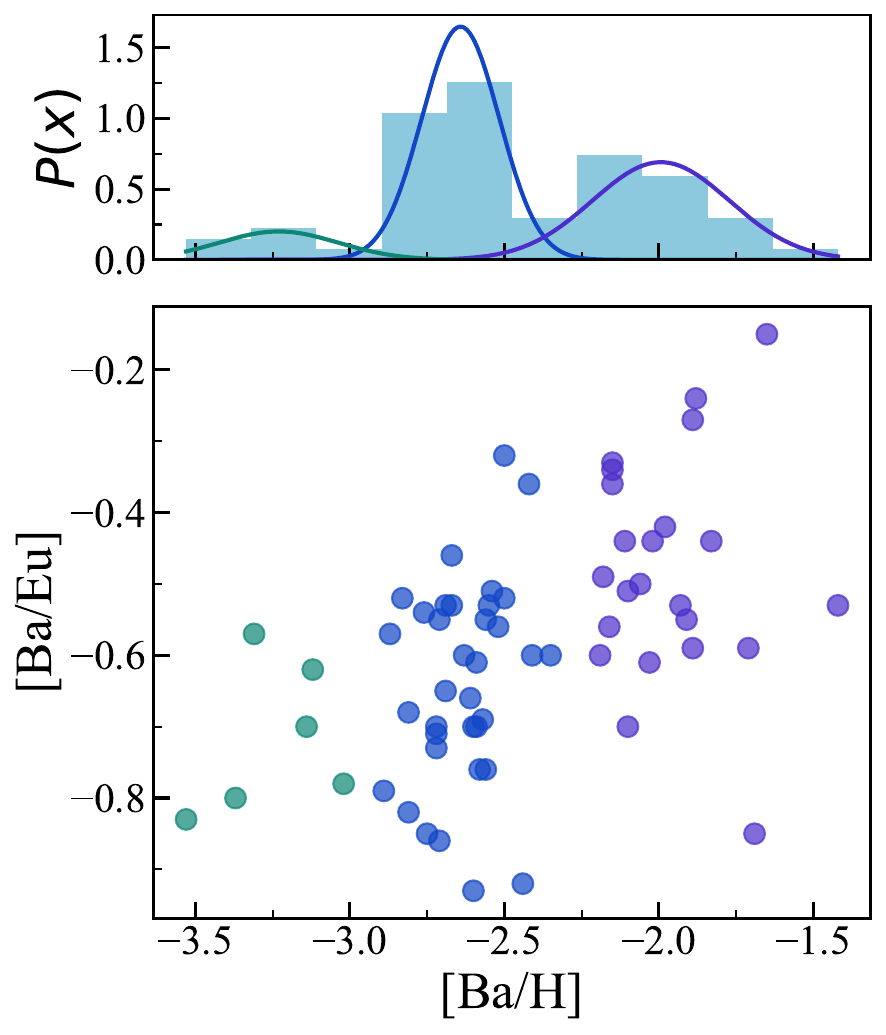}
\caption{Top panel: Histogram of the $\rm \left [ \mathrm{Ba}/\mathrm{H} \right ]$ distribution in \ac{$r$-process}-enhanced stars, overlaid with a three-component \ac{GMM} fit. Bottom panel: The $\rm \left [ \mathrm{Ba}/\mathrm{Eu} \right ]$ versus $\rm \left [ \mathrm{Ba}/\mathrm{H} \right ]$ distribution, where the sample stars are classified into three distinct clusters (color-coded) using mean shift clustering algorithm.
\label{fig:GMM_meanshift}}
\end{figure}

To establish the metallicity threshold for significant \ac{$s$-process} enrichment, we apply the method described in \citet{2025A&A...693A.293L}, utilizing the mean shift clustering algorithm with kernel density estimation \autofootlink{https://scikit-learn.org/stable/index.html}to analyze $\rm \left [ \mathrm{Ba}/\mathrm{Eu} \right ]$ versus $\rm \left [ \mathrm{Ba}/\mathrm{H} \right ]$ distribution of our sample stars.  The results reveal three distinct groups in this abundance space (Figure \ref{fig:GMM_meanshift}).  Specifically, (1) the group with low $[\mathrm{Ba/Eu}]$ and low $[\mathrm{Ba/H}]$  (green) corresponds to stars dominated by the pure \ac{$r$-process}, with median $[\mathrm{Ba/Eu}]$ and $[\mathrm{Fe/H}]$ values of approximately -0.74 and -2.98, respectively;  (2) the intermediate group (blue), located between the other two, exhibits transitional characteristics in both $[\mathrm{Ba/Eu}]$ and $[\mathrm{Ba/H}]$, with median $[\mathrm{Ba/Eu}]$ and $[\mathrm{Fe/H}]$ values of -0.61 and -2.58;  and (3) the group with high $[\mathrm{Ba/Eu}]$ and relatively higher $[\mathrm{Ba/H}]$ (purple) shows clear $r$+$s$ mixed signatures, with median $[\mathrm{Ba/Eu}]$ and $[\mathrm{Fe/H}]$ values of approximately -0.50 and -2.18.  To assess the reliability of the clustering results, we fit the $[\mathrm{Ba/H}]$ distribution using \ac{GMM} with 1 to 4 components and evaluate model performance using the Akaike Information Criterion (AIC; \citealp{10.1093/biomet/60.2.255}) to balance goodness-of-fit against model complexity, thereby avoiding overfitting or underfitting.  The analysis shows that a three-component \ac{GMM} provides the optimal fit (AIC = 73.17; see the upper panel of Figure \ref{fig:GMM_meanshift}), likewise supporting the presence of three physically distinct groups in the $[\mathrm{Ba/H}]$ distribution. Among these, the intermediate group (blue cluster) plays a particularly crucial role. This group not only displays a large dispersion in $[\mathrm{Ba/Eu}]$ (-0.61 $\pm$ 0.14 dex), but its median [Fe/H] and [Ba/H] values also lie between those of the pure \ac{$r$-process} and $r$+$s$ mixed groups, exhibiting the expected transitional behavior. Furthermore, its median \ac{$s$-process} contribution fraction, $F_s$, reaches 10.7$\%$, indicating that once the metallicity increases to $[\mathrm{Fe/H}]$ $\geqslant$ -2.6, \ac{$s$-process} yields from AGB stars begin to exert a non-negligible influence on the heavy-element abundances.  Consequently, this intermediate group naturally delineates the key metallicity interval in which \ac{$s$-process} nucleosynthesis transitions from a minor to a significant contributor to chemical enrichment in the early Galaxy.

\subsubsection{The differences between the $r$-I and $r$-II stars in $\rm \left [ \mathrm{X}/\mathrm{Eu} \right ]$} \label{subsec: the difference of the r-I and r-II stars in [X/Eu]}


This section investigates the elemental abundance ratios of $r$-I and $r$-II stars, revealing differences in their nucleosynthetic origins. Correlation analyses of $\rm \left [ \mathrm{X}/\mathrm{Eu} \right ]$ versus $\rm \left [ \mathrm{Eu}/\mathrm{Fe} \right ]$ ratios demonstrate statistically significant distributional divergences between these two groups, with Ho, Hf, and Pb excluded from the graphical analysis due to insufficient data points ($N$ $<$ 5).  

\begin{figure}[h]
\centering
\includegraphics[width=0.48\textwidth,keepaspectratio]{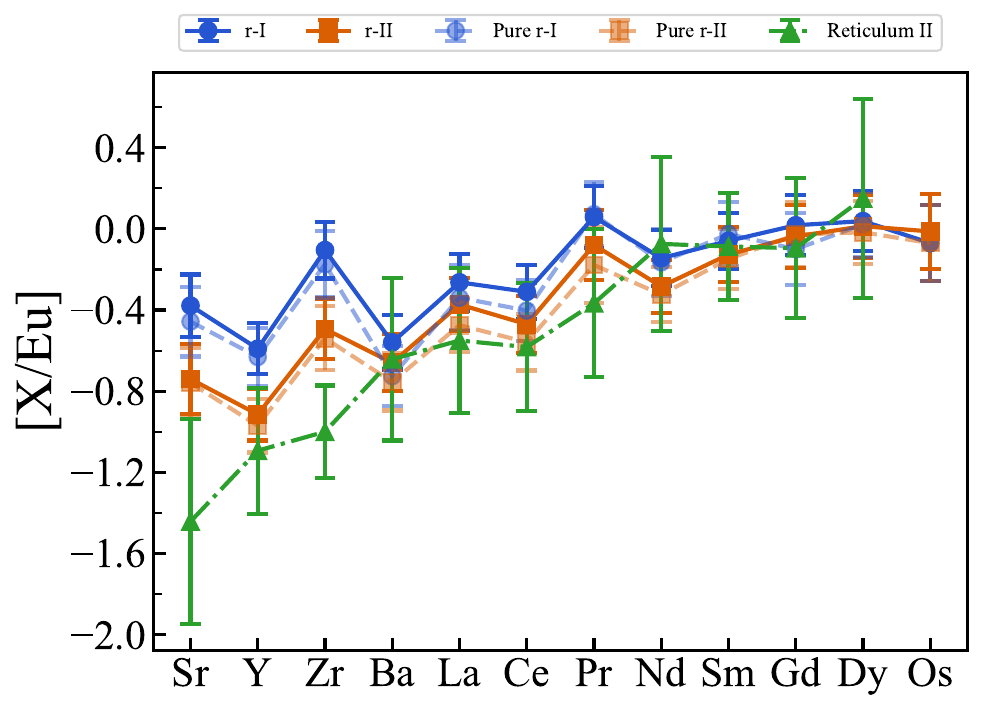}
\caption{$\rm \left [ \mathrm{X}/\mathrm{Eu} \right ]$ ratios in $r$-I and $r$-II stars. The solid line connects data points representing all $r$-I and $r$-II stars, while the dashed line corresponds to $r$-I and $r$-II stars exhibiting pure \ac{$r$-process} enhancement ($F_s <$ 10$\%$). The green dash-dotted line shows the $\rm \left [ \mathrm{X}/\mathrm{Eu} \right ]$ ratios of stars in the Reticulum II dwarf galaxy \citep{2016ApJ...830...93J} for comparison.
\label{fig:Mean_XEu_EuFe}}
\end{figure}

The most striking disparity emerges in the light neutron-capture elements $\rm \left [ \mathrm{Sr, Y, Zr}/\mathrm{Eu} \right ]$, as shown in Figure \ref{fig:Mean_XEu_EuFe}, where $r$-I stars exhibit systematically elevated $[\mathrm{X/Eu}]$ ratios exceeding those of $r$-II stars by 0.36 dex on average. These results align closely with the earlier findings by \citet{2010A&A...516A..46M}, who reported comparable enhancements of 0.36 dex for $[\mathrm{Sr/Eu}]$ and 0.33 dex for $[\mathrm{Zr/Eu}]$ in $r$-I stars. Notably, this discrepancy cannot be entirely attributed to the mean offset in $\rm \left [ \mathrm{Eu}/\mathrm{Fe} \right ]$ between the two stellar types ($\Delta$ $\rm \left [ \mathrm{Eu}/\mathrm{Fe} \right ]$ = 0.53 dex). Even when controlling for \ac{$s$-process} contamination through strict selection of "Pure $r$" samples ($F_{s}$ $<$ 10 $\%$), $r$-I stars still exhibit significantly elevated $\rm \left [\mathrm{Sr, Y, Zr}/\mathrm{Eu} \right ]$ values by 0.29 dex. For the heavy neutron-capture element range (Z $\geq$ 56), the difference in $\rm \left [ \mathrm{X}/\mathrm{Eu} \right ]$ between $r$-I and $r$-II stars becomes substantially smaller, with a mean offset of only 0.1 dex. Furthermore, these differences progressively decrease with increasing atomic number. Although $[\mathrm{Os/Eu}]$ ratios appear slightly lower in $r$-I stars compared to $r$-II stars, this observation lacks statistical significance due to the limited sample size (one $r$-I star versus eight $r$-II stars with Os abundance measurements) and requires verification through expanded datasets. 


The systematic differences in the $\mathrm{[X/Eu]}$ ratios between $r$-I and $r$-II stars, particularly the notably lower $\mathrm{[Sr, Y, Zr/Eu]}$ in $r$-II stars, may be attributed to differences in the relative contributions of the main and weak $r$-processes in these two subclasses. A plausible scenario is that $r$-II stars experience stronger enrichment from the main \ac{$r$-process} compared to $r$-I stars. At the same time, $r$-II stars may receive a level of weak \ac{$r$-process} contribution comparable to that of $r$-I stars, or they may experience more limited, possibly negligible, weak \ac{$r$-process} pollution. Furthermore, the differences in the balance between the main and weak \ac{$r$-process} contributions may also be closely linked to the environments in which these stars form. As shown in Figure \ref{fig:Mean_XEu_EuFe}, \ac{$r$-process}-enhanced stars in the ultra-faint dwarf galaxy Reticulum II exhibit systematically lower abundances of light neutron-capture elements compared to $r$-II stars in the Galactic halo \citep{2016Natur.531..610J, 2016ApJ...830...93J, 2018ApJ...856..138J}. Due to its low mass, the \ac{$r$-process} elements in Reticulum II are thought to originate predominantly from a single main \ac{$r$-process} event \citep{2016ApJ...832..149B, 2017MNRAS.471.2088S}. This suggests that $r$-II stars with high Eu abundances likely formed in dwarf galaxies similar to Reticulum II, where contamination from the weak \ac{$r$-process} in core-collapse supernovae is relatively limited, naturally leading to their lower $\mathrm{[Sr, Y, Zr/Eu]}$ ratios.

To investigate the relationship between the surface Eu abundance of stars and their birth environments, and to address whether $r$-II stars preferentially originate in dwarf galaxies, we compute the kinematic properties of \ac{$r$-process}-enhanced stars in our sample \citep{2024ApJ...966..174Z}. To improve statistical robustness, we incorporate additional \ac{$r$-process}-enhanced stars from the literature, primarily from the compilation by \citet{2023ApJ...943...23S} based on the R-Process Alliance and other sources. Photo-astrometric distances from \citet{2022A_A...658A..91A} and astrometric data from Gaia DR3 \citep{2023A_A...674A...1G}  are used to calculate the dynamical and orbital parameters of these stars. Only stars with reliable astrometry (RUWE $<$ 1.4) and relatively precise distances ($\sigma_d/d < 0.4$) are retained, resulting in a final sample of 1602 stars. We adopt a Galactocentric Cartesian coordinate system and assume the position and velocity of the Sun to be $(X,Y,Z)_{\odot} = (-8.21, 0.0, 0.0208)\; \mathrm{kpc}$ \citep{2017MNRAS.465...76M,2019MNRAS.482.1417B} and $(U, V, W)_\odot = (11.1, 12.24, 7.25)\; \mathrm{km\; s^{-1}}$ \citep{2010MNRAS.403.1829S}, respectively, with a local standard of rest velocity $V_{\mathrm{LSR}} = 233.1\; \mathrm{km\; s^{-1}}$. Using the \texttt{AGAMA} software \citep{2019MNRAS.482.1525V} and the Milky Way gravitational potential model from \citet{2017MNRAS.465...76M}, dynamical and orbital properties are computed, including orbital energy ($E$), orbital pericentric radius ($r_{\mathrm{peri}}$), orbital apocentric radius ($r_{\mathrm{apo}}$), orbital eccentricity ($e$), maximum height from the Galactic plane ($Z_{\mathrm{max}}$), and actions ($J_r$, $J_\phi$, $J_z$).

\begin{figure}[h!]
\centering
\includegraphics[width=0.45\textwidth,keepaspectratio]{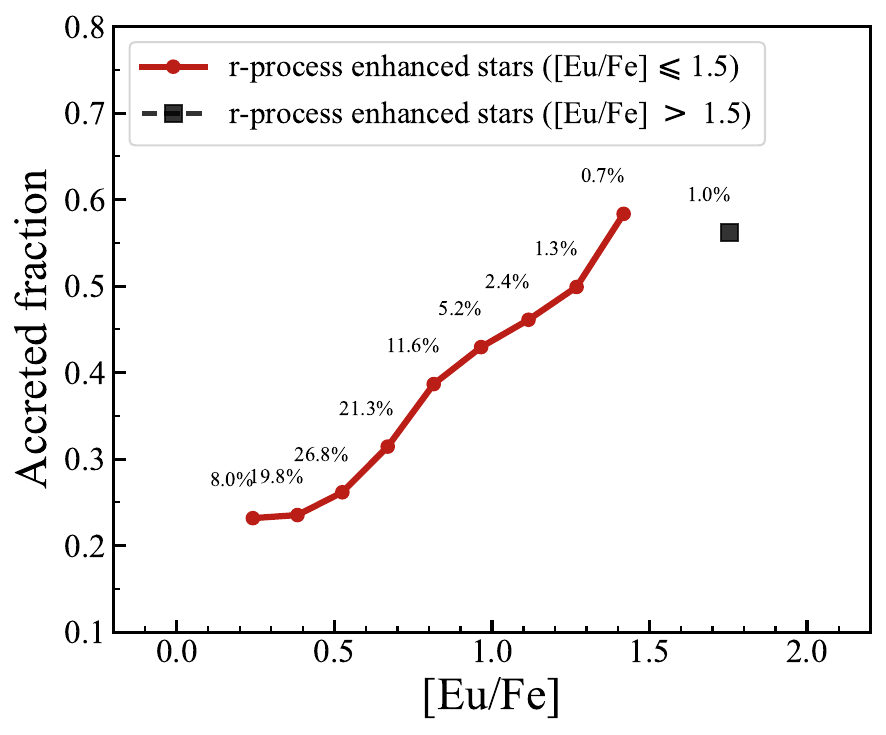}
\caption{Accreted fraction of stars as a function of $\mathrm{[Eu/Fe]}$. The red solid line represents the ex-situ fraction across different $\mathrm{[Eu/Fe]}$ bins, derived using a machine-learning based approach. Stars with [Eu/Fe] $> 1.5$ are grouped into a single bin, denoted by a black solid square. The percentage above each point indicates the proportion of samples in that bin relative to the total sample size. The sample combines r-process-enhanced stars from this work with the literature compilation from \citet{2023ApJ...943...23S}.
\label{fig:Accreted_Fraction_Eu_Fe}}
\end{figure}

Based on the kinematic and orbital properties of the sample stars and machine-learning techniques \citep{2024MNRAS.527.9767L}, we classify them into two categories: $\textit{in-situ}$ stars, which are more likely to have formed within the Milky Way, and $\textit{ex-situ}$ stars, which are more likely to have originated in accreted dwarf galaxies. Figure \ref{fig:Accreted_Fraction_Eu_Fe} shows the fraction of $\textit{ex-situ}$ stars as a function of $\mathrm{[Eu/Fe]}$. To account for measurement uncertainties in $\mathrm{[Eu/Fe]}$, we perform 100 times Monte Carlo resampling within $\pm 0.15$ dex for each star. The proportion of stars in each bin, relative to the total sample size, is indicated above the corresponding data point. Due to the small number of stars with $\mathrm{[Eu/Fe]}$ $> 1.5$ ($N = 14$ in the original sample), they are merged into a single bin, represented by a black solid square. The results show that the fraction of $\textit{ex-situ}$ stars increases with $\mathrm{[Eu/Fe]}$, and this trend persists up to $\mathrm{[Eu/Fe]}$ $\approx 1.5$. However, for stars with $\mathrm{[Eu/Fe]}$ $> 1.5$, the $\textit{ex-situ}$ fraction does not increase further, which may be due to the limited sample size.

These results suggest that stars with higher $\mathrm{[Eu/Fe]}$ are more likely to have formed in dwarf galaxies later accreted by the Milky Way \citep{2025ApJ...990..125H}. Compared to the rapid chemical evolution and prolonged star formation history of the Milky Way, stars in dwarf galaxies have a lower probability of being contaminated by core-collapse supernovae. This results in lower abundances of light neutron-capture elements and Fe in their birth gas clouds, leading to lower $\mathrm{[Sr, Y, Zr/Eu]}$  and higher $\mathrm{[Eu/Fe]}$. 

The above discussion is based on a fundamental assumption that the abundance patterns produced by the weak \ac{$r$-process} and the main \ac{$r$-process} do not exhibit significant variation. However, it is important to note that even within the same astrophysical site, intrinsic variations in \ac{$r$-process} yields may occur, which can likewise lead to differences in the abundance ratios between light and heavy neutron-capture elements. For example, \citet{2017ApJ...836L..21N} demonstrate with \ac{MRDSN} models that in the case of $\hat{L}_\nu = 0.6$, the yields of $\mathrm{[Sr, Y, Zr/Eu]}$ are higher by approximately 1.44 dex compared to the $\hat{L}_\nu = 0.1$ scenario.

\subsubsection{Zn-rich star: J1158+0734} \label{subsec:High Zn star}

\begin{figure}[h!]
\centering
\includegraphics[width=0.45\textwidth,keepaspectratio]{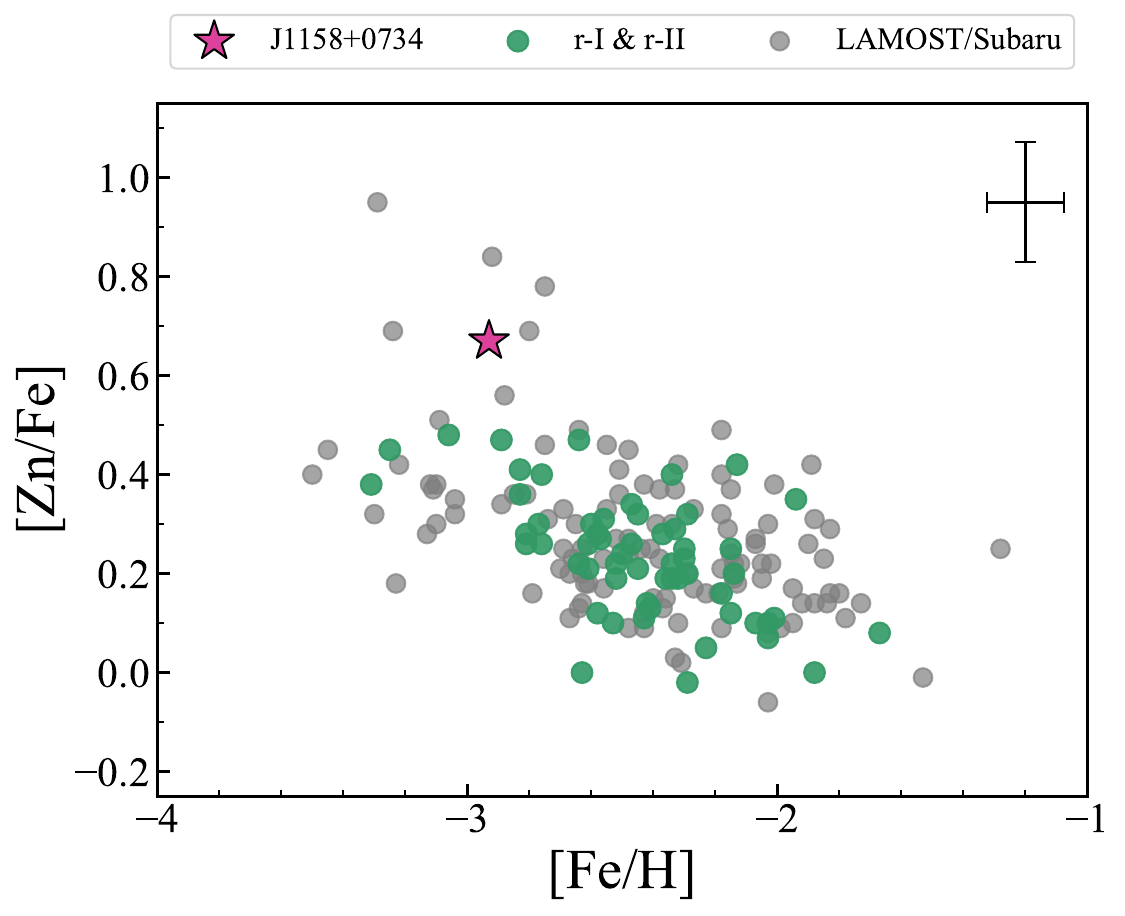}
\caption{The relationship between $\rm \left [ \mathrm{Zn}/\mathrm{Fe} \right ]$ and $\rm \left [ \mathrm{Fe}/\mathrm{H} \right ]$. The gray circles represent all metal-poor stars in the \ac{LAMOST}/Subaru sample \citep{2022ApJ...931..147L}, while the green circles denote \ac{$r$-process}-enhanced stars within the sample. J1158+0734 is highlighted with a magenta pentagon. The mean abundance errors for the entire dataset are indicated in the upper right corner.
\label{fig:ZnFe_FeH}}
\end{figure}

\begin{figure*}[t]
\centering
\includegraphics[width=0.6\textwidth,keepaspectratio]{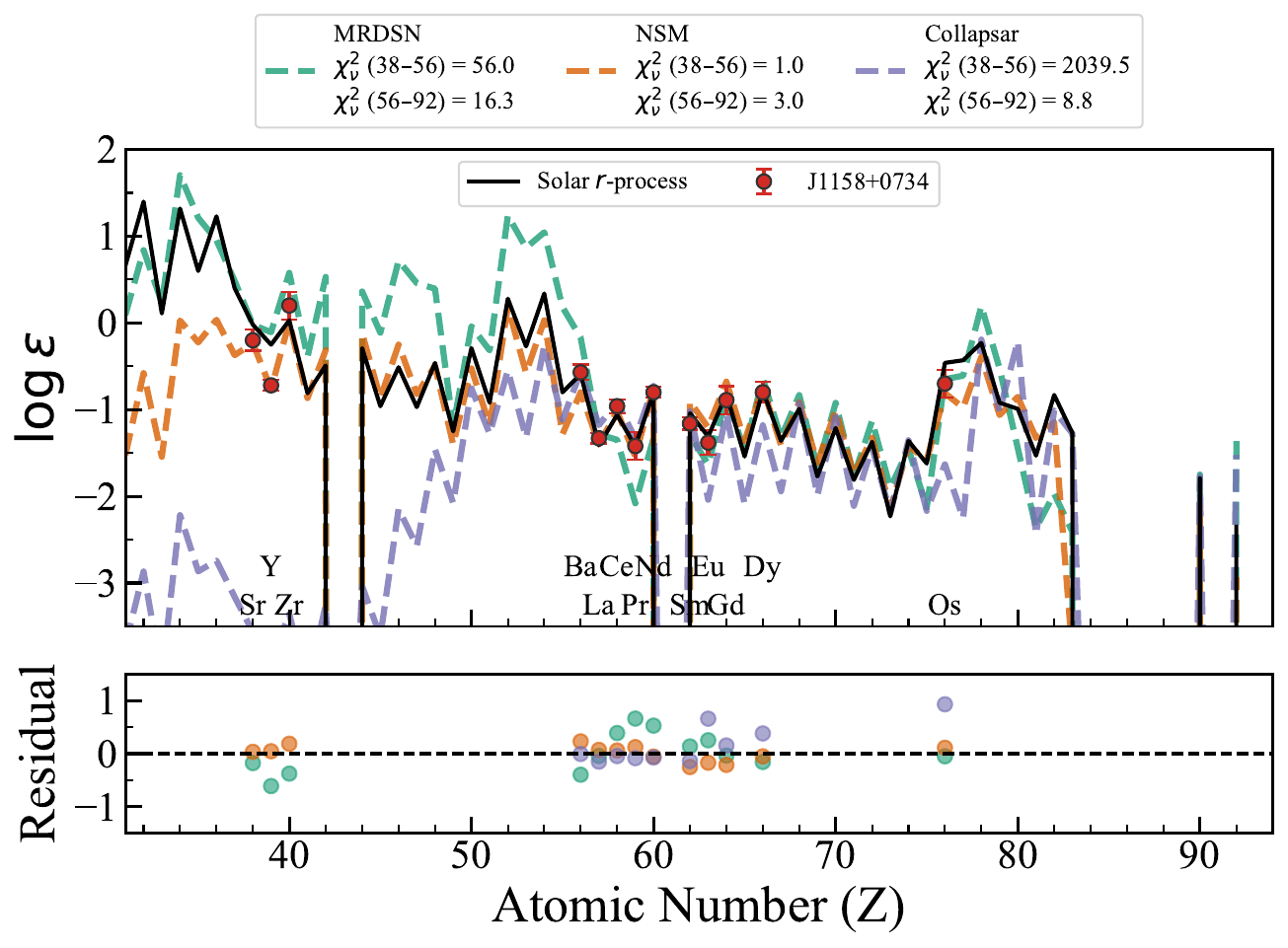}
\caption{Top panel: Neutron-capture element abundance pattern of J1158+0734. Observational data (red points with error bars) are compared with the solar \ac{$r$-process} (black) and three theoretical \ac{$r$-process} nucleosynthesis models: \ac{NSM} (orange), \ac{MRDSN} (green), and Collapsar (dusty purple). Bottom panel: Residuals ($\Delta\log\epsilon$) between observed abundances and three \ac{$r$-process} theoretical models.
\label{fig:J1158_0734_neutron_capture_elements_pattern}}
\end{figure*}

In our sample, we identify an extremely metal-poor ($\rm \left [ \mathrm{Fe}/\mathrm{H} \right ]$ = -2.93) $r$-II star ($\rm \left [ \mathrm{Eu}/\mathrm{Fe} \right ]$ = 1.03), J1158+0734, which exhibits significant Zn enrichment ($\rm \left [ \mathrm{Zn}/\mathrm{Fe} \right ]$ = 0.67) along with moderate Ni enhancement ($\rm \left [ \mathrm{Ni}/\mathrm{Fe} \right ]$ = 0.20), as illustrated in Figure \ref{fig:ZnFe_FeH}. The abundance pattern of this star aligns well with the solar r-process pattern ($F_s$ = 0, $\rm \left [ \mathrm{Ba}/\mathrm{Eu} \right ]$ = -0.85), as shown in Figure \ref{fig:J1158_0734_neutron_capture_elements_pattern}, indicating that its heavy element originated predominantly from a single, vigorous \ac{$r$-process} event. Only several dozen stars with similar characteristics have been identified to date \citep{2005A&A...439..129B, 2012A&A...548A..34A, 2013ApJ...778...56C, 2014AJ....147..136R, 2015ApJ...807..173H, 2015ApJ...807..171J, 2021Natur.595..223Y}.

The nucleosynthetic origin of Zn, the heaviest iron-peak element, remains incompletely understood. Observations reveal that halo stars with -2.5 $\leq$ $\rm \left [ \mathrm{Fe}/\mathrm{H} \right ]$ $\leq$ -1 typically exhibit $\rm \left [ \mathrm{Zn}/\mathrm{Fe} \right ]$ $\approx$ 0.2 \citep{2020ApJ...900..179K}. However, at lower metallicities ($\rm \left [ \mathrm{Fe}/\mathrm{H} \right ] < $ -3), $\rm \left [ \mathrm{Zn}/\mathrm{Fe} \right ]$ shows a pronounced enhancement, increasing as metallicity decreases \citep{2000fist.conf...51P}. In extreme cases, such as a star with $\rm \left [ \mathrm{Fe}/\mathrm{H} \right ]$ = -5.2, $\rm \left [ \mathrm{Zn}/\mathrm{Fe} \right ]$ can reach 0.8 \citep{2019ApJ...876...97E}, suggesting high Zn production efficiency in the early Universe. Theoretical models predict significant Zn yields from the deepest ejecta of hypernovae, which are characterized by high explosion energies and entropies \citep{2006ApJ...653.1145K}. Additionally, massive stars formed in the early universe are expected to have rapid rotation \citep{2006A&A...447..623M}. Recent studies propose that \ac{MRDSNe} --- supernovae with strong magnetic fields and rapid rotation --- can produce extreme Zn enhancements up to $\rm \left [ \mathrm{Zn}/\mathrm{Fe} \right ]$ $\sim$ +1.5 \citep{2015ApJ...810..109N, 2017ApJ...836L..21N, 2018ApJ...863L..27T}. The strength of the magnetic fields in these events further modulates the degree of \ac{$r$-process} enrichment, a scenario that is supported by the \ac{$r$-process} enhanced \ac{EMP} star SMSS 2003-1142 \citep{2021Natur.595..223Y}.

To investigate the origin of the light elements on the surface of J1158+0734, we employ the Bayesian inference framework proposed by \citet{2025arXiv251005712J} to model its surface light-element abundances. For the supernova models, we use the zero-metallicity \ac{CCSN} yields calculated by \citet{2010ApJ...724..341H}, which cover progenitor masses ($M$) in the range of 10--100 $M_\odot$ and explosion energies ($E$) from $0.3\times10^{51}$ to $10^{52}$ erg. Following the description of \citet{2025arXiv251005712J}, we consider only models with rotation-induced mixing (i.e., $\log f_{mix} >-3.0$) and further exclude low-mass hypernova models ($M <$ 20 $M_\odot$ with $E \geqslant$ $10^{51}$ erg) to avoid unrealistic nucleosynthetic yields. The set of light elements selected for abundance pattern fitting also follows their methodology, including C, Na, $\alpha$-elements (Mg, Si, Ca, and Ti), and iron-peak elements (V, Mn, Fe, Co, and Ni). Because of significant discrepancies between observed abundances and theoretical predictions, Sc and Cr are excluded. Furthermore, we extend the element set to include Zn, aiming to explore the progenitor properties of stars with enhanced Zn abundances. The fitting quality is assessed using the reduced chi-square statistic ($\chi^{2}_{\nu}$=$\chi^{2}$/($N$-1)), where $N$ represents the number of neutron-capture elements involved in the fitting. This approach reduces the dependence of the $\chi^{2}$ value on the number of measured abundances, preventing misleading assessments of the fit quality.

\begin{figure*}[t]
\centering
\includegraphics[width=0.8\textwidth,keepaspectratio]{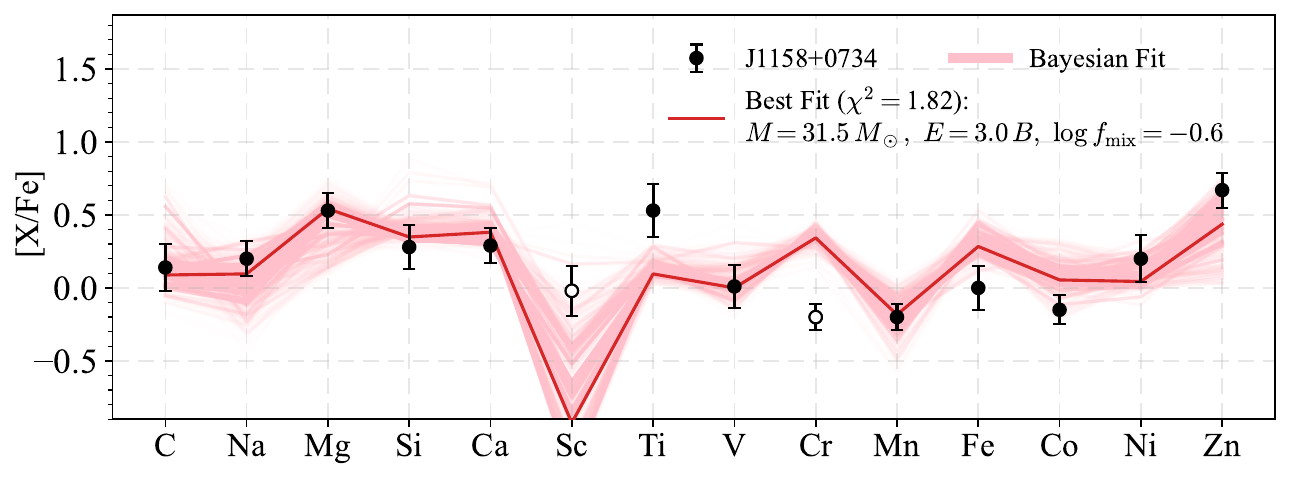}
\caption{Light-element abundance pattern of J1158+0734. The observed abundances are shown as black data points with error bars, where the open black points denote elements excluded from the fitting process, as described in the main text. The dark red solid line represents the best-fit \ac{CCSN} model with a progenitor mass of $M$ = $31.5\,M_\odot$ and an explosion energy of $E$ = $3.0 \times10^{51}$ erg. In addition, the light red solid lines illustrate the \ac{CCSN} abundance patterns corresponding to progenitor properties with posterior probability density $> 1\%$. 
\label{fig:J1158_0734_light_elements_pattern}}
\end{figure*}

Our analysis indicates that the light-element abundances of J1158+0734 are best reproduced by a progenitor star with a mass of 31.5 $M_\odot$ and an explosion energy of 3.0 $B$ (where $B$ = $10^{51}$ erg), as shown by the dark red solid line in Figure \ref{fig:J1158_0734_light_elements_pattern}. This suggests that its chemical enrichment originated from a massive and highly energetic astrophysical source. This result is consistent with the findings of \citet{2019ApJ...876...97E} and \citet{2021Natur.595..223Y}, who demonstrate that \ac{CCSN} models with progenitor masses of 25 $M_\odot$ and explosion energies of 5 $\times 10^{51}$ erg or $10^{52}$ erg successfully reproduce the light-element patterns observed in Zn-enhanced stars.

\begin{figure*}[t]
\centering
\includegraphics[width=0.9\textwidth,keepaspectratio]{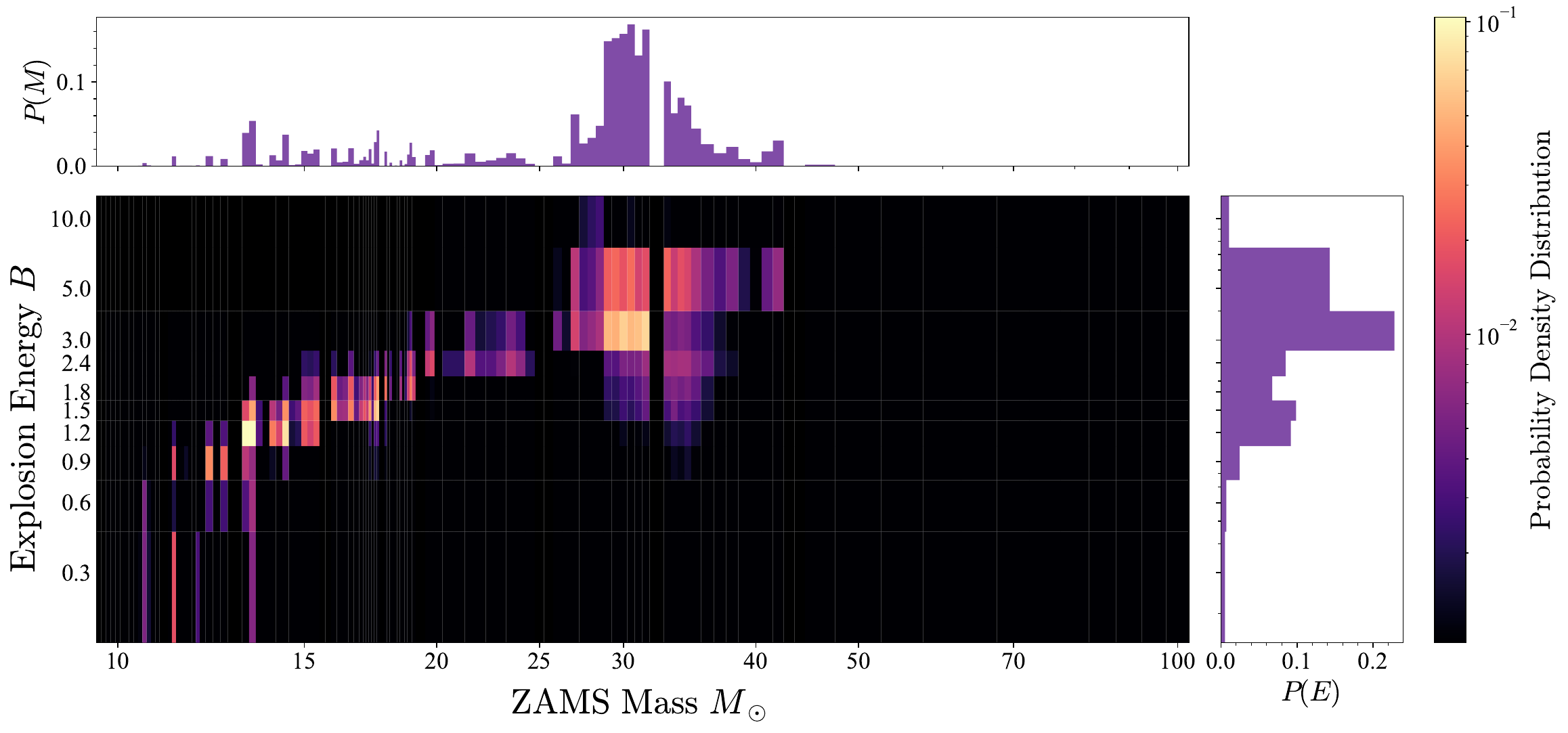}
\caption{Probability density distribution of the progenitor derived from the light-element abundances of J1158+0734. The color scale indicates the joint probability density across different combinations of progenitor mass and explosion energy. The histogram at the top (right) shows the marginal probability distribution of progenitor mass (explosion energy), obtained by integrating over all possible explosion energies (masses).
\label{fig:J1158_0734_Progenitor_distribution}}
\end{figure*}

We note that the best-fit model does not fully reproduce the high Zn abundance observed in J1158+0734 and slightly overestimates the abundances of Fe and Co. Based on a systematic analysis of light-element abundances in 40 \ac{EMP} stars, \citet{2025arXiv251005712J} show that progenitor properties are highly sensitive to the abundance pattern: even small variations within the observational uncertainties can lead to substantial changes in the inferred progenitor mass and explosion energy. Therefore, in addition to reporting the progenitor mass and explosion energy corresponding to the best-fit solution, we also provide the posterior probability density distributions of these properties (Figure \ref{fig:J1158_0734_Progenitor_distribution}) to more comprehensively reflect the relative probabilities across the property space. In Figure \ref{fig:J1158_0734_light_elements_pattern}, we present the abundance patterns of \ac{CCSNe} corresponding to progenitor properties with posterior probability density $> 1\%$ (approximately consistent with a 3$\sigma$ confidence region), shown as light-red solid curves. The two-dimensional posterior probability density distribution indicates that the progenitor capable of reproducing the overall light-element abundances of J1158+0734 is more likely to lie in the massive ($M$ $\sim$ 20--40 $M_\odot$) and high-energy ($E$ $\sim$ 5 $\times 10^{51}$ erg) region.  However, a non-negligible posterior probability also persists in the lower-mass (M $\sim$ 15 $M_\odot$) and lower-energy (E $\sim$ 1.2 $\times 10^{51}$ erg) region, even though progenitors in this region cannot adequately reproduce the star’s elevated Zn abundance. This result suggests that, while the best-fit solution favors a massive, high-energy explosion, weaker explosion scenarios cannot be statistically excluded. Further tests indicate that excluding the Zn abundance from the fitting procedure does not significantly alter the best-fit progenitor mass and explosion energy. However, it does affect the resulting two-dimensional posterior probability density distribution in the mass–energy plane. When Zn is not used as a fitting constraint, the posterior probability density increases in the lower-mass and lower-energy region of the parameter space. This behavior demonstrates that the Zn abundance provides substantial constraining power on the progenitor properties.

To investigate the nucleosynthetic origin of the neutron-capture elements on the surface of J1158+0734, we compare its observed abundance pattern with theoretical predictions from three distinct \ac{$r$-process} models: NSMs \citep{2015MNRAS.448..541J}, \ac{MRDSNe} \citep{2017ApJ...836L..21N}, and collapsars \citep{2024ApJ...966L..37H}, as shown in Figure \ref{fig:J1158_0734_neutron_capture_elements_pattern}. In the heavy neutron-capture element region, \ac{EMP} stars ($\mathrm{[Fe/H]}$ $\sim$ -3) exhibit striking agreement with the solar \ac{$r$-process} residuals, indicating that early-Universe \ac{$r$-process} events already produce a highly stable and universal abundance pattern across the lanthanide region. 

Among the three proposed \ac{$r$-process} production sites, the \ac{NSM} model provides the best match to the observed abundance patterns of the light neutron-capture elements (Sr, Y, Zr), the stable lanthanides (Ba-Dy), and the third \ac{$r$-process} peak (e.g., Os). The best-fitting NSM model corresponds to a 1.35 + 1.8 $M_{\odot}$ neutron-star binary merger computed with the SFHO equation of state. The merger produces a rapidly spinning black-hole–accretion-disk system (spin parameter $A_{BH} \approx 0.8$), with a disk mass of $\sim 0.1\,M_{\odot}$ and a viscosity parameter of $\alpha_{vis} \sim 0.05$. This system ejects a total of $\sim 0.03\,M_{\odot}$ of \ac{$r$-process} material. However, a substantial fraction of NSMs requires several hundred Myr to multiple Gyr of the delay time\citep{2012ApJ...759...52D, 2019MNRAS.487.4847B}. This characteristic makes it difficult for NSMs to contribute widely to \ac{$r$-process} enrichment in the early Galaxy \citep{2004A&A...416..997A, 2015MNRAS.452.1970W}. Consequently, although NSMs are capable of producing \ac{$r$-process} yields consistent with observations, attributing the abundances observed at [Fe/H] $\sim$ -3 solely to standard NSM events generally requires exceptionally short delay times or special host environments, such as that inferred for Reticulum II.


For the collapsar scenario, the predicted abundances of the light neutron-capture elements are systematically lower than the observed values. This discrepancy arises because current collapsar models consider only the relatively narrow jet-nucleosynthesis channel within the black-hole accretion-disk system, while neglecting the contribution from disk winds. In the heavy neutron-capture region (Z $\geqslant$ 63), the collapsar model also exhibits significant deviations from the observations. This behavior results from the production of abundant fission fragments and free neutrons through $\beta$-decay chains and neutron-induced fission, which activates \ac{$i$-process} and \ac{$s$-process} reaction pathways and leads to an abundance pattern with pronounced odd–even structures in the range 140 $<$ A $<$ 170 \citep{2024ApJ...966L..37H}.

For the \ac{MRDSN} scenario, the abundance pattern that best matches the observations corresponds to the case with $\hat{L}v = 0.4$. This model reproduces the overall \ac{$r$-process} abundance trend, though notable discrepancies persist in the Z = 56--60 (Ba–Nd) region. Under the $\hat{L}v = 0.4$ condition, the \ac{MRDSN} jet ejecta maintain a low electron fraction and high entropy, enabling the synthesis of nuclei across a broad mass range---from the light neutron-capture elements to the third \ac{$r$-process} peak. Consequently, such events are considered potential \ac{$r$-process} enrichment sources in the early Universe. Moreover, given that this star exhibits a relatively high Zn abundance together with extremely low metallicity---both characteristic nucleosynthetic signatures predicted by \ac{MRDSN} models---we consider an \ac{MRDSN} a likely source of the elemental enrichment observed on its surface, provided that both the light and heavy elements in J1158+0734 originate from the same progenitor.

It is important to note that the differences among \ac{$r$-process} models in the lanthanide region arise not only from astrophysical conditions but also from significant uncertainties in nuclear physics inputs \citep{2019PrPNP.107..109K}. Variations in nuclear mass models, $\beta$-decay rates, and fission fragment distributions all influence the resulting abundance patterns. Consequently, uncertainties in nuclear physics inputs constitute a major limitation in precisely identifying the astrophysical origin of the \ac{$r$-process}. It is also noteworthy that observational constraints in key diagnostic regions of the \ac{$r$-process} remain limited. These regions typically exhibit pronounced abundance variations across different \ac{$r$-process} models, particularly between the first and second \ac{$r$-process} peaks (e.g., Ru, Rh, Pd, Ag), within the lanthanides (Tm, Lu), and in the third \ac{$r$-process} peak (Ir, Pt). The spectral lines of these elements lie predominantly in the blue/ultraviolet wavelength range ($\lambda < 4000$~$\angstrom$), which makes accurate abundance measurements especially challenging.  Future spectroscopic observations with high signal-to-noise ratios (S/N $>$ 100) in the blue–UV range will enable a more comprehensive determination of the neutron-capture element abundances in J1158+0734.

\subsection{CEMP stars} \label{subsec:CEMP stars}


\ac{CEMP} stars are conventionally defined as stars exhibiting $\rm \left [ \mathrm{Fe}/\mathrm{H} \right ] <$ -1.0 and $\rm \left [ \mathrm{C}/\mathrm{Fe} \right ] >$ 1.0 \citep{2005ARA_A..43..531B}. A more detailed classification criterion was subsequently developed by \citet{2007ApJ...655..492A} through their analysis of 64 \ac{CEMP} stars, incorporating both carbon abundance thresholds ($\rm \left [ \mathrm{C}/\mathrm{Fe} \right ]$ $\geq$ +0.7) and luminosity criteria. Application of this refined classification criterion to our sample identifies 12 confirmed \ac{CEMP} stars, with their specific subclasses listed in Column (14) of Table \ref{tab:CEMP_stars}. Particularly noteworthy are several objects marked with asterisks in Table \ref{tab:CEMP_stars}, which exhibit carbon abundances slightly below the enhancement threshold after accounting for evolutionary effects. They demonstrate heavy-element patterns similar to their categorized analogues, as described in Section \ref{subsec:classification of the sample star}. Such cases may indicate that their surface carbon has undergone additional mixing, leading to carbon depletion, as described in Section \ref{subsec:CEMP-s stars}.

It is worth noting that the criterion for classifying \ac{CEMP} stars may be significantly influenced by the initial carbon abundances of a stellar system and its subsequent chemical evolution history. Recent observations reveal that dwarf galaxies possess global carbon abundances distinct from those of the Milky Way \citep{2013A&A...549A..88S, 2015A&A...583A..67J, 2015A&A...574A.129S, 2023MNRAS.526.2620V, 2024A&A...681A..44S, 2024A&A...686A.266L, 2024A&A...690A.333S}. Consequently, in systems with intrinsically low mean $[ \mathrm{C/Fe}]$, signatures of carbon enrichment are expected to manifest below the conventional +0.7 threshold, consistent with the moderately carbon-enhanced (\ac{CEMP}$^{*}$) stars identified in our sample. Despite these potential environmental differences, the \ac{CEMP}$^{*}$ stars do not display the retrograde motions or extreme $J_\phi$ values typically associated with accreted dwarf-galaxy debris; instead, they exhibit orbital kinematics consistent with in-situ Milky Way halo membership. To further clarify this issue, future systematic studies of the kinematics of stars with moderate carbon enhancement (e.g., $0.4 < \mathrm{[C/Fe]} < 0.7$) are essential to investigate whether they originate from specific dwarf galaxy systems, thereby providing deeper insights into their formation mechanisms.

\subsubsection{CEMP-s stars} \label{subsec:CEMP-s stars}

Our sample contains eight \ac{CEMP}-$s$ stars, which exhibit high effective temperatures ($T_{\rm eff}$  $>$ 6000 K), substantial surface gravities ($\log \textit{g} $  $>$ 3.5), and faint magnitudes (V$_{\mathrm{average}}$ $\approx$ 14.09). These characteristics lead to significantly weakened absorption features and an average \ac{S/N} $\approx$ 43 at 4500$\angstrom$. These observational constraints hinder precise measurements of neutron-capture element abundances, restricting our analysis to only limited number of elements in most \ac{CEMP}-$s$ stars. For instance, in J1020+4046, only Sr and Ba abundances are detectable. Notably, we successfully measured the abundance of Pb, a characteristic third-peak \ac{$s$-process} element, in only two \ac{CEMP}-$s$ stars. Our sample contains J2256+0215, a star with $\rm \left [\mathrm{C}/\mathrm{Fe} \right ]$ = 0.07 but exhibits a significant \ac{$s$-process} enhancement, suggesting a common origin with \ac{CEMP}-$s$ stars but requiring additional carbon depletion mechanisms to explain its current composition.



To investigate the origin of surface elements in our \ac{CEMP}-$s$ sample, we evaluate binarity through radial velocity measurements. For each star, we compile radial velocity data determined from high-resolution Subaru spectroscopic observations \citep{2022ApJ...931..146A}, supplemented  by Gaia DR3 \citep{2023A_A...674A...1G}, LAMOST DR11 \autofootlink{https://www.lamost.org/dr11/v1.0}, APOGEE DR17 \citep{2022ApJS..259...35A}, and RAVE DR6 \citep{2020AJ....160...82S} measurements for comparison. It is important to emphasize that, when using LAMOST DR11 data, we retain only observations with $FIBMASK$ = 0 in order to avoid radial velocity biases introduced by fiber anomalies. For stars with multi-epoch observations, we first compute the weighted mean radial velocity, using the inverse square of the measurement uncertainties as weights, which serves as the reference baseline. We then compute a chi-square statistic from the deviations of each epochal measurement from this weighted mean to quantify the level of radial-velocity scatter. Finally, a chi-square test is applied to determine whether these deviations can be reasonably explained by the reported observational uncertainties, thereby assessing whether the star exhibits significant radial-velocity variability. In particular, we calculate the corresponding $p$-value from the chi-square distribution, and stars with $p < 0.01$ are classified as radial-velocity variable candidates. For stars with only two radial-velocity measurements ($N$ = 2), the chi-square test becomes statistically unreliable because the degrees of freedom are limited to one. In such cases, we additionally employ an error-weighted velocity-difference metric, defined as $z = |\Delta v| / \sigma_{\Delta v}$, to assess whether the observed radial-velocity difference is significant relative to the combined measurement uncertainties. Stars with $z > 3$ are considered to exhibit significant radial-velocity variations. Table \ref{tab:CEMP_stars} presents the weighted mean and standard deviation of the $N$ radial-velocity measurements for these stars, together with the corresponding $p$-values (or $z$-values for $N$ = 2) and the assessment of their binary nature based on the available data. The individual radial-velocity measurements from different surveys for each star are provided in Table \ref{tab:rv_measurements}.

Among the nine \ac{CEMP}-$s$ stars in our sample (including J2256+0215), five exhibit significant radial velocity variations. This strongly supports the conclusion that their surface abundance patterns originate from mass transfer from more massive \ac{AGB} companions, which have since evolved into undetectable white dwarf remnants \citep{2011MNRAS.418..284B, 2012MNRAS.422..849B}. The remaining four stars show no detectable variations across two to three radial-velocity measurements. With the exception of J2256+0215, all CEMP-s stars in our sample display elevated carbon abundances (A(C) $>$ 7.1) --- a characteristic strongly correlated with binarity in the study of \citet{2016ApJ...833...20Y}, who analyzed 57 \ac{CEMP} stars with established companion status. This suggests that even these seemingly single \ac{CEMP}-$s$ stars may have acquired their surface composition through binary interaction, possibly with their current orbital configurations preventing the detection of radial velocity variations.


Our examination of $\rm \left [ \mathrm{X}/\mathrm{Fe} \right ]$ versus $\rm \left [ \mathrm{Fe}/\mathrm{H} \right ]$ trends in Section \ref{subsec:comparison with Literature} reveals substantial abundance dispersion among \ac{CEMP}-$s$ stars, particularly for \ac{$s$-process} elements. The $\rm \left [ \mathrm{Sr}/\mathrm{Fe} \right ]$ ratios span over 0.9 dex, while $\rm \left [ \mathrm{Ba}/\mathrm{Fe} \right ]$ variations exceed 1.8 dex, which suggests that these stars are enriched by progenitor \ac{AGB} stars exhibiting diverse properties. Additionally, the observed abundance patterns may further depend on the amount of material transferred from the \ac{AGB} companions, as well as the degree of dilution and mixing within the recipient stars. To constrain the nature of these progenitors, we employ a grid of \ac{AGB} nucleosynthesis models \citep{2010MNRAS.404.1529B} spanning stellar masses from 1.3 M$_\odot$ to 2.0 M$_\odot$, [Fe/H] from -3.6 to -1.0, and s-process efficiencies ranging from ST/150 to 2$\times$ST, in order to fit the observed abundance patterns of \ac{CEMP}-$s$ stars. Our fitting methodology incorporates several constraints: including consistency in the overall neutron-capture element distribution, the star's $[\mathrm{Fe/H}]$ as a metallicity indicator, and key light element ratios ($[\mathrm{Na/Fe}]$, $[\mathrm{Mg/Fe}]$) that sensitively trace progenitor mass. Table \ref{tab:agb_models} summarizes the optimal model parameters for each \ac{CEMP}-$s$ star, along with diagnostic abundance ratios (columns (2)--(8)) that collectively characterize the properties of their \ac{AGB} progenitors.

Our sample contains two notable stars, J0119-0121 and J0949+2707, both of which exhibit exceptional enhancements in heavy $s$-process element ($\rm \left [ \mathrm{hs}/\mathrm{Fe} \right ]$ $>$ 2.5) along with significantly elevated Na and Mg abundances ($\rm \left [ \mathrm{Na}/\mathrm{Fe} \right ]$ $>$ 1.5; $\rm \left [ \mathrm{Mg}/\mathrm{Fe} \right ]$ $>$ 1.0). These chemical signatures indicate enrichment from relatively massive \ac{AGB} progenitors, as higher-mass \ac{AGB} stars experience more frequent \ac{TDU} events during thermal pulses, resulting in increased production of Na, Mg, and \ac{$s$-process} elements. In addition, metallicity plays a crucial role in AGB \ac{$s$-process} nucleosynthesis: in low‑metallicity environments, the reduced number of initial iron seed nuclei allows more neutrons to be captured per seed, thereby significantly enhancing the efficiency of the \ac{$s$-process} and favoring the production of heavy $s$-peak elements (e.g., Ba, La, Ce) as well as Pb.

For J0119-0121, the best-fitting model ($\chi^{2}_{\nu}$ = 4.38, considering only elements with Z $\geq$ 38) corresponds to an \ac{AGB} progenitor with a mass of 1.3 M$_\odot$ and a metallicity of $[\mathrm{Fe/H}]$ = -3.0, adopting an \ac{$s$-process} efficiency of ST/12 and a dilution factor of 0.56 dex. A positive dilution factor may indicate that the natal gas cloud of J0119-0121 is already mildly enriched in \ac{$s$-process} material prior to its formation, or it may simply reflect uncertainties arising from the simplified treatment of \ac{AGB} nucleosynthetic yields and mass-transfer/mixing processes. The star’s relatively high Na, Mg, and heavy \ac{$s$-process} element abundances likely represent characteristic signatures of \ac{$s$-process} nucleosynthesis operating in low-metallicity environments. \citet{2019PASJ...71...89Z} reported that the best-fitting AGB progenitor model for J0119-0121 is 1.4 M$_\odot$ with $\mathrm{[Fe/H]}$ = -2.6 and ST/4.5. This small difference in the parameter set compared to \citet{2019PASJ...71...89Z} primarily arises from differences in the adopted observational abundance data. \citet{2019PASJ...71...89Z} derived additional heavy-element abundances, including Eu, Er, and Yb, from ultraviolet--blue spectra. Moreover, their adopted Mg abundance ($\mathrm{[Mg/Fe]}$ = 0.94) is lower than the value reported by \citet{2022ApJ...931..147L}, which reduces the need for invoking a relatively higher-mass \ac{AGB} progenitor. Using their observational data, we also obtain a good fit with the 1.4 M$_\odot$, $\mathrm{[Fe/H]}$ = -2.6, ST/4.5 model ($\chi^{2}_{\nu}$ = 4.28), consistent with the results of \citet{2019PASJ...71...89Z}. The abundance pattern of J0949+2707 is best reproduced by a model with $\rm \left [ \mathrm{Fe}/\mathrm{H} \right ]$ = -2.6, a progenitor mass of 2.0 M$_\odot$, and an s-process efficiency of ST/6 with a dilution factor of -0.20 dex, although models employing 1.4 and 2.0 M$_\odot$ (ST/9, $\rm \left [ \mathrm{Fe}/\mathrm{H} \right ]$= -2.6) also yield reasonable fits. Constraints for this object remain limited, as only Ba and La among the heavy \ac{$s$-process} elements are currently detected.

\begin{figure}[h]
\centering
\includegraphics[width=0.45\textwidth,keepaspectratio]{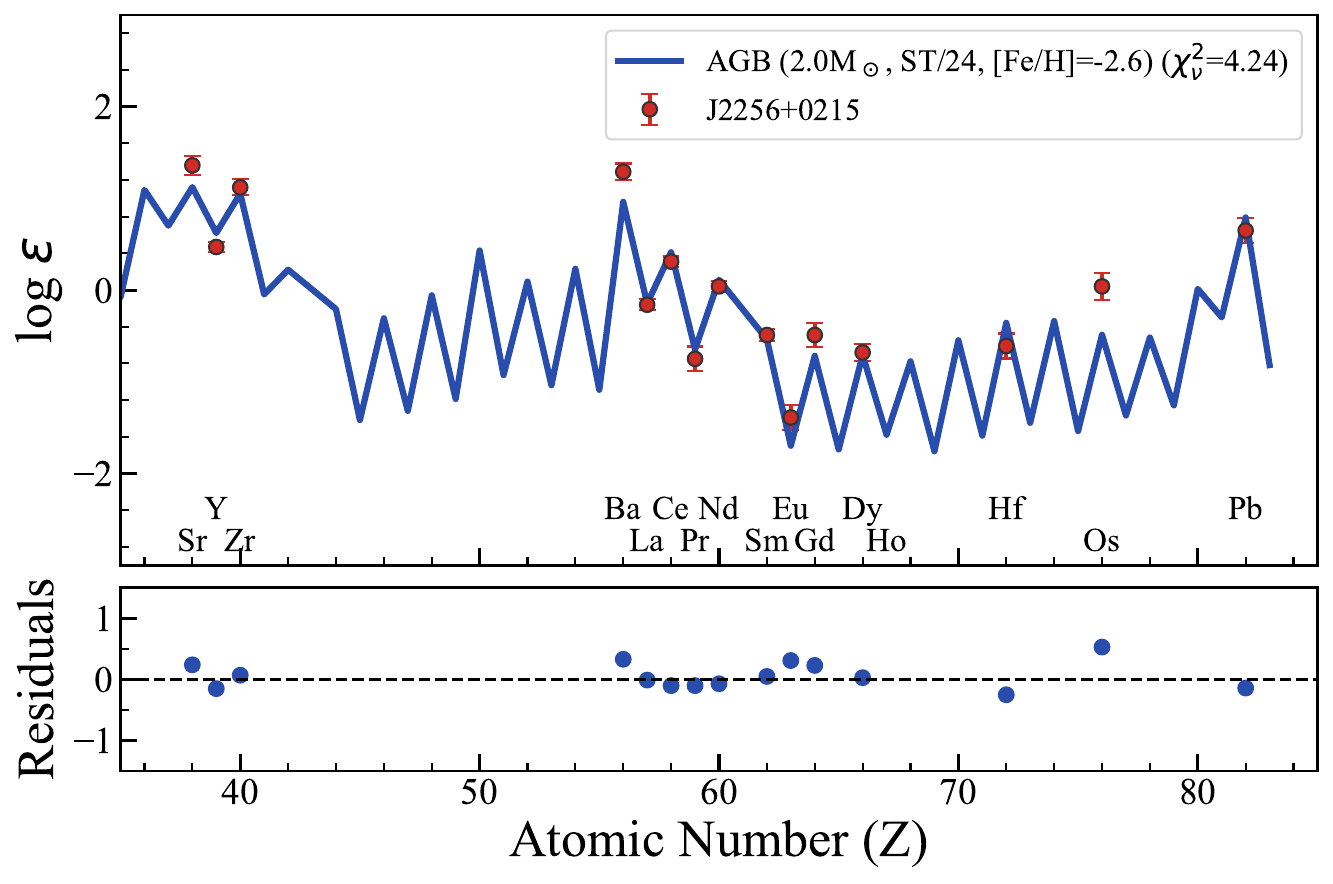}
\caption{Top panel: The abundance pattern of J2256+0215, where solid red circles with error bars represent the observed abundances, and the blue solid line illustrates the theoretical \ac{AGB} \ac{$s$-process} model predictions for a 2.0 M$_\odot$ star with $\rm \left [ \mathrm{Fe}/\mathrm{H} \right ]$ = -2.6 and ST/24. Bottom panel: Residuals between the observed abundances and theoretical model predictions.
\label{fig:CEMP-s abundance pattern}}
\end{figure}

J2256+0215 is a \ac{VMP} giant star (${\mathrm{[Fe/H]}}$ = -2.27; $T_{eff}$ = 5044 K; $\log g$ = 1.68) that exhibits pronounced s-process enrichment (${\mathrm{[Ba/Fe]}}$ = +1.38; ${\mathrm{[Ba/Eu]}}$ = +1.02). Its observed heavy-element abundance pattern is well reproduced by a 2.0 M$_{\odot}$ AGB model at ${\mathrm{[Fe/H]}}$ = -2.6 with \ac{$s$-process} efficiency of ST/24, as illustrated in Figure \ref{fig:CEMP-s abundance pattern}. The model requires a dilution factor of -1.40, which aligns with \ac{FDU} predictions. This star also shows a markedly enhanced ${\mathrm{[C+N/Fe]}}$ ratio of 1.6, a characteristic commonly found in \ac{CEMP}-$s$ stars (${\mathrm{[C+N/Fe]}} \gtrsim $ 1) \citep{2010A&A...509A..93M, 2015ApJ...807..173H}. Notably, its evolution-corrected carbon abundance is only moderately enhanced (${\mathrm{[C/Fe]}}_{corr}$ = 0.44), indicating that the surface carbon abundance has undergone additional depletion during stellar evolution, accompanied by a corresponding increase in nitrogen. The carbon isotopic ratio ($^{12}\mathrm{C}/^{13}\mathrm{C}$) serves as a key diagnostic of internal mixing processes. Primordial gas generally exhibits high values of $^{12}\mathrm{C}/^{13}\mathrm{C} > $ 70 \citep{2006A&A...455..291S}. Standard convective models predict that the FDU reduces this ratio from ($\sim$ 60--80) to ($\sim$ 20--30). As the star evolves toward the red giant branch (RGB) bump, a molecular weight inversion triggered by $^{3}{\rm He}$ burning initiates thermohaline mixing \citep{2007A&A...467L..15C}. This process rapidly transports material into layers where CN cycling is efficient, thereby reducing the $^{12}\mathrm{C}/^{13}\mathrm{C}$ ratio toward the CNO-cycle equilibrium value of $\sim$ 3--4 \citep{2000A&A...354..169G, 2006A&A...455..291S, 2023MNRAS.524.4418M, 2024A&A...684A..70L}. For J2256+0215, the measured $^{12}\mathrm{C}/^{13}\mathrm{C}$ ratio is only 2.3 $\pm$ 1.5, and its luminosity ($\log(L/L_{\odot}) = 2.426$) is comparable to both the RGB-bump luminosity reported by \citealt{2000A&A...354..169G} ($\log(L/L_{\odot}) \approx 2$) and the extrapolated RGB-bump luminosity for EMP stars from \citealt{2006A&A...455..291S} ($\log(L/L_{\odot}) \approx 2.6$). These properties together strongly suggest that the star has undergone deep extra mixing driven by thermohaline instability. Furthermore, the mixing efficiency is likely enhanced by additional processes---such as rotation-induced \citep{2000A&A...361..159M, 2010A&A...522A..10C, 2012A&A...543A.108L} or magnetically mediated mixing \citep{2004A&A...422..225M, 2007ApJ...671..802B} ---which accelerate surface carbon depletion and promote nitrogen enrichment.

The remaining \ac{CEMP}-$s$ stars in our sample show no significant Na or Mg enhancements ($\rm \left [ \mathrm{Na}/\mathrm{Fe} \right ]$ $<$ 1.0, $\rm \left [ \mathrm{Mg}/\mathrm{Fe} \right ]$ $<$ 1.0), consistent with enrichment from lower-mass \ac{AGB} progenitors (1.3 M$_\odot$). However, our modeling reveals substantial variation in the $^{13}$C pocket sizes --- indicative of differences in \ac{$s$-process} efficiencies --- among these progenitors, ranging from ST to ST/18. These variations are clearly reflected in the corresponding light $s$-process ($\rm \left [ \mathrm{ls}/\mathrm{Fe} \right ]$) and heavy $s$-process ($\rm \left [ \mathrm{hs}/\mathrm{Fe} \right ]$) abundance ratios. Notably, J0824+3025 exhibits the lowest \ac{$s$-process} efficiency (ST/18) while displaying the highest $\rm \left [ \mathrm{ls}/\mathrm{Fe} \right ]$ and $\rm \left [ \mathrm{hs}/\mathrm{Fe} \right ]$ values. For 1.3 M$_\odot$ \ac{AGB} models at $\rm \left [ \mathrm{Fe}/\mathrm{H} \right ]$ = -2.6, both $\rm \left [ \mathrm{ls}/\mathrm{Fe} \right ]$ and $\rm \left [ \mathrm{hs}/\mathrm{Fe} \right ]$ increase
as the \ac{$s$-process} efficiency decreases across the ST/1.5 to ST/30 range. Although the $\rm \left [ \mathrm{Pb}/\mathrm{hs} \right ]$ ratio could provide additional constraints on \ac{$s$-process} efficiency, Pb is detected only in J1434+3540. This star shows a relatively low $\rm \left [ \mathrm{Pb}/\mathrm{hs} \right ]$ ratio, consistent with its higher metallicity, where the reduced neutron-to-seed ratio suppresses production at the third \ac{$s$-process} peak. However, our best-fit \ac{AGB} model yields a $\rm \left [ \mathrm{Pb}/\mathrm{hs} \right ]$ ratio that is 0.61 dex lower than the observed value. While low-metallicity or high \ac{$s$-process} efficiency models can better reproduce $\rm \left [ \mathrm{Pb}/\mathrm{hs} \right ]$, they still have difficulty in simultaneously matching the observed $\rm \left [ \mathrm{ls}/\mathrm{Fe} \right ]$ and $\rm \left [ \mathrm{hs}/\mathrm{Fe} \right ]$ ratios.

\begin{table*}[ht]
\centering
\caption{Best-fit AGB $s$-process models for CEMP-$s$ and CEMP-$r$+$s$ stars.}
\label{tab:agb_models}
\begin{tabular}{cccccccccccc}
\toprule
Name & [Fe/H]$_{\rm Star}$ & [Na/Fe] & [Mg/Fe] & [ls/Fe] & [hs/Fe] & [Pb/hs] & [hs/ls] & $M_{\rm AGB}$ ($M_\odot$) & [Fe/H]$_{\rm AGB}$ & ST & $O_s$ \\
\midrule
\multicolumn{11}{c}{CEMP-$s$} \\  
\hline
J0119-0121 & -3.19 & 1.94 & 1.12 & 1.39 & 2.59 & 1.23 & 1.20 & 1.3 & -3.00 & ST/12 & 0.56 \\
J0824+3025 & -2.72 & ... & 0.51 & 1.27 & 2.26 & ... & 0.99 & 1.3 & -2.60 & ST/18 &  0.10 \\
J0949+2707 & -2.90 & 1.83 & 1.11 & 1.97 & 2.98 & ... & 1.01 & 2.0 & -2.60 & ST/6 & -0.20 \\
J1003-0358 & -2.18 & 0.38 & 0.30 & 0.78 & 1.58 & ... & 0.80 & 1.3 & -2.30 & ST/12 & -0.40 \\
J1011+3524 & -2.25 & 0.41 & 0.45 & 0.49 & 1.49 & ... & 1.00 & 1.3 & -2.60 & ST/9 & 0.10 \\
J1020+4046 & -2.94 & 0.73 & 0.51 & 0.35 & 1.16 & ... & 0.81 & 1.3 & -2.60 & ST/3 & -0.20 \\
J1044-0358 & -2.13 & 0.59 & 0.38 & 0.70 & 1.33 & ... & 0.63 & 1.3 & -2.00 & ST & 0.30 \\
J1434+3540 & -1.91 & 0.28 & 0.40 & 0.58 & 1.29 & 0.84 & 0.71 & 1.3 & -2.00 & ST/12 & -0.80 \\
J2256+0215 & -2.27 & 0.82 & 0.38 & 0.70 & 1.13 & 0.43 & 0.04 & 2.0 & -2.60 & ST/24 & -1.40 \\
\hline
\multicolumn{11}{c}{CEMP-$r$+$s$} \\  
\hline
J0653+4735 & -2.41 & 0.19 & 0.48 & 0.32 & 1.09 & 0.85 & 0.77 & 1.4 & -2.60 & ST/9 & -1.30 \\
J2114-0616 & -2.55 & 0.61 & 0.34 & 0.21 & 1.27 & 0.78 & 1.06 & 1.5 & -2.60 & ST/6 & -1.60 \\
J2257+3859 & -2.65 & 0.38 & 0.38 & 0.38 & 1.17 & 1.17 & 0.79 & 1.3 & -3.00 & ST/12 & -0.30 \\
\bottomrule
\end{tabular}
\begin{tablenotes}
\item Note. 
\footnotesize{$\mathrm{[ls/Fe]}$ = $\frac{1}{3}$($\mathrm{[Sr/Fe]}$ + $\mathrm{[Y/Fe]}$ + $\mathrm{[Zr/Fe]}$);
$\mathrm{[hs/Fe]}$ = $\frac{1}{3}$($\mathrm{[Ba/Fe]}$ + $\mathrm{[La/Fe]}$ + $\mathrm{[Ce/Fe]}$)}
\end{tablenotes}
\end{table*}

\subsubsection{CEMP-r/s and CEMP-r+s stars} \label{subsec:CEMP-r/s and CEMP-r+s stars}

\begin{figure*}[t]
    \centering
        \includegraphics[width=0.48\linewidth]{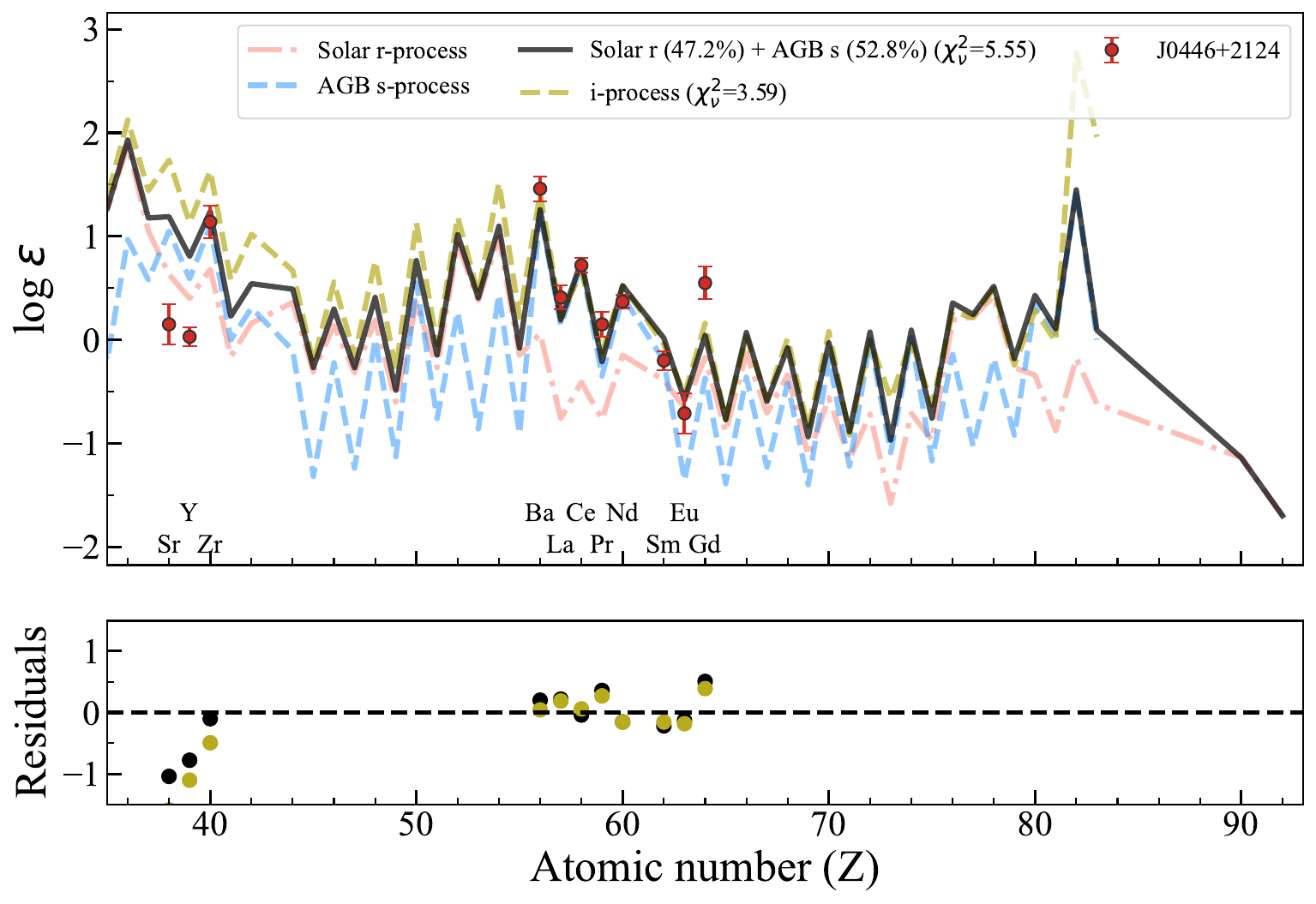}
        \hspace{0.1cm} 
        \includegraphics[width=0.48\linewidth]{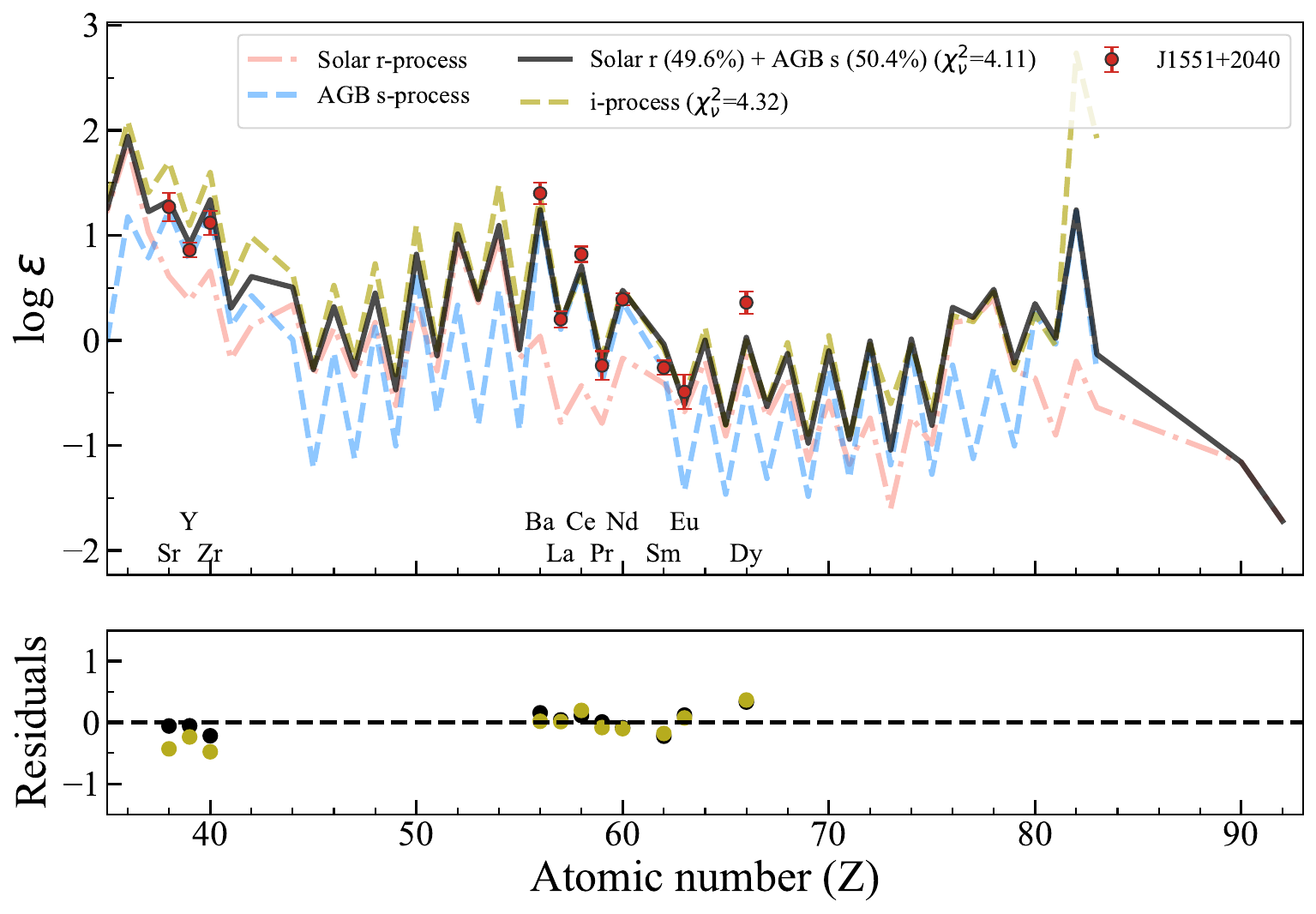}
    \caption{Abundance patterns of \ac{CEMP}-$r$/$s$ stars in our sample. The observed abundances are represented by yellow solid circle with error bars. Theoretical predictions are shown as follows: the solar \ac{$r$-process} abundance pattern (red dotted line), \ac{AGB} \ac{$s$-process} abundance pattern (blue dashed line), combined solar \ac{$r$-process} with \ac{AGB} \ac{$s$-process} abundance pattern (black solid line), and \ac{$i$-process} abundance pattern (yellow solid line). The bottom panel of each plot displays the corresponding residuals between the observed abundances and both the $r$+$s$-process and \ac{$i$-process} model predictions.
    }
    \label{fig:CEMP-r/s abundance pattern}
\end{figure*}

\begin{figure*}[t]
    \centering
    \includegraphics[width=0.7\linewidth]{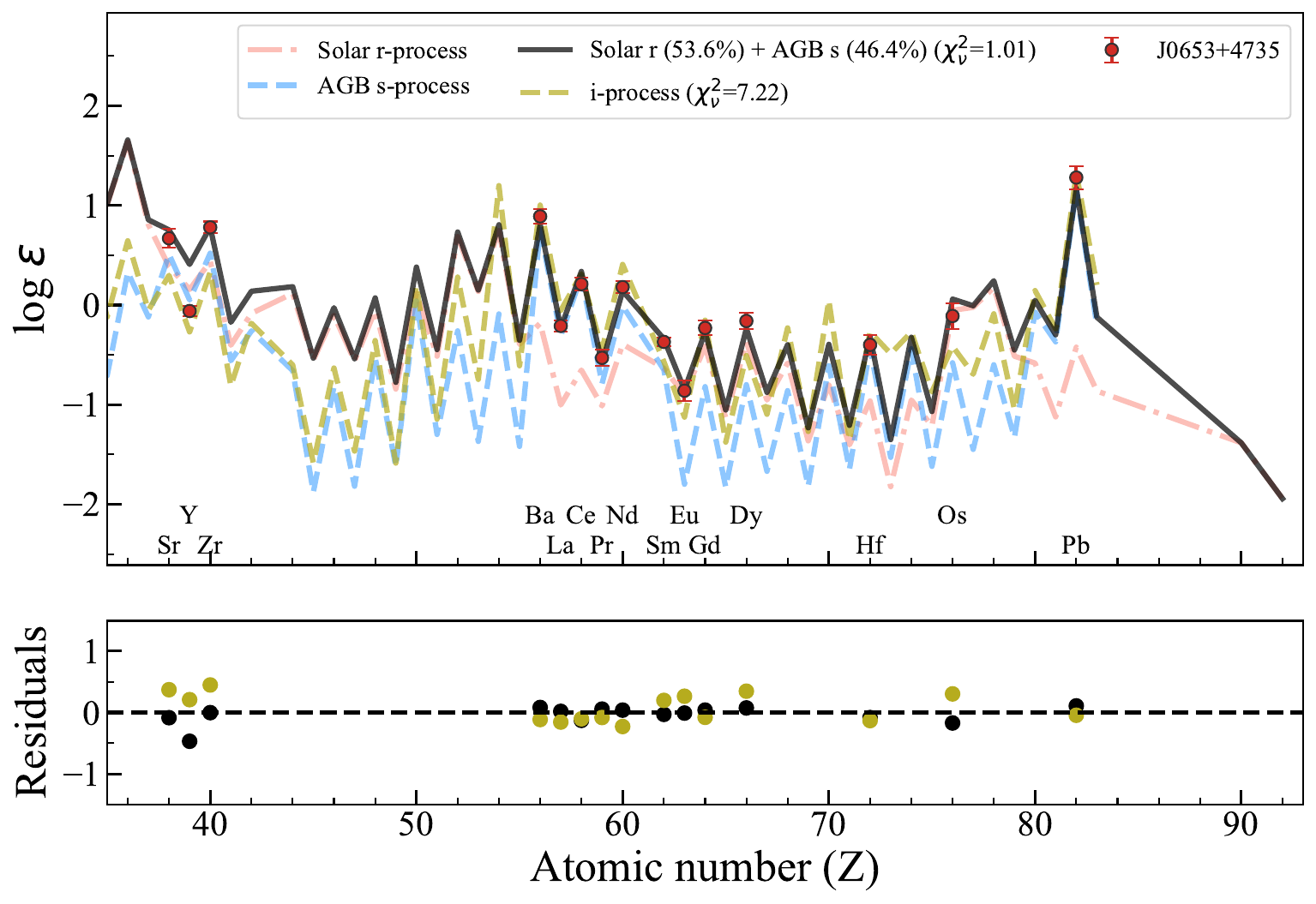}
    \vspace{0.5cm}
    \caption{The abundance pattern of J0653+4735, a representative \ac{CEMP}-$r$+$s$ star in our sample, with symbols and line styles following the same conventions as in Figure \ref{fig:CEMP-r/s abundance pattern}.
    }
    \label{fig:CEMP-r+s abundance pattern}
\end{figure*}

In contrast to \ac{CEMP}-$s$ stars, \ac{CEMP}-$r$/$s$ and \ac{CEMP}-$r$+$s$ stars exhibit enhancements in both \ac{$s$-process} elements (e.g., Ba, La, Ce) and \ac{$r$-process} elements (e.g., Eu, Gd, Dy). We identified five such stars in our sample, as presented in Table \ref{tab:CEMP_stars}. Among them, J0446+2124 and J0653+4735 display only moderate carbon enrichment ($\rm \left [ \mathrm{C}/\mathrm{Fe} \right ]_{corr} $ = 0.57 $\pm$ 0.30 and 0.6 $\pm$ 0.10, respectively). To investigate their origins, we employ composite models combining solar \ac{$r$-process} and \ac{AGB} \ac{$s$-process} abundance patterns \citep{2010MNRAS.404.1529B}, as well as \ac{$i$-process} models \citep{2024A&A...684A.206C} to fit their surface abundance patterns, as illustrated in Figures \ref{fig:CEMP-r/s abundance pattern}, \ref{fig:CEMP-r+s abundance pattern}, and \ref{fig:Other CEMP-r+s abundance pattern}. It is important to note that the fitting procedure excludes the light neutron-capture elements (e.g., Sr, Y, Zr), as their abundances could be significantly affected by additional nucleosynthetic processes, particularly the weak \ac{$r$-process}. The results demonstrate systematic overabundances of 0.4--1.56 dex relative to solar \ac{$r$-process} predictions in the Ba-Nd region (Z=56-60) and excesses of 0.2-1.0 dex compared to AGB \ac{$s$-process} model predictions in the Eu-Dy region (Z=63-71) for all investigated stars. These systematic discrepancies strongly indicate the necessity of adopting more complicated nucleosynthesis mechanisms to fully explain the observed abundance patterns.


J0446+2124 and J1551+2040 exhibit distinctive heavy-element abundance patterns that cannot be fully reproduced by either $r$+\ac{$s$-process} composite models or \ac{$i$-process} nucleosynthesis models (Figure \ref{fig:CEMP-r/s abundance pattern}). In particular, the observed Gd abundance in J0446+2124 and the Dy abundance in J1551+2040 are underpredicted by approximately 0.35 dex in all theoretical models. This discrepancy suggests either contributions from unrecognized nucleosynthetic processes or the need for refinement in model parameters. It should be emphasized that the relatively low \ac{S/N} $\sim$ 20 of these spectra introduce significant uncertainties of up to $\pm$0.25 dex in the derived neutron-capture element abundances, potentially affecting the interpretation of such model--data mismatches. 

In contrast, J0653+4735, J2114-0616, and J2257+3859 exhibit markedly different abundance pattern signatures that are well reproduced by $r$+\ac{$s$-process} composite models, as shown in Figure \ref{fig:CEMP-r+s abundance pattern} and \ref{fig:Other CEMP-r+s abundance pattern}. These fits yield a mean reduced chi-square value of $\chi^{2}_{\nu}$ = 1.79 (range: 1.01-2.54), with residuals generally below 0.15 dex in the Ba-Pb region (Z=56-82). Among these, J0653+4735 is particularly noteworthy, showing the best fit ($\chi^{2}_{\nu}$ = 1.01), which significantly outperforms the corresponding \ac{$i$-process} model ($\chi^{2}_{\nu}$ = 7.22), while also exhibiting the highest \ac{$r$-process} contribution fraction ($F_r$=54$\%$). These results strongly support an origin involving the mixing of \ac{AGB} \ac{$s$-process} material with \ac{$r$-process}-enriched gas, with the best-fit \ac{AGB} progenitor parameters presented in Table \ref{tab:agb_models}. The best-fit \ac{AGB} model for J0653+4735 corresponds to a 1.4 M$_\odot$ AGB star with $\rm \left [ \mathrm{Fe}/\mathrm{H} \right ] $=-2.6 and an \ac{$s$-process} efficiency of ST/9. It is crucial to recognize that these conclusions are drawn from currently available abundance data, and significant discrepancies remain between $r$+\ac{$s$-process} and \ac{$i$-process} predictions in the heavier lanthanide region (Z=68, 70; Er, Yb) and the third \ac{$r$-process} peak (Z=76-78; Os, Ir, Pt). Due to observational limitations, reliable abundance measurements in these regions are only available for Os (Z=76). Future blue-end spectroscopic observations will be essential for obtaining additional neutron-capture element abundances to further constrain these nucleosynthesis models.


Given the extreme rarity of \ac{CEMP}-$r$+$s$ stars, we further compile \ac{CEMP}-$r$+$s$ and \ac{CEMP}-$r$/$s$ stars reported in the literature and fit their abundance patterns using the same method. To ensure the robustness of the results, we select only stars that satisfy both $F_{r} > 30\%$ and $F_{s} > 30\%$, and we prioritize samples with a larger number of measured elemental abundances ($N \geq 9$).
Furthermore, we require the reduced chi-square of the fit to satisfy $\chi^{2}_{\nu} < 2.6$, which indicates that the model predictions account for the observed abundances within approximately 1.5$\sigma$. In addition, the fit must be superior to that obtained under the \ac{$i$-process} scenario. After applying these strict criteria, we have identified 12 \ac{CEMP}-$r$+$s$ candidates, including two previously confirmed cases reported in the literature: RAVE J094921.8-161722 \citep{2018ApJ...862..174G} and GIU J190734.24-315102.1 \citep{2020A&A...641A.135S}.

The identification of \ac{CEMP}-$r$+$s$ stars based solely on elemental abundance ratios remains challenging, as highlighted in previous studies \citep{2018ARNPS..68..237F, 2018ApJ...862..174G}. This difficulty largely arises from the limited sample size of currently known \ac{CEMP}-$r$+$s$ stars. Based on the only \ac{CEMP}-$r$+$s$ star known at the time, RAVE J094921.8-161722 \citep{2018ApJ...862..174G}, \citet{2018ARNPS..68..237F} proposed empirical criteria suggesting the use of 0.0 $ < \rm \left [ \mathrm{Ba}/\mathrm{Eu} \right ] < $ +0.5 and -1.0 $ < \rm \left [ \mathrm{Ba}/\mathrm{Pb} \right ] < $ -0.5 as thresholds for classifying these stars. However, our observations reveal limitations in the practical application of these criteria. Within our sample, only J0653+4735 fully satisfies the proposed conditions ($ \rm \left [ \mathrm{Ba}/\mathrm{Eu} \right ] $ = 0.09$\pm$0.12, $ \rm \left [ \mathrm{Ba}/\mathrm{Pb} \right ] $ = -0.82$\pm$0.14). Notably, J2114-0616 exhibits a marginally elevated Ba excess ($ \rm \left [ \mathrm{Ba}/\mathrm{Eu} \right ] $ = 0.62$\pm$0.11), whereas J2257+3859 shows that its $ \rm \left [ \mathrm{Ba}/\mathrm{Pb} \right ] $ ratio falls slightly below the designated range ($ \rm \left [ \mathrm{Ba}/\mathrm{Pb} \right ] $ = -1.20$\pm$0.14). Although these values deviate from the standard criteria, they remain within the 2$\sigma$ uncertainty range. Moreover, among the twelve literature-reported \ac{CEMP}-$r$+$s$ stars re-confirmed in this study, only three satisfy the criteria. These results collectively suggest that the existing empirical thresholds may not fully capture the chemical characteristics of \ac{CEMP}-$r$+$s$ stars, indicating a need for reassessment and potential revision.

To explore more robust diagnostic indicators, we perform a dispersion analysis of various heavy-element abundance ratios across the 15 \ac{CEMP}-$r$+$s$ stars from this study and the literature. The results show that the $\mathrm{[La/Nd]}$ (MAD = 0.05, std = 0.09) and $\mathrm{[La/Sm]}$ (MAD = 0.075, std = 0.09) ratios exhibit the smallest median absolute deviations and standard deviations, indicating relatively high internal consistency within the \ac{CEMP}-$r$+$s$ sample. In contrast, the $\mathrm{[Ba/Eu]}$ (MAD = 0.13, std = 0.24) and $\mathrm{[Ba/Pb]}$ (MAD = 0.18, std = 0.29) ratios display significantly larger star-to-star dispersion. However, in the $\mathrm{[La/Nd]}$ versus $\mathrm{[La/Sm]}$ diagram, \ac{CEMP}-$r$+$s$ stars do not separate clearly from \ac{CEMP}-$s$ stars, largely because the \ac{$s$-process} significantly contributes to La, Nd, and Sm, thereby diminishing the diagnostic power of these ratios. In light of this limitation, we introduce the $\mathrm{[Eu/Nd]}$ ratio (MAD = 0.08, std = 0.13 for 15 \ac{CEMP}-$r$+$s$ stars), which also shows relatively low dispersion. In solar abundances, La and Eu are representative elements of the $s$- and $r$-processes, respectively, while Nd receives comparable contributions from both processes. Our choice of these three elements is motivated not only by the relatively small dispersion observed in the abundance ratio space of \ac{CEMP}-$r$+$s$ stars, but more importantly by the fact that, in most high-resolution optical spectra of \ac{CEMP}-$r$+$s$ stars, all three elements are relatively easy to detect, thereby ensuring the reliability and comparability of the analysis. 

\begin{figure}[h]
\centering
\includegraphics[width=0.45\textwidth,keepaspectratio]{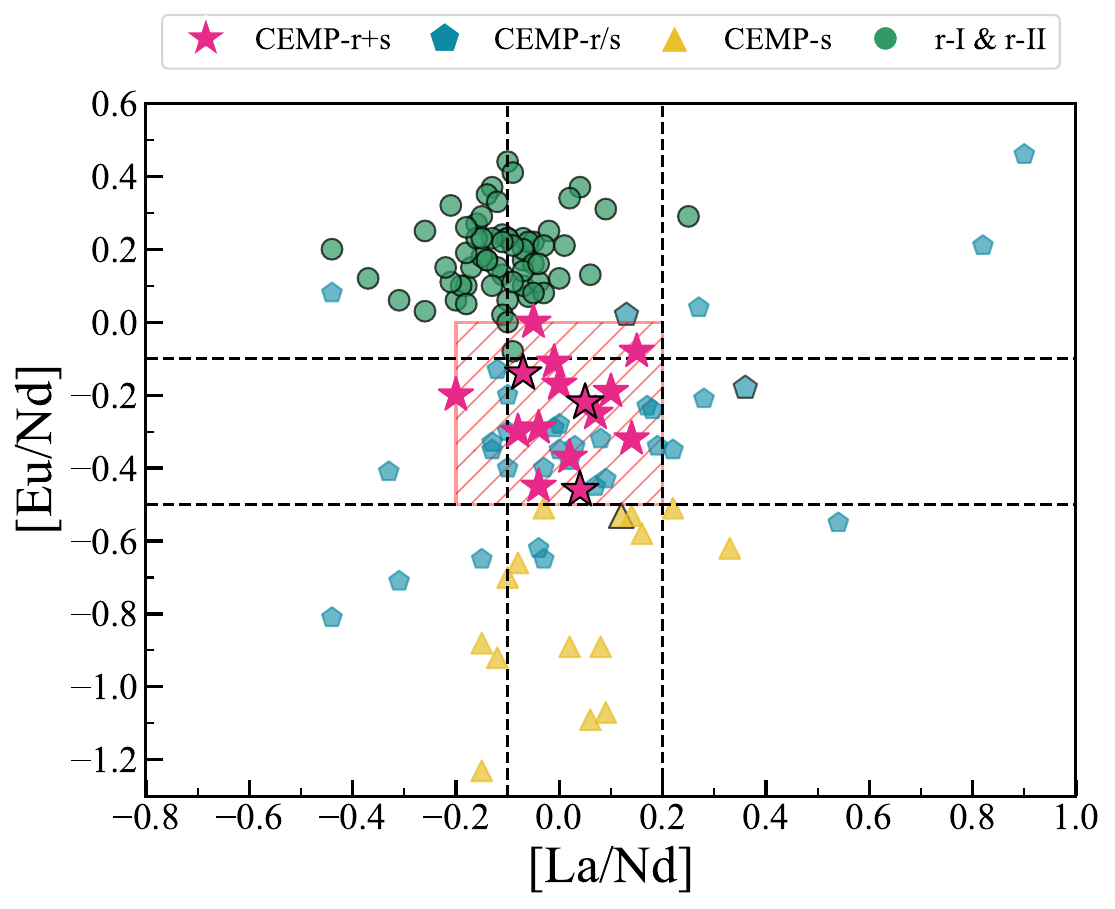}
\caption{The relation between $\mathrm{[La/Nd]}$ and $\mathrm{[Eu/Nd]}$ for different types of heavy-element enhanced stars. Red pentagrams denote \ac{CEMP}-$r$+$s$ stars, cyan pentagons denote \ac{CEMP}-$r$/$s$ stars, yellow triangles denote \ac{CEMP}-$s$ stars, and green filled circles denote $r$-I and $r$-II stars. The stars analyzed in this work are highlighted with black borders. Literature data for \ac{CEMP}-$r$+$s$, \ac{CEMP}-$r$/$s$, and \ac{CEMP}-$s$ stars are primarily taken from the compilation of \citep{2021A_A...649A..49G}, with additional contributions from \citep{2013A_A...558A..36C, 2013ApJ...770..104P, 2015ApJ...807..173H, 2023A&A...674A.180H, 2015ApJ...814..121H, 2018ApJ...862..174G, 2020A&A...641A.135S, 2021A&A...645A..61K, 2022ApJ...934..110S}. The black dashed lines outline the region where -0.1 $< \mathrm{[La/Nd]} < $ 0.2 and -0.5 $ < \mathrm{[Eu/Nd]} < $ -0.1, highlighting the concentration of \ac{CEMP}-$r$+$s$ stars. The red dashed grid indicates the region where \ac{CEMP}-$r$+$s$ stars are concentrated after accounting for the 0.1 dex uncertainties in the left boundary of $\mathrm{[La/Nd]}$ and the upper boundary of $\mathrm{[Eu/Nd]}$.
\label{fig:CEMP_rs_NdEu_NdLa}}
\end{figure}

As shown in Figure \ref{fig:CEMP_rs_NdEu_NdLa}, in the $\mathrm{[La/Nd]}$--$\mathrm{[Eu/Nd]}$ distribution of various heavy-element-enhanced stars, \ac{CEMP}-$r$+$s$ stars are mostly concentrated in the region defined by -0.1$< \mathrm{[La/Nd]} < $0.2 and -0.5$<\mathrm{[Eu/Nd]}<$-0.1. Only three stars lie on the boundary grid lines offset by 0.1 dex, corresponding to $\mathrm{[La/Nd]}$ = -0.2 and $\mathrm{[Eu/Nd]}$ = 0.0, respectively. This concentration suggests that these stars have experienced efficient mixing of $s$- and $r$-process materials, with their heavy-element abundance patterns reflecting contributions from both neutron-capture processes. 

In chemical abundance space, this region is clearly separated from the majority of \ac{CEMP}-$s$ stars and \ac{$r$-process}-enhanced stars, although a few objects are still located near the boundary. For instance, the $r$-I star J0159+1219, situated in the red shaded area, exhibits an \ac{$s$-process} contribution of $F_s$ = 25$\%$ to its surface abundances---the highest $F_s$ value among all \ac{$r$-process}-enhanced stars in our sample. However, due to its high Pr abundance, the $r$+$s$ model fails to adequately fit its surface abundances. Another $r$-I star, J1730+4143, also shows a relatively high $s$-process contribution ($F_s$ = 22$\%$). Considering the uncertainties in the boundary criteria, we propose that stars satisfying -0.2 $\leqslant \mathrm{[La/Nd]} \leqslant$ 0.2 and -0.5 $\leqslant \mathrm{[Eu/Nd]} \leqslant$ 0.0 be classified as \ac{CEMP}-$r$+$s$ stars.

However, we note a significant overlap between \ac{CEMP}-$r$/$s$ and \ac{CEMP}-$r$+$s$ stars in this parameter space, further indicating that one or a few abundance ratios alone are insufficient to unambiguously distinguish the two subclasses. More diagnostic nucleosynthetic features or independent observational constraints are still needed. Notably, we find that $\mathrm{[Eu/Nd]}$ in \ac{CEMP}-$r$/$s$ stars increases with $\mathrm{[La/Nd]}$, revealing a correlation between the \ac{$r$-process} element (Eu) and the \ac{$s$-process} element (La), which suggests that these elements may originate from the same astrophysical site. In contrast, such a correlation is not evident in \ac{CEMP}-$r$+$s$ stars. It is important to emphasize that the current classification criteria for \ac{CEMP}-$r$+$s$ remain empirical, based on available data. Future work should incorporate physical models to establish more robust standards, particularly for clearly distinguishing between \ac{CEMP}-$r$/$s$ and \ac{CEMP}-$r$+$s$ stars.

In the LAMOST/Subaru sample, the occurrence probability of \ac{$r$-process}-enhanced stars ($r$-I, $r$-II, and \ac{CEMP}-$r$) is $f_{r} = 66/385 \approx 17.1\%$, while that of \ac{$s$-process}-enhanced stars is $f_{s} = 9/385 \approx 2.3\%$. The fraction of stars that simultaneously exhibit both $r$- and $s$-process signatures is $f_{rs} = 3/385 \approx 0.8\%$, a value close to the simple product $f_{r} \times f_{s}$. This similarity suggests that the surface enhancements of $r$- and $s$-process elements in $r$+$s$ stars may originate from two independent physical events. From this interpretation, variations of radial velocities are expected because s-process excesses are attributed to the mass transfer from the companion AGB stars. The results show that, for J2114-0616, four radial-velocity observations over an eight-year baseline reveal a clear variation. It is noteworthy that this apparent variation arises from a single measurement obtained with a LAMOST low-resolution spectrum. By contrast, over a 3--4 year monitoring baseline, J0653+4735 and J0446+2124 exhibit no statistically significant variations, and thus cannot be confirmed as members of binary systems. For J1551+2040 and J2257+3859, the lack of long-term radial-velocity data from different surveys prevents any assessment of their possible binarity. Among the twelve \ac{CEMP}-$r$+$s$ stars from the literature that are re-confirmed in this work, eight exhibit significant radial-velocity variations, indicating potential binarity. These results suggest that the formation scenario of \ac{CEMP}-$r$+$s$ stars may be diverse. \citet{2006A&A...451..651J} and \citet{2021A_A...649A..49G} discuss multiple formation scenarios for \ac{CEMP}-$r$+$s$ stars. In these frameworks, the \ac{$r$-process} elements are usually attributed to the pollution of the natal gas clouds by neutron star mergers or special types of supernovae in the early Universe, whereas the origins of the \ac{$s$-process} elements are more diverse. For \ac{CEMP}-$r$+$s$ stars that exhibit clear radial-velocity variations, the surface \ac{$s$-process} enrichment is most likely acquired through mass transfer in binary systems, including Roche-lobe overflow (RLOF), wind Roche-lobe overflow (WRLOF), or stellar wind accretion \citep{2016A&A...588A...3H}. 




One possible explanation for the origin of \ac{CEMP}-$r$+$s$ stars without detection of radial velocity variation is that these stars belong to binary systems with extremely long orbital periods (on the order of $10^{3}$--$10^{4}$ days) and very low velocity amplitudes (semi-amplitude $< 100\ \mathrm{m\ s^{-1}}$). \citet{2016A&A...588A...3H} find in their orbital study of 19 \ac{CEMP}-$s$ and 3 \ac{CEMP}-$r$/$s$ stars that more than 35$\%$ of the sample have orbital periods longer than 3000 days, with HE 2201-0345 reaching as long as 10093 days. At present, however, the observational precision and time baselines of our sample are insufficient to confirm the existence of such long-period binaries, and future high-precision, long-term monitoring is essential to test this hypothesis. Moreover, in such wide binary systems, achieving sufficiently efficient mass transfer to reproduce the observed \ac{$s$-process} abundances of \ac{CEMP}-$r$+$s$ stars remains a theoretical challenge \citep{2015A&A...576A.118A}.

Another scenario assumes that the \ac{$s$-process} elements originate from stellar winds ejected by TP-AGB stars during intense thermal-pulse phases. The material, enriched in carbon and \ac{$s$-process} products, can locally pollute the surrounding interstellar medium (ISM), thereby producing significant \ac{$s$-process} enhancements on small spatial scales. However, the efficiency of this mechanism depends on several factors, including the nucleosynthetic yields and total mass ejected by TP-AGB stars, the dilution mass of the injected material, and the inhomogeneity and turbulent mixing timescales of the local ISM. If the enriched material is rapidly accreted by subsequent stars before turbulent mixing occurs (i.e., $t_{\text{acc}} \ll t_{\text{mix}}$), for example within dense clusters or local gravitational potential wells, the local enrichment can be preserved and imprinted in the chemical signatures of the next stellar generation. Conversely, if the mixing timescale is short, the enrichment signal is quickly diluted and becomes difficult to retain.  

A third possibility involves rapidly rotating, metal-poor massive stars, which in their late evolutionary stages inject \ac{$s$-process} products into the early ISM through rotation-induced mixing and stellar wind ejection \citep{2006A&A...447..623M, 2008ApJ...687L..95P, 2012A&A...538L...2F, 2016MNRAS.456.1803F, 2017A&A...607L...3C}. The models of \citet{2017A&A...607L...3C} show that rotation-induced mixing in stars of 25 $M_{\odot}$ with $v_{\mathrm{ini}} \approx 500$ km s$^{-1}$ can reproduce abundance patterns similar to those of \ac{CEMP}-$s$ stars. However, current models still fail to simultaneously account for the observed abundances of Na, Mg, and Pb, indicating that the nucleosynthesis processes in spinstars require more refined modeling and observational constraints.  

Although the above mechanisms can, in principle, provide plausible explanations for the formation of \ac{CEMP}-$r$+$s$ stars, quantitative validation requires more complex stellar evolution and nucleosynthesis models, as well as binary evolution simulations that include long orbital periods and wind accretion processes. A detailed exploration of these issues lies beyond the scope of this study.

\section{Conclusion} \label{subsec: conclusion}

This study presents a systematic investigation of heavy-element-enhanced metal-poor stars selected from the LAMOST/Subaru sample. The final sample comprises 64 $r$-process-enhanced stars (45 $r$-I and 19 $r$-II), as well as 2 limited-$r$ stars. In addition,
the sample contains 12 \ac{CEMP} stars (8 \ac{CEMP}-$s$, 1 \ac{CEMP}-$r$, 1 \ac{CEMP}-$r$/$s$, 2 \ac{CEMP}-$r$+$s$), alongside 4 moderately carbon-enhanced stars (0.4 $ < \rm \left [ \mathrm{C}/\mathrm{Fe} \right ] < $ 0.7), which exhibit heavy-element abundance patterns consistent with those of the $s$, $r$, $r$/$s$, and $r$+$s$ subclasses, respectively. The identification of these heavy-element-enhanced metal-poor stars, combined with a homogeneous analysis of this statistically significant sample, provides crucial observational constraints on nucleosynthetic processes in the early Galaxy. 

The abundance patterns of \ac{$r$-process}-enhanced stars reveal significant deviations of light neutron-capture elements (Sr, Y, Zr) from the solar \ac{$r$-process} abundance pattern, exhibiting pronounced star-to-star scatter exceeding 1 dex. Notably, Y and Zr exhibit a tight correlation with heavy \ac{$r$-process} elements (\ac{MAD} $\sim$ 0.13 dex), while Sr displays substantially greater dispersion (\ac{MAD} $>$ 0.20 dex). In contrast, heavy neutron-capture elements (56 $\leq$ Z $\leq$ 76) generally align well with the solar \ac{$r$-process} pattern, with several notable exceptions. Pr shows anomalous enhancement ($\mathrm{[Pr/Eu]} \approx 0.47$) in eight \ac{$r$-process}-enhanced stars, potentially reflecting unidentified mixing effects or additional nucleosynthetic channels. Ba exhibits a marked departure from the solar \ac{$r$-process} pattern, suggesting that the sample stars may experience varying degrees of \ac{$s$-process} contamination ($ F_s \leq 0.25$). Furthermore, cluster analysis reveals that \ac{$s$-process} signatures become increasingly prominent at $\mathrm{[Fe/H]} > $ -2.6.

A comparison of $\mathrm{[X/Eu]}$ abundance ratios reveals systematic differences between $r$-I and $r$-II stars. Specifically, $r$-I stars exhibit significantly elevated $\mathrm{[X/Eu]}$ ratios for light neutron-capture elements (e.g., Sr, Y, Zr), with an average offset of $\sim$ 0.36 dex, whereas for heavy neutron-capture elements (Z $\geqslant$ 56), the discrepancy between $r$-I and $r$-II stars diminishes to $\sim$ 0.1 dex and decreases further with increasing atomic number. Dynamical evidence indicates that $r$-II stars with higher $\mathrm{[Eu/Fe]}$ are more likely to have originated from low-mass dwarf galaxies accreted by the Milky Way. Consequently, they experienced less contamination from additional nucleosynthetic processes---particularly the weak \ac{$r$-process}---and thus retained a more "pure" signature of the main \ac{$r$-process}.



Among \ac{$r$-process}-enhanced stars, we identify an \ac{EMP} ($\mathrm{[Fe/H]}$ = -2.93) $r$-II star, J1158+0734, which exhibits significant Zn enrichment ($\mathrm{[Zn/Fe]}$ = 0.67). Its light-element abundance pattern is best reproduced by the yields of a \ac{CCSN}  with a progenitor mass of $31.5\,M_\odot$ and an explosion energy of $3 \times 10^{51}$\,erg.
The probability density distribution of its progenitor properties also favors high-energy, massive supernovae, indicating that such events efficiently synthesize Zn in the early Galactic environment. At the same time, the theoretical nucleosynthesis yields from NSMs and \ac{MRDSNe} are both able to reproduce the observed neutron-capture element abundance pattern of J1158+0734 reasonably well. However, given the substantial uncertainties that persist in current \ac{$r$-process} models, together with the extremely low metallicity of this star and its significantly enhanced Zn abundance, we favor \ac{MRDSNe} as the predominant origin of its surface heavy elements.



The \ac{CEMP}-$s$ stars in our sample show significant dispersion in their \ac{$s$-process} element abundances. This variation likely originates from enrichment by a range of AGB progenitors with differing properties, including masses of 1.3--2.0\,M$_\odot$, metallicities of -3.0 to -2.3, and \ac{$s$-process} efficiencies from ST/24 to ST. Among these stars, J2256+0215 is particularly notable, as its surface neutron-capture abundance pattern is well reproduced by an AGB model with a progenitor mass of  $2.0\,M_\odot$, $\mathrm{[Fe/H]}=-2.6$ and an \ac{$s$-process} efficiency of ST/24. However, this star shows an anomalously low carbon abundance ($\mathrm{[C/Fe]}$ = 0.08), and its extremely low $^{12}\mathrm{C}/^{13}\mathrm{C}$ ratio of 2.33 provides strong evidence that additional mixing occurs during its evolution. Such mixing is likely driven by thermohaline instability and may be further enhanced through rotation-induced or magnetically mediated mechanisms.

In our sample, five stars exhibit simultaneous enhancements of both \ac{$s$-process} and \ac{$r$-process} elements, accompanied by moderate to significant carbon enrichment. Among them, three stars show heavy-element abundance patterns that are well reproduced by a combination of solar \ac{$r$-process} and \ac{AGB} \ac{$s$-process} contributions, indicating that their enrichment is more consistent with $r$+$s$ dual contributions rather than the \ac{$i$-process}. Using the same analysis approach, we re-confirm twelve \ac{CEMP}-$r$+$s$ candidates previously reported in the literature. Taken together with our results, nine stars display significant radial-velocity variations, supporting the interpretation that their carbon and \ac{$s$-process} elements originate from binary mass transfer. However, six \ac{CEMP}-$r$+$s$ stars show no detectable radial-velocity variations, suggesting alternative enrichment mechanisms such as long-period binary systems, local pollution from nearby \ac{AGB} winds, or \ac{$s$-process} nucleosynthesis in rapidly rotating massive stars. Nevertheless, high-precision, long-baseline radial-velocity monitoring remains essential for constraining their formation pathways. Based on the available abundance data, we propose that $\mathrm{[La/Nd]}$ and $\mathrm{[Eu/Nd]}$ serve as more stable diagnostic ratios for identifying \ac{CEMP}-$r$+$s$ stars, and we recommend a revised definition: stars satisfying -0.2 $\leqslant$ $\mathrm{[La/Nd]}$ $\leqslant$ 0.2 and -0.5 $\leqslant$ $\mathrm{[Eu/Nd]}$ $\leqslant$ 0.0 could be classified as \ac{CEMP}-$r$+$s$. Even so, the significant overlap with \ac{CEMP}-$r$/$s$ stars in this parameter space demonstrates that abundance ratios alone are insufficient for a clear observational distinction between the two subclasses, and future work must incorporate additional nucleosynthetic or dynamical diagnostics to establish more robust classification criteria.

In the future, with the continued progress of ongoing spectroscopic surveys such as LAMOST, GALAH, Gaia-ESO, and APOGEE, together with forthcoming next-generation projects including 4MOST, WEAVE, and SDSS-V, the availability of larger stellar samples will enable the systematic identification and investigation of heavy-element-enhanced stars, particularly extremely \ac{$r$-process} enhanced stars and \ac{CEMP}-$r$+$s$ stars. At the same time, detailed analyses of high-resolution ultraviolet–blue spectra for these peculiar stars (to be presented in forthcoming work) will provide more complete heavy-element abundance patterns and place critical constraints on the nucleosynthetic origins of their surface compositions. By combining large-scale survey data with targeted high-resolution follow-up observations, this research program will establish a solid foundation for advancing our understanding of the early chemical evolution of the Galaxy and the astrophysical sites of multiple neutron-capture processes.

\section{Acknowledgments}
We thank the anonymous referee for the constructive comments and efficient review, which have significantly improved the manuscript.
This work was supported by the National Key $R\&D$ Program of China Nos. 2024YFA1611903 and 2023YFE0107800, the National Natural Science Foundation of China grant No. 12222305, the Strategic Priority Research Program of Chinese Academy of Sciences grant No. XDB1160103, the International Partnership Program of Chinese Academy of Sciences grant No. 178GJHZ2022040GC, and the science research grants from the China Manned Space Project, the CAS Project for Young Scientists in Basic Research No. YSBR-092. This work was also supported by JSPS KAKENHI grant Nos. 20H05855, 21H04499, 22K03688, and 23HP8014.

This research is based on data collected at Subaru Telescope, which is operated by the National Astronomical Observatory of Japan. We are honored and grateful for the opportunity of observing the Universe from Maunakea, which has cultural, historical, and natural significance in Hawaii. Guoshoujing Telescope (the Large Sky Area Multi-Object Fiber Spectroscopic Telescope, LAMOST) is a National Major Scientific Project built by the Chinese Academy of Sciences. Funding for the project has been provided by the National Development and Reform Commission. It is operated and managed by the National Astronomical Observatories, Chinese Academy of Sciences. This work has made use of data from the European Space Agency (ESA) mission Gaia (\href{https://www.cosmos.esa.int/gaia}{https://www.cosmos.esa.int/gaia}), processed by the Gaia Data Processing and Analysis Consortium (DPAC, \href{https://www.cosmos.esa.int/web/gaia/dpac/consortium}{https://www.cosmos.esa.int/web/gaia/dpac/consortium}).   Funding for the DPAC has been provided by national institutions, in particular the institutions participating in the Gaia Multilateral Agreement. This work has made use of the VALD database (\href{https://vald.astro.uu.se/vald.php}{https://vald.astro.uu.se/vald.php}), operated at Uppsala University, the Institute of Astronomy RAS in Moscow, and the University of Vienna.

\vspace{5mm}
\facilities{LAMOST, Subaru}

\software{MOOG \citep{1973PhDT.......180S},  
          TAME \citep{2012MNRAS.425.3162K}, 
          Linemake \citep{2021RNAAS...5...92P},
          iSpec \citep{2014A&A...569A.111B},
          Agama \citep{2019MNRAS.482.1525V},
          TOPCAT \citep{2005ASPC..347...29T}
          }
          
\appendix
\section{Appendix information}
\setcounter{figure}{0}
\renewcommand{\thefigure}{A\arabic{figure}}
\setcounter{table}{0}
\renewcommand{\thetable}{A\arabic{table}}

\begin{figure*}[t]
    \begin{minipage}{\linewidth}
        \centering
        \includegraphics[width=0.48\linewidth]{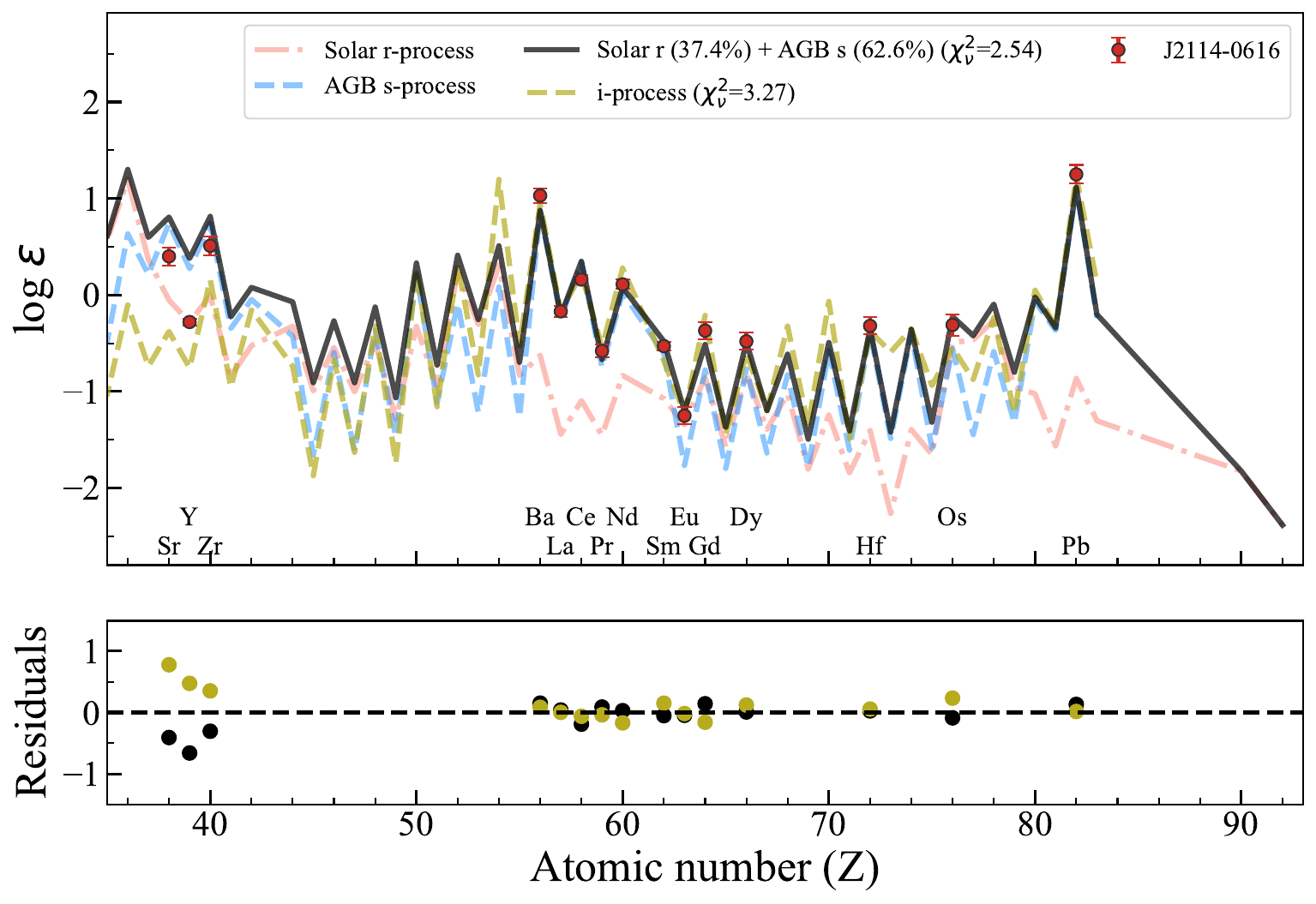}
        \hspace{0.1cm} 
        \includegraphics[width=0.48\linewidth]{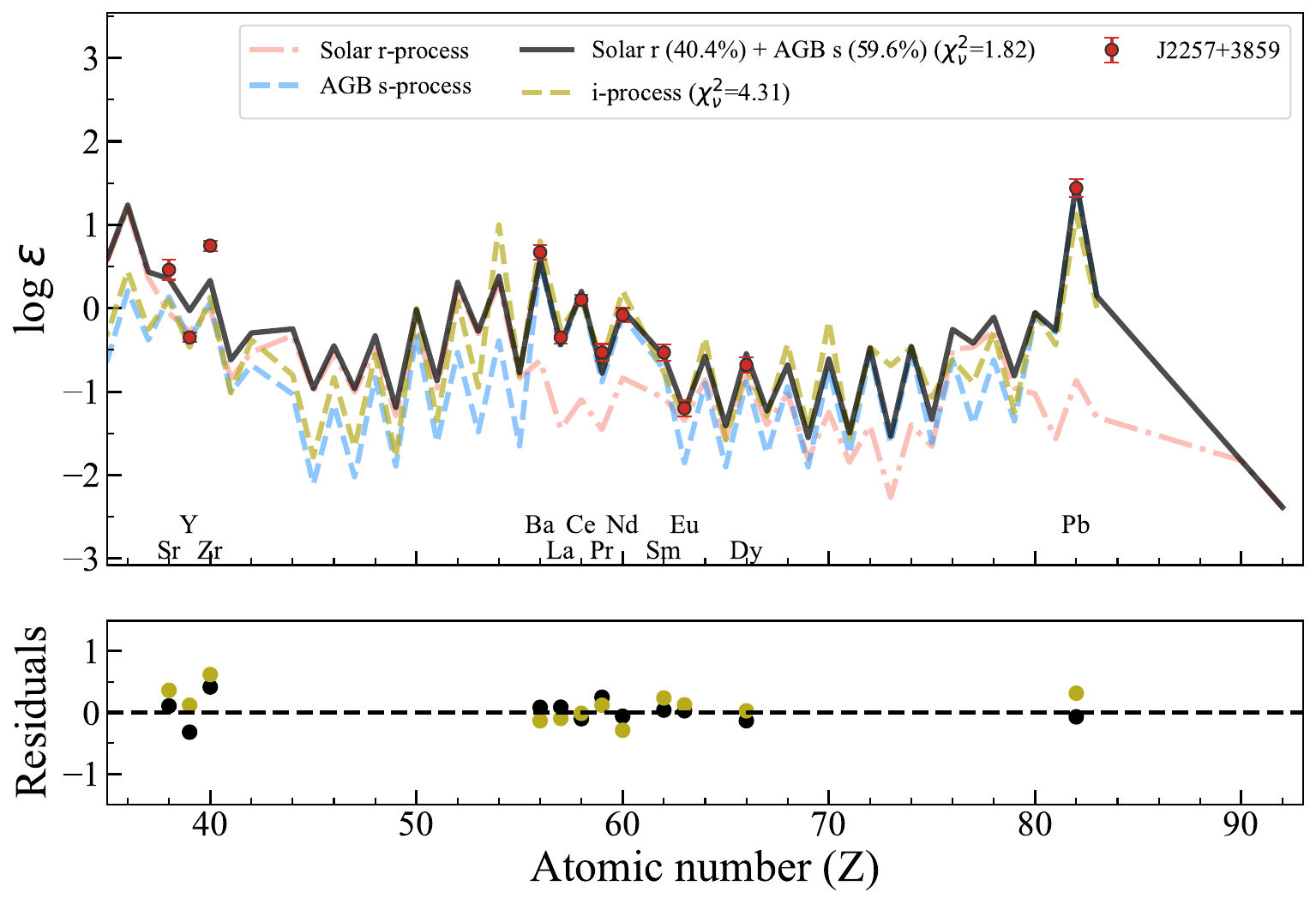}
    \end{minipage}
    \caption{The abundance pattern of two \ac{CEMP}-$r$+$s$ stars J2114-0616 and J2257+3859 in our sample, with symbols and line styles following the same conventions as in Figure \ref{fig:CEMP-r/s abundance pattern}.
    }
    \label{fig:Other CEMP-r+s abundance pattern}
\end{figure*}

Table \ref{tab:rv_measurements} summarizes the heliocentric radial velocities of the \ac{CEMP} stars compiled from various sources. Figure \ref{fig:Other CEMP-r+s abundance pattern} presents the abundance patterns of two additional \ac{CEMP}-$r$+$s$ stars in our sample. 

\begin{table}[ht]
\centering
\footnotesize
\caption{Heliocentric radial velocities of CEMP stars from different surveys and high-resolution spectroscopic measurements in the literature.}
\label{tab:rv_measurements}
\begin{tabular}{@{}lcccl@{}}
\toprule
Name & Observation Date & $V_{\rm helio}$ (km\,s$^{-1}$) & Class & Source \\
\midrule
\multirow{3}{*}{J0006+0123} 
  & 2016/11/19 & $-65.90 \pm 0.20$ & CEMP-$r$ & (1) \\
  & 2012/10/12 & $-68.60 \pm 3.85$ & CEMP-$r$ & (2) \\
  & 2016/01/01 & $-63.80 \pm 3.10$ & CEMP-$r$ & (3) \\
\midrule
\multirow{2}{*}{J2333+0505} 
  & 2016/11/16 & $ -136.9 \pm 0.20 $ & CEMP*-$r$ & (1) \\
  & 2016/01/01 & $ -137.18 \pm 0.66 $ & CEMP*-$r$ & (3) \\
\midrule
\end{tabular}
\begin{minipage}{\linewidth}
\vspace{0.5em}
\footnotesize
References: (1) Subaru high-resolution spectroscopy \citep{2022ApJ...931..146A}; (2) LAMOST DR11; (3) Gaia DR3; (4) RAVE DR6; (5) APOGEE DR17

(This table is available in its entirety in machine-readable form.)
\end{minipage}
\label{tab:rv_data}
\end{table}

\bibliography{reference}{}
\bibliographystyle{aasjournal}
\end{document}